\documentclass[twoside,twocolumn,9pt]{article}
\usepackage{extsizes}
\usepackage[super,sort&compress,comma]{natbib} 
\usepackage[version=4]{mhchem}
\usepackage{chemformula}
\usepackage[left=1.5cm, right=1.5cm, top=1.785cm, bottom=2.0cm]{geometry}
\usepackage{balance}
\usepackage[mathscr]{euscript}
\usepackage{times,mathptmx}
\usepackage{dblfloatfix}
\usepackage{sectsty}
\usepackage{graphicx} 
\usepackage{lastpage}
\usepackage[format=plain,justification=justified,singlelinecheck=false,font={stretch=1.125,small,sf},labelfont=bf,labelsep=space]{caption}
\usepackage{float}
\usepackage{fancyhdr}
\usepackage{fnpos}
\usepackage{dirtytalk}
\usepackage[english]{babel}
\usepackage[export]{adjustbox}% http://ctan.org/pkg/adjustbox
\addto{\captionsenglish}{%
  
}
\usepackage{array}
\usepackage{amsmath}
\usepackage{amssymb}
\usepackage{multirow}
\usepackage{droidsans}
\usepackage{charter}
\usepackage[T1]{fontenc}
\usepackage{setspace}
\usepackage[compact]{titlesec}
\usepackage{hyperref}
\usepackage{blindtext}
\usepackage{xcolor}

\definecolor{mypink3}{cmyk}{0, 0.7808, 0.4429, 0.1412}
\usepackage{epstopdf}%This line makes .eps figures into .pdf - please comment out if not required.

\definecolor{cream}{RGB}{222,217,201}

\begin{document}

\pagestyle{fancy}
\thispagestyle{plain}
%\fancypagestyle{plain}{

%%%HEADER%%%
%\fancyhead[C]{\includegraphics[width=18.5cm]{head_foot/header_bar}}
%\fancyhead[L]{\hspace{0cm}\vspace{1.5cm}\includegraphics[height=30pt]{head_foot/journal_name}}
%\fancyhead[R]{\hspace{0cm}\vspace{1.7cm}\includegraphics[height=55pt]{head_foot/RSC_LOGO_CMYK}}
%\renewcommand{\headrulewidth}{0pt}
%}
%%%END OF HEADER%%%

%%%PAGE set-up - Please do not change any commands within this section%%%
\makeFNbottom
\makeatletter
\renewcommand\LARGE{\@setfontsize\LARGE{15pt}{17}}
\renewcommand\Large{\@setfontsize\Large{12pt}{14}}
\renewcommand\large{\@setfontsize\large{10pt}{12}}
\renewcommand\footnotesize{\@setfontsize\footnotesize{7pt}{10}}
\makeatother
% Paragraph formatting - VAZE
\setlength{\parindent}{5ex}
\setlength{\parskip}{1em}

\renewcommand{\thefootnote}{\fnsymbol{footnote}}
\renewcommand\footnoterule{\vspace*{1pt}% 
\color{cream}\hrule width 3.5in height 0.4pt \color{black}\vspace*{5pt}} 
\setcounter{secnumdepth}{5}

\makeatletter 
\renewcommand\@biblabel[1]{#1}            
\renewcommand\@makefntext[1]% 
{\noindent\makebox[0pt][r]{\@thefnmark\,}#1}
\makeatother 
\renewcommand{\figurename}{\small{Fig.}~}
\sectionfont{\sffamily\Large}
\subsectionfont{\normalsize}
\subsubsectionfont{\bf}
\setstretch{1.125} %In particular, please do not alter this line.
\setlength{\skip\footins}{0.8cm}
\setlength{\footnotesep}{0.25cm}
\setlength{\jot}{10pt}
\titlespacing*{\section}{0pt}{4pt}{4pt}
\titlespacing*{\subsection}{0pt}{15pt}{1pt}
%%%END OF PAGE set-up%%%

%%%FOOTER%%%
\fancyfoot{}
\fancyfoot[LO,RE]{\vspace{-7.1pt}\includegraphics[height=9pt]{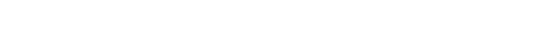}}
\fancyfoot[CO]{\vspace{-7.1pt}\hspace{13.2cm}\includegraphics{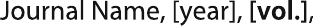}}
\fancyfoot[CE]{\vspace{-7.2pt}\hspace{-14.2cm}\includegraphics{head_foot/RF}}
\fancyfoot[RO]{\footnotesize{\sffamily{1--\pageref{LastPage} ~\textbar  \hspace{2pt}\thepage}}}
\fancyfoot[LE]{\footnotesize{\sffamily{\thepage~\textbar\hspace{3.45cm} 1--\pageref{LastPage}}}}
\fancyhead{}
\renewcommand{\headrulewidth}{0pt} 
\renewcommand{\footrulewidth}{0pt}
\setlength{\arrayrulewidth}{1pt}
\setlength{\columnsep}{6.5mm}
\setlength\bibsep{1pt}
%%%END OF FOOTER%%%

%%%FIGURE set-up - please do not change any commands within this section%%%
\makeatletter 
\newlength{\figrulesep} 
\setlength{\figrulesep}{0.5\textfloatsep} 

\newcommand{\topfigrule}{\vspace*{-1pt}% 
\noindent{\color{cream}\rule[-\figrulesep]{\columnwidth}{1.5pt}} }

\newcommand{\botfigrule}{\vspace*{-2pt}% 
\noindent{\color{cream}\rule[\figrulesep]{\columnwidth}{1.5pt}} }

\newcommand{\tj}[6]{ \begin{pmatrix}
#1 & #2 & #3 \\
#4 & #5 & #6 
\end{pmatrix}}

\newcommand{\dblfigrule}{\vspace*{-1pt}% 
\noindent{\color{cream}\rule[-\figrulesep]{\textwidth}{1.5pt}} }

\makeatother
%%%END OF FIGURE set-up%%%

%%%TITLE, AUTHORS AND ABSTRACT%%%
\twocolumn[
  \begin{@twocolumnfalse}
\vspace{3cm}
\sffamily
\begin{tabular}{m{4.5cm} p{13.5cm} }

& \noindent\LARGE{\textbf{Multipole-moment effects in ion-molecule reactions at low
temperatures: Part I - Ion-dipole enhancement of the rate coefficients of the \ce{He+ + NH3} and \ce{He+ + ND3} reactions at collisional energies $E_{\mathrm{coll}}/k_{\mathrm{B}}$ near 0 K}} \\%Article title goes here instead of the text "This is the title"
\vspace{0.3cm} & \vspace{0.3cm} \\

 & \noindent\large{Valentina Zhelyazkova, Fernanda B. V. Martins, Josef A. Agner, Hansj\"{u}rg Schmutz and Fr\'{e}d\'{e}ric Merkt$^{\ast}$}
\\%Author names go here instead of "Full name", etc.

& \noindent\normalsize{The energy dependence of the rates of the reactions between He$^+$ and ammonia (NY$_3$, Y= \{H,D\}), forming NY$_2^+$,  Y and He as well as NY$^+$, Y$_2$ and He, and the corresponding product branching ratios have been measured at low collision energies $E_{\rm coll}$ between 0 and $k_{\mathrm{B}}\cdot40$~K using a recently developed merged-beam technique [Allmendinger {\it et al.}, ChemPhysChem {\bf 17}, 3596 (2016)]. To avoid heating of the ions by stray electric fields, the reactions are observed within the large orbit of a highly excited Rydberg electron. A beam of He Rydberg atoms was merged with a supersonic beam of ammonia using a curved surface-electrode Rydberg-Stark deflector, which is also used for adjusting the final velocity of the He Rydberg atoms, and thus the collision energy. A collision-energy resolution of about 200~mK was reached at the lowest $E_{\rm coll}$ values. The reaction rate coefficients exhibit a sharp increase at collision energies below $\sim k_{\mathrm{B}}\cdot5$ K and pronounced deviations from Langevin-capture behaviour. The experimental results are interpreted in terms of an adiabatic capture model describing the rotational-state-dependent orientation of the ammonia molecules by the electric field of the He$^+$ atom. The model faithfully describes the experimental observations and enables the identification of three classes of $|JKMp\rangle$ rotational states of the ammonia molecules showing different low-energy capture behaviour: (A) high-field-seeking states with $|KM|\ge 1$ correlating to the lower component of the umbrella-motion tunnelling doublet at low fields. These states undergo a negative linear Stark shift, which leads to strongly enhanced rate coefficients; (B) high-field-seeking states subject to a quadratic Stark shift at low fields and which exhibit only weak rate enhancements; and (C) low-field-seeking states with $|KM|\ge 1$. These states exhibit a positive Stark shift at low fields, which completely suppresses the reactions at low collision energies. Marked differences in the low-energy reactivity of NH$_3$ and ND$_3$ --- the rate enhancements in ND$_3$ are more pronounced than in NH$_3$ --- are quantitatively explained by the model.  They result from the reduced magnitudes of the tunnelling splitting and rotational intervals in ND$_3$ and the different occupations of the rotational levels in the supersonic beam caused by the different nuclear-spin statistical weights. Thermal capture rate constants are derived from the model for the temperature range between 0 and 10~K relevant for astrochemistry. Comparison of the calculated thermal capture rate coefficients with the absolute reaction rates measured above 27~K by Marquette {\it et al.} (Chem. Phys. Lett., 1985, {\bf 122}, 431) suggests that only 40\% of the close collisions are reactive.}
\end{tabular}

 \end{@twocolumnfalse} \vspace{0.6cm}

  ]

%%%FONT set-up - please do not change any commands within this section
\renewcommand*\rmdefault{bch}\normalfont\upshape
\rmfamily
\section*{}
\vspace{-1cm}

\section{Overview of the article series}

This article is the first of a three-article series devoted to experimental studies of barrier-free ion-molecule reactions at low collision energies. The article series addresses fundamental aspects of chemical reactivity in low-energy ion-molecule collisions which play a crucial role in the understanding and modelling of ion-molecule reactions at low temperatures.\cite{bowers79a,ausloos78a,clary85a,troe87a,clary90a,troe96a,ng92a,gerlich08a,willitsch08a,heazlewood21a} The focus is placed on the measurement and theoretical analysis of the deviations of the collision-energy-dependent reaction rates $k(E_{\rm coll})$ from the Langevin rate $k_{\rm L}$ at very low collision energies resulting from the electrostatic interaction between the charge of the ion and the electric dipole, quadrupole and octupole moments of the neutral molecule. 

For these studies, we have chosen He$^+$ as a reactant ion because the high ionisation energy of He guarantees strongly exothermic barrier-free reactions with a broad range of neutral molecules. The neutral molecules we have selected are (i) ammonia (NH$_3$ and ND$_3$), which is a polar molecule with a dipole moment of $4.91\times 10^{-30}$~C\,m,\cite{marshall97a} (ii) N$_2$, which does not have a permanent dipole moment but has a quadrupole moment of $-4.65\times 10^{-40}$~C\,m$^2$,\cite{graham98} and (iii) methane (CH$_4$ and CD$_4$), which has neither a permanent dipole moment nor a quadrupole moment, but an octupole moment of $7.41\times 10^{-50}$ C\,m$^3$.\cite{birnbaum75} The reactions of all three molecules are known from previous work to proceed with high rates.\cite{albritton78,anicich86} With these three molecules, we can examine the convergence of the capture rate coefficients at low energies towards the Langevin rate as the lowest non-vanishing member of the multipole expansion series of the neutral molecule gradually increases in the sequence dipole ($n_{\rm mp}=2$), quadrupole ($n_{\rm mp}=4$), and octupole ($n_{\rm mp}=8$). This sequence also corresponds to the order of this three-article series, which presents our results on the $n_{\rm mp}=2$ case in article I, and those on the $n_{\rm mp}=4$ and $n_{\rm mp}=8$ cases in articles II and III, respectively. 

The experimental observations are analysed in the realm of an adiabatic capture model inspired by the earlier theoretical work of Clary and co-workers~\cite{clary85a,clary90a,stoecklin92a} and Troe and co-workers.\cite{troe87a,troe96a,dashevskaya05a,auzinsh13a,auzinsh13b,dashevskaya16a} Because the existence or non-existence of a given $n_{\rm mp}$-pole moment is dictated by the molecular symmetry, the interpretation of the experimental observations requires consideration of the corresponding molecular-symmetry groups, in particular when determining the nuclear-spin symmetries and occupations of the rotational quantum states of the neutral molecules. We follow here earlier treatments, in particular those presented in Refs.~\cite{hougen76a,bunker06,snels00,wichmann20} and compare reactions of undeuterated and fully deuterated ammonia and methane to highlight specific aspects. The overall analysis gives insights into the dependence of the rate coefficients on the rotational quantum states of the neutral molecules and provides access to the thermal rate coefficients $k(T)$ through statistical averaging.

The overall structure of this three-article series is as follows: In this first article, we present a general introduction to fast ion-molecule reactions and detail the experimental approach and set-up we have used, which are the same for all three studies. We also summarise the main aspects of the capture model we have implemented to interpret the experimental results in a general form, adequate to treat the effects of an arbitrary $n_{\rm mp}$ moment. We then describe the experimental results we have obtained in our study of the He$^+$~+~NH$_3$ and  He$^+$~+~ND$_3$ reactions and their analysis using the capture model for the $n_{\rm mp}=2$ case. In the second article, we focus on the results obtained on the $n_{\rm mp}=4$ case and present our studies of the He$^+$~+~N$_2$ reaction. To clarify the role of the sign of the quadrupole, we also compare the results obtained for N$_2$, which has a negative quadrupole moment, with those obtained for the reactions of D$_2^+$ with H$_2$, which has positive quadrupole moment. The last article is devoted to the study of the $n_{\rm mp}=8$ case with the example of the He$^+$~+~CH$_4$ and  He$^+$~+~CD$_4$ reactions. It also presents our main conclusions concerning the convergence of the low-energy capture rate coefficients towards Langevin rates. This structure offers the advantage that the introductory, experimental and theoretical-modelling sections of articles II and III could be reduced to a minimum, and that the general conclusions could be condensed in the last section of article III.

\section{Introduction}
 
Ion-molecule reactions play a key role in the chemistry of dilute environments, such as low-density plasmas, interstellar clouds and the upper layers of planetary atmospheres. Their rate coefficients and the branching ratios for different product channels are required as input in global kinetics models designed to describe and predict the chemical compositions of these environments.\cite{herbst73a,roueff87,smith92a,herbst01a,snow08a,smith11a,wakelam10a,dishoeck17} In particular,   
barrierless, exothermic ion-molecule reactions proceed with high rate coefficients even at very low temperatures and several of these reactions represent essential steps in the synthesis of molecules in the interstellar medium, enabling the formation of polyatomic molecules through reaction chains in cold ($3-50$~K) dark molecular clouds. 

Measurements of the rates of ion-molecule reactions at temperatures below 50~K are challenging because ions are easily heated up by stray electric fields and space-charge effects. An electric-potential difference of only 1 mV is indeed sufficient to accelerate ions to kinetic energies corresponding to about 12~K. Studies of ion-molecule reactions in ion guides and traps~\cite{gerlich08a,wester09a,markus20a} and in supersonic flows~\cite{marquette85a,rowe95a} enable measurements of reaction rates down to $\approx 10$~K, but the range below 10~K remains largely unexplored experimentally. 

Advances in the experimental studies of ion-molecule reactions near 0~K have been recently made that rely on the use of laser-cooled and sympathetically cooled ions in ion traps and Coulomb crystals.\cite{willitsch08a,willitsch17a,petralia20a,toscano20} Our own approach to study ion-molecule reactions at low energies consists of suppressing the stray-electric-field-induced heating of the ions  by replacing the ion by an atom or molecule in a Rydberg state of high principal quantum number $n$.\cite{allmendinger16a,allmendinger16b,zhelyazkova20,hoeveler21a} In such states, the Rydberg electron moves on an orbit of large radius ($\propto n^2$, e.g., $\approx 50$~nm for $n= 30$), without significantly affecting the reactions involving the ion core.\cite{pratt94a,wrede05a,matsuzawa10a} The distant Rydberg electron, however, effectively shields the reaction from stray electric fields. In addition, one can exploit the large induced dipole moments of Rydberg atoms and molecules ($\mu_{\mathrm{max}}\approx3400$ D for $n=30$) to manipulate their translational motion with inhomogeneous electric fields of modest magnitude. 
We have used this property to deflect, accelerate and decelerate beams of Rydberg atoms and molecules using on-chip Rydberg-Stark decelerators and deflectors.\cite{hogan12b,allmendinger14a,hogan16a,zhelyazkova19a} With such devices, supersonic beams of Rydberg atoms and molecules can be merged with supersonic beams of ground-state neutral molecules, allowing the study of ion-molecule reactions at very low collisional energies. 

This merged Rydberg-packet--ground-state-beam approach exploits the large velocity dispersion of supersonic beams produced with short-pulse valves, as in recent elegant work on Penning-ionisation and associative-ionisation reactions at low collision energies by Narevicius and co-workers~\cite{shagam13a,shagam15a,bibelnik19a} and Osterwalder and co-workers,\cite{jankunas14a,jankunas15c,gordon20a} and the well-defined spatial extent and mean velocity of the deflected Rydberg-atom cloud. This combination makes it possible to achieve a collision-energy resolution as low as $\sim 100$ mK.\cite{hoeveler21a} In the last years, we have observed deviations from Langevin capture rates at low collisional energies, in particular: (a) a small enhancement ($\sim15$\%) of the $\mathrm{H}_2^+ + \mathrm{H}_2$ reaction rate constant compared to $k_{\mathrm{L}}$ at $E_{\mathrm{coll}}/k_\mathrm{B}<1$ K,\cite{allmendinger16b} and (b) a strong enhancement by more than an order of magnitude of the $\mathrm{He}^++\mathrm{CH}_3\mathrm{F}$ capture rate constant at the lowest collisional energies.\cite{zhelyazkova20} These enhancements were attributed to ion-quadrupole interactions for the reaction between H$_2^+$ and ortho-H$_2$ molecules in the $J = 1$ rotational state, and  to rotational-state-dependent Stark shifts of the polar CH$_3$F molecules in the Coulomb field of the helium ion for the $\mathrm{He}^++\mathrm{CH}_3\mathrm{F}$ reaction. 
Despite these advances, experimental data on the rates of ion-molecule reactions in the range below 10~K remain extremely scarce. 

In contrast, theoretical models describing ion-molecule reactions are advanced and represent the main source of rate constants for these reactions at low temperatures. 
At low temperatures, the dominant ion-molecule reactions are barrier-free reactions and their rates can be modelled by ion-neutral capture models. In the simplest such model, introduced by Langevin,\cite{langevin05a} the rates are determined by considering the charge-induced dipole interaction between the ion and the polarisable molecule and it is assumed that the reaction takes place with unit probability provided that the collision energy is larger than the centrifugal barrier in the effective intermolecular potential. In this regime, the rate constants are independent of the temperature or the collision energy $E_{\rm coll}$ and are given by the Langevin rate constant (in SI units)
\begin{equation}\label{eq_langevin}
k_{\rm L}=\sigma v_{\rm rel} =  2\sqrt{\frac{\pi^2\alpha^\prime (Ze)^2}{4\pi\epsilon_0\mu}}.
\end{equation}
In Eq.~(\ref{eq_langevin}), $\sigma$ is the reaction cross-section, $ v_{\rm rel}= \sqrt{2E_{\rm coll}/\mu}$ is the asymptotic relative velocity, $\alpha'$ is the polarisability volume of the neutral molecule, $Ze$ is the charge of the ion, $\mu$ is the reduced mass of the reactants and $\epsilon_0$ is the permittivity of vacuum.
Whereas this treatment is often adequate above 100~K, it fails at low temperatures because long-range forces between the charge of the ion and the dipole, quadrupole, octupole, etc. moments of the neutral molecules affect their rotational motion.\cite{bowers79a,ausloos78a,ng92a} Specifically, the orientation of the molecular axes and the alignment of the rotational angular-momentum vector tend to be locked to the collision axis, leading to strong quantum-state(index $i$)- and collision-energy-dependent rate coefficients $k_i(E_{\rm coll})$.\cite{clary90a,troe87a,troe96a} In this range, thermal capture rate constants are determined from the $k_i(E_{\rm coll})$ values by averaging over the thermally populated translational and rovibrational states $i$ of the reactants.

This article summarises the results of our studies of the reaction of \ce{He+} with \ce{NH3} and \ce{ND3}. These reactions are of astrophysical importance. Ammonia was the first polyatomic molecule to be detected in the interstellar medium.\cite{cheung68} \ce{NH3} and several of its isotopologues (\ce{NH2D}, \ce{ND2H} and \ce{ND3} and \ce{^{15}NH3}) were later found in external galaxies and interstellar ices, proto-planetary disks, and in the atmosphere of Jupiter.\cite{roueff05,hermsen85,martin79,salinas16,irwin18} Ammonia can be detected through transitions in the radio-frequency (transitions between the tunnelling components of the umbrella-inversion mode), microwave (rotational transitions), IR (vibrational transitions) and the visible/UV (electronic transitions) ranges of the electromagnetic spectrum. The ease of its detection makes it one of the most important molecules for the study of molecular clouds (ammonia is dubbed "the interstellar thermometer")\cite{danby88}. Helium is the second most abundant element in the universe ($\sim25$\% by mass)\cite{izotov10} and can be ionised by cosmic rays in the interstellar medium to form He$^+$.

The reaction between \ce{He+} and \ce{NH3} has been studied in the 1970s at temperatures at and above $\sim300$~K in flowing-afterglow,\cite{bolden70} flow-drift tube~\cite{lindinger75} and ion-cyclotron resonance~\cite{kim75} set-ups. The thermal reaction rate constant at 300 K in these studies was found to be in the range between $1.15(\pm20\%)\times10^{-15}$ m$^3$/s and $2.20(\pm10\%)\times10^{-15}$ m$^3$/s ($k_{\mathrm{L}}^{\mathrm{He^+ + NH_3}} = 1.887\times10^{-15}$ m$^3$/s). An increase of the reaction rate constant with decreasing temperature in the $300-2900$ K range was observed in Ref. \cite{lindinger75}. Marquette {\it et al.} \cite{marquette85a} have measured the thermal rate coefficient in uniform supersonic flows and reported values of $4.5\times 10^{-15}$~m$^3$/s at 27~K, $3.0\times 10^{-15}$~m$^3$/s 68~K, and $1.65\times 10^{-15}$~m$^3$/s at 300~K. Several reaction channels were identified with the following branching ratios:\cite{kim75}
\begin{align} 
\ce{He^+} + \ce{NH3} &\ce{->}\;  \ce{NH3+} + \ce{He}  & \qquad (\sim12\%) \tag{R1}\\
                                   &\ce{->}\;  \ce{NH2^+} +  \ce{H} + \ce{He}  & \qquad (\sim80\%) \tag{R2}\\
                                   &\ce{->}\;  \ce{NH^+} + \ce{H2} + \ce{He} & \qquad (\sim8\%) \tag{R3}.                           
\end{align}
The estimated released energies in each of these reactions are: $ E_{\mathrm{R1}} = 14.401$ eV, $ E_{\mathrm{R2}} = 8.822$ eV and $ E_{\mathrm{R3}} = 6.993$ eV.\cite{ruscic05}

Low-temperature and low-collisional-energy studies of the reaction between He$^+$ and ammonia are, however, lacking, although this energy regime is particularly relevant for astrochemistry. The current work presents experimental and theoretical studies of the \ce{He+ + NH3} and \ce{He+ + ND3} reactions in the range of collisional energies $E_{\mathrm{coll}}$ between 0 and $k_{\mathrm{B}}\cdot 40$ K at a resolution of $k_{\mathrm{B}}\cdot 200$ mK at $E_{\mathrm{coll}} = 0$.

\section{The effect of the molecular multipole moment on capture rate coefficients}
%{\it I moved this section here but have not made any of the corrections I had marked per hand in red on the pdf-file}
\label{sec2}
\subsection{The modified Langevin interaction potential}
In the Langevin model, the interaction potential $V_{\mathrm{L}}(R)$ between an ion and a molecule is described only through the centrifugal potential and the ion-induced dipole interaction as:
\begin{equation}
V_{\mathrm{L}}(R) = \frac{L^2}{2\mu R^2} - \frac{\alpha'(Ze)^2}{8\pi\epsilon_0R^4}.
\label{eq:VintL}
\end{equation}
In Eq.~(\ref{eq:VintL}), $L$ is the angular momentum of the collision (which can be expressed as $L = \mu v_{\mathrm{rel}}b$, where $b$ is the impact parameter) and $R$ is the ion-molecule separation. If the molecule possesses multipole moments, however, additional rotational-state-dependent terms need to be added to Eq.~(\ref{eq:VintL}) to account for the Stark-shift experienced by the molecule in rotational state $i$ in the electric field of the ion. The modified potential energy is:
\begin{equation}
V_{\mathrm{int},i}(R) = V_{\mathrm{L}}(R) + \Delta E_i(R),
\label{eq:Vint}
\end{equation}
where the energy shift of state $i$, $\Delta E_i(R)$, depends on $R$ through the electric field $Ze/(4\pi\epsilon_0 R^2)$ of the ion. Eq.~(\ref{eq:Vint}) implies that the molecule adiabatically remains in its original rotational state (at $R\to\infty$) as it approaches the ion.  

The electrostatic potential at a point ${\bf R}$ in space resulting from the molecular charge distribution, $\rho({\bf r}')$, can be expressed as:\cite{wong98}
\begin{equation}
\phi({\bf R}) = \frac{1}{4\pi\epsilon_0}\int_{\tau}\frac{\rho({\bf r}')}{|{\bf R}-{\bf r}'|}\mathrm{d^3}{\bf r}'.
\label{ESpot}
\end{equation}
In the region where ${\bf R} > {\bf r}'$, one can expand the $1/|{\bf R}-{\bf r}'|$ term in Eq.~(\ref{ESpot}) as:\cite{zare88a}
\begin{equation}
\frac{1}{|{\bf R}-{\bf r}'|} = \frac{1}{\sqrt{R^2 + r'^2 - 2Rr'\cos\theta}} = \sum_{\lambda=0}^{\infty}\frac{r'^{\lambda}}{R^{\lambda+1}}P_{\lambda}(\cos\theta), 
\label{eq:SHexp}
\end{equation}
where $\theta$ is the angle between ${\bf R}$ and ${\bf r}'$, and $P_{\lambda}(\cos\theta)$ is a Legendre polynomial:
\begin{equation}
P_{\lambda}(\cos\theta) = \sqrt{\frac{4\pi}{2\lambda+1}}Y_{\lambda 0}(\theta,\phi),
\end{equation}
and $Y_{\lambda 0}(\theta,\phi)$ is a spherical harmonic. If one defines the multipole coefficients characterising the nuclear charge distribution, as:\cite{wong98}
\begin{equation}
Q_{\lambda} \equiv  \int_{\tau} P_{\lambda}(\cos\theta)\rho({\bf r}')r'^{\lambda}\mathrm{d^3}{\bf r}',
\label{mult_coeff}
\end{equation}
the value of the electrostatic potential at any given point ${\bf R}$ can be expanded as:
\begin{equation}
\phi(R) = \frac{1}{4\pi\epsilon_0}\frac{1}{R} \sum_{\lambda=0}^{\infty}\frac{Q_{\lambda}}{R^{\lambda}}.
\label{ESexp2}
\end{equation}
For a nearly-spherical charge distribution, Eq.~(\ref{ESexp2}) is a fast-converging series, and only  a few terms are needed. The terms for $\lambda = 1,2$ and $3$ correspond to dipole-, quadrupole- and octupole-moment expansion terms. 

In the symmetric-top wavefunction basis, $|JKM\rangle$, the matrix elements of the multipole operator are given by:\cite{wong98}
\begin{align}
\langle J'K'M'|Q_{\lambda}|JKM \rangle & =  \left[\int_{\tau}\rho({\bf r}')r'^{\lambda}\mathrm{d^3}{\bf r}'\right] \times(-1)^{M^\prime-K^\prime}\sqrt{(2J+1)(2J'+1)}\nonumber\\
& \times \tj{J}{\lambda}{J'}{M}{0}{-M'}\tj{J}{\lambda}{J'}{K}{0}{-K'},
\label{Multipole_op_ME}      
\end{align}
where the integral is the multipole moment in the molecular reference frame. The symmetric-top wavefunctions, $|JKM\rangle$, are given by:\cite{zare88a}
\begin{eqnarray}
|JKM\rangle &=& \sqrt{\frac{2J+1}{8\pi^2}}\mathscr{D}^{J*}_{MK}(\theta,\phi,\chi) \nonumber\\
&=& (-1)^{M-K}\sqrt{\frac{2J+1}{8\pi^2}}\mathscr{D}^{J}_{-M-K}(\theta,\phi,\chi),
\label{SymTop_wf}
\end{eqnarray}   
where $K$ and $M$ are the quantum numbers associated with the projections of the rotational angular momentum $\vec{J}$ on the molecular symmetry axis and a space-fixed axis, respectively, and $\mathscr{D}^{J*}_{MK}(\theta,\phi,\chi)$ is a Wigner rotation matrix. 

The $R$-dependent Stark-shifts, $\Delta E_{i=JKM}(R)$, which describe the interaction between the ion charge and the multipole moments of the molecule, are obtained by diagonalising the Hamiltonian:
\begin{equation}
\widehat{H}(R) = \widehat{H}_{\mathrm{rot}} + \widehat{H}_{\mathrm{Stark}}(R),
\label{Stark_Ham}
\end{equation}
where $\widehat{H}_{\mathrm{rot}}$ is the rotational Hamiltonian --- a diagonal matrix consisting of the molecular rotational energies $E_{JKM}$ in zero electric field, and $\widehat{H}_{\mathrm{Stark}}(R)$ is the Stark Hamiltonian describing the multipole-moment-induced mixing of the rotational levels in the presence of an external electric field. The matrix elements of $\widehat{H}_{\mathrm{Stark}}(R)$ in the electric field of the ion are given by:
\begin{align}
\label{Stark_H_ME}
\langle J'K'M'|\widehat{H}^{\lambda}_{\mathrm{Stark}}(R)|JKM\rangle & = \frac{1}{4\pi\epsilon_0R^{\lambda+1}}\langle J'K'M'|Q_{\lambda}|JKM \rangle.
\end{align}
The rotational-level mixing is stronger for a dipole moment ($\widehat{H}_{\mathrm{Stark}}^{\lambda = 1}$ scales as $R^{-2}$) than for quadrupole ($\widehat{H}_{\mathrm{Stark}}^{\lambda=2}\propto R^{-3}$) and octupole ($\widehat{H}_{\mathrm{Stark}}^{\lambda=3}\propto R^{-4}$)  moments. In addition, its strength also depends on the molecule-fixed multipole moment. After the diagonalisation of $\widehat{H}(R)$, the eigenenergies $E_i + \Delta E_i(R)$ are assigned a value of the $J$, $K$ and $M$ quantum numbers through an adiabatic correlation to the field-free $i = |JKM\rangle$ states at $R\rightarrow\infty$. 

\subsection{Determination of the rotational-state-dependent capture rate coefficients}
From the total interaction potential $V_{\mathrm{int},i}(R)$ associated with a given molecular $i = |JKM\rangle$ state, one can determine for each collision energy $E_{\mathrm{coll}}=\frac{1}{2}\mu v_{\mathrm{rel}}^2$ the classically allowed maximal value of the angular momentum, $L_{\mathrm{max},i} = \mu v_{\mathrm{rel}} b_{\mathrm{max},i}$ for which the reaction can take place. 

The reaction rate coefficient for state $i$ is determined through the reaction cross-section $\sigma_i = \pi b_{\mathrm{max},i}^2$ as:
\begin{align}
\label{expr_k}
k_{i}(E_{\mathrm{coll}}) & = \frac{\pi L_{\mathrm{max},i}^2}{\sqrt{2\mu^3 E_{\mathrm{coll}}}}.
\end{align}
The effect of the multipole moment is to significantly modify the interaction potential for each molecular rotational state. This is particularly true in the case of an ion-dipole interaction, as will be discussed in Sec.~\ref{sec:Vint}. The addition of the $\Delta E_{i}(R)$ terms in Eq.~(\ref{eq:Vint}) creates an interaction potential for state $i$ which is either more or less attractive than the Langevin potential, $V_{\mathrm{L}}$, so that either higher or lower values of $L_{\mathrm{max},i}$ are allowed. The exact dependence of $k_i$ on the collisional energy is thus related to the $R$ dependence of $\Delta E_i$.

Adiabatic capture models such the one described in this section and used to interpret our experimental results include the effects of long-range interactions in the entrance channel but cannot predict the branching ratios between competing reaction channels, nor the product state distributions, nor angular distributions, nor vector correlations. Moreover, they predict the total cross section, whereas our experiments only measure the part of the capture cross section that leads to reactions products, which we estimate to be about 40\% for the He$^+$ + NH$_3$ reaction, see Section~\ref{sec:concl}. Calculations of the (nonadiabatic) dynamics on the relevant multidimensional potential energy surfaces would be needed for a full interpretation of the experimental results and a full description of the reaction dynamics. 

\section{Experimental approach}
\label{sec:Exp}
\subsection{The merged-beam set-up}
\begin{figure*}[!h]
	\includegraphics[width = \textwidth]{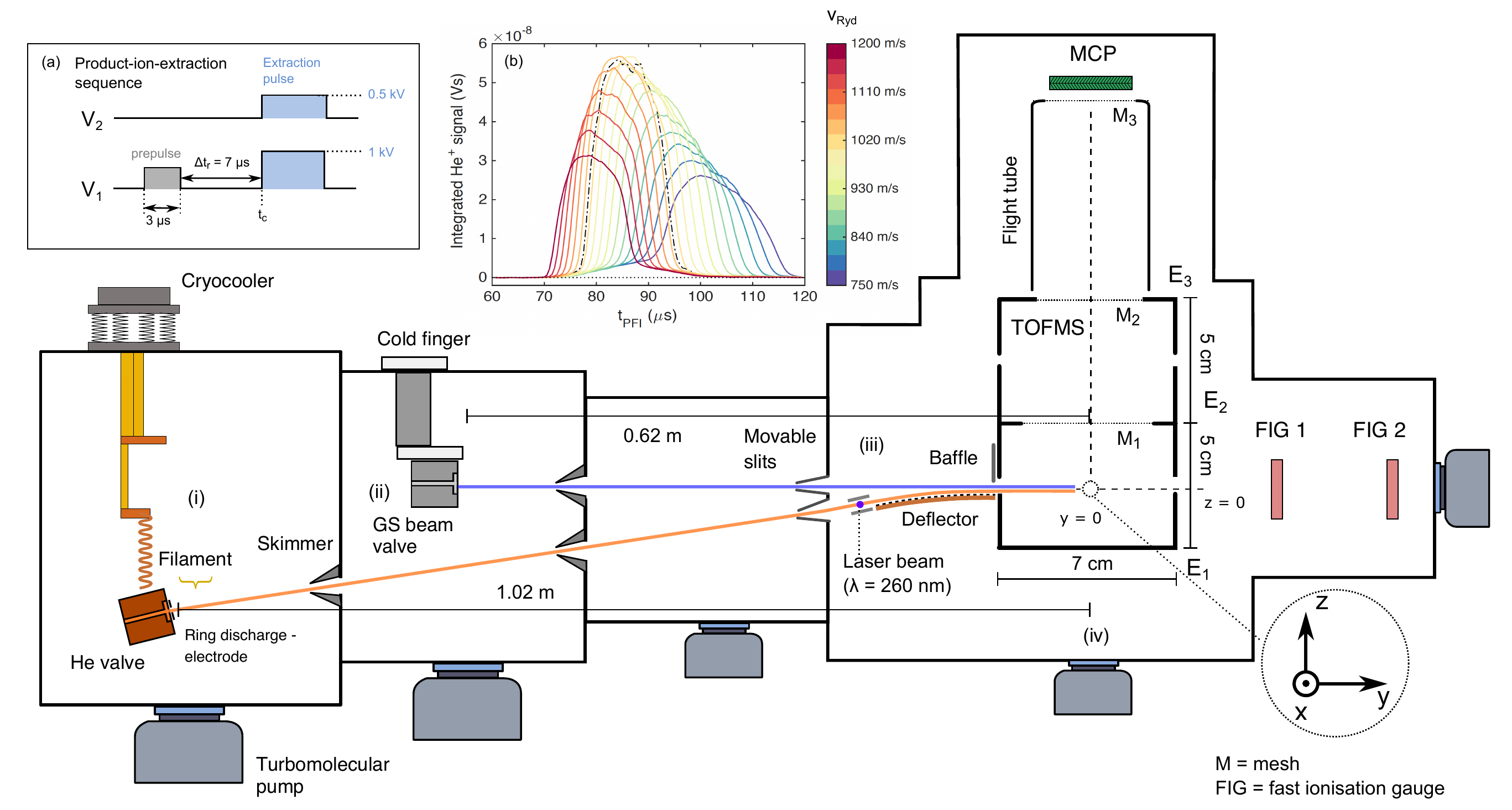}
	\caption{\label{Fig4} Schematic overview of the merged-beam experimental setup to study the reactions of \ce{He+} and ammonia. The He and ground-state beams are produced using home-built valves generating pulses of $\approx 20$ $\mu$s duration. (i) Helium source chamber. (ii) Ground-state-beam source chamber. (iii) Region where He$^*$ is excited to a Rydberg-Stark state and the Rydberg-He beam is deflected and merged with the ammonia beam. (iv) Time-of-flight mass spectrometer (TOFMS) used to monitor the product ions. Inset (a): Timing and pulse characteristics of the potentials applied to electrodes E$_1$ and E$_2$ when the TOFMS is operated in product-ion-extraction mode. Inset (b): Pulsed-field-ionisation signal of the He$(n)$ atoms deflected by the surface deflector, after acceleration/deceleration to final velocities $v_{\mathrm{Ryd}}$ in the $750-1200$ m/s range, as a function of the time of application of the ionisation/extraction pulse, $t_{\mathrm{PFI}}$. The dash-dotted black line presents the results obtained when guiding the He$(n)$ atoms at the initial central beam velocity of $1040$ m/s.}
\end{figure*}
The experimental set-up has been briefly described in Ref.~\cite{zhelyazkova20} and is outlined schematically in Fig.~\ref{Fig4}. The merged-beam apparatus consists of two supersonic beams produced by home-built valves (nozzle-opening duration $\sim20$ $\mu$s, repetition rate of 25 Hz) - one for the He atoms subsequently excited to a Rydberg state, and the other for the ground-state (GS) beam consisting of either pure NH$_3$ or ND$_3$. The two beams propagate along axes initially separated by a 5$^{\circ}$ angle. The helium atoms are photoexcited  to a Rydberg-Stark state (referred to as He$(n)$ from here on) between two planar electrodes used to generate an electric field. They are then deflected by a $50$-electrode Rydberg-Stark surface deflector and merged with the ground-state beam. The merged beams then enter a time-of-flight mass spectrometer (TOFMS), where the reaction takes place and the product ions are extracted toward a microchannel plate (MCP) detector in a direction perpendicular to the merged-beam propagation axis. The TOFMS consists of three cylindrical electrodes (labelled E$_{1-3}$ in Fig.~\ref{Fig4}, with inner radius of 35.5 mm). 

The central point of the TOFMS along the beam-propagation axis defines the origin of the space-fixed coordinate system. The helium valve is positioned at ($0,-102.5,-8.2$) cm (see Fig.~\ref{Fig4}). Its copper body is tilted by a 5$^{\circ}$ angle with respect to the vertical ($z$) axis and is thermally connected by copper braids to the cold heads of a two-stage pulse-tube cryo-cooler. The He valve is temperature-stabilised to $100.0\pm0.1$ K and operated at a stagnation pressure of 2.5 bar. At the valve opening, a pulsed potential ($\sim+300$ V, pulse duration $70$ $\mu$s) applied to a ring electrode attracts a beam of electrons emitted by a tungsten filament, creating an electric discharge at the valve orifice. In this discharge, the metastable (1s)(2s) $^3$S$_1$ state of helium (lifetime of $7.859\times10^3$ s)~\cite{lach01} is populated, amongst other states.\cite{halfmann00a} The metastable helium beam (referred to as He$^*$) travels at a velocity of $\sim1040$ m/s. After passing two skimmers, the He$^*$ beam intersects a pulsed laser beam ($\lambda_{\mathrm{exc}} \cong 260$ nm, pulse duration $\sim10$ ns, pulse energy $\sim$ 800 $\mu$J, beam diameter $\sim0.2$ cm), produced by frequency tripling the output of a Nd:YAG-pumped commercial dye laser using two $\beta$-barium-borate crystals. The excitation takes place between two parallel metallic plates separated by 0.25 cm used to generate a dc electric field, $\vec{F}_{\mathrm{exc}}$. The laser is polarised parallel to $\vec{F}_{\mathrm{exc}}$; its frequency and the value of $|\vec{F}_{\mathrm{exc}}|$ are tuned such that a one-photon excitation is driven to a low-field-seeking Rydberg-Stark state $(n,k,m_{\ell}) = (30,21,0)$ just below the Inglis-Teller limit [$n$ and $m_{\ell}$ are the principal and magnetic quantum numbers, respectively, and $k = -(n-m_{\ell}-1):2:(n-m_{\ell}-1)$].\cite{gallagher94a}

After excitation, the He$(n)$ atoms are loaded into an electric quadrupole trap formed by applying time-dependent potentials to several sets of electrodes located at the surface of a $\sim5$-cm-long, $50$-electrode surface Rydberg-Stark deflector (electrode dimensions $d_y = 0.5$~mm, $d_ x = 30$~mm, separated by $\Delta d_y=0.5$~mm in the $y$-dimension). The deflector electrodes are patterned onto the surface of a printed circuit board, which is attached to a curved substrate (see Fig. 1 of Ref.~\cite{zhelyazkova20}). By applying a sinusoidally-varying potential to each set of electrodes (see Ref.~\cite{allmendinger14a,zhelyazkova19a}), the trapped cloud of Rydberg atoms can be guided in a direction parallel to the surface of the chip. By setting a frequency chirp to these sinusoidally-varying potentials, one can also accelerate or decelerate the Rydberg atoms and precisely adjust their final velocity, $v_{\mathrm{Ryd}}$, after release from the trap at the end of the decelerator. When it leaves the quadrupole trap, the Rydberg-atom cloud has a cigar shape of radius $\sim0.5$ mm~\cite{zhelyazkova20} and length $\sim3$ mm in the $x$-dimension (estimated from Monte Carlo simulations), and travels in a direction parallel to the $y$-axis, which is also the axis of propagation of the GS beam. We estimate the temperature of the Rydberg cloud from Monte Carlo simulations and from experiments to be in the $0.1-0.3$ K range.\cite{zhelyazkova20} A baffle at the end of the deflector blocks the He$^*$ beam and all Rydberg atoms which have not been captured by the quadrupole trap but have travelled over the deflector in a straight line. We estimate that approximately 100 Rydberg atoms exit the surface deflector and enter the TOFMS at every experimental cycle (repetition rate 25 Hz). The He$(n)$ atom cloud has a width of $\approx1$ mm in the $y$-direction in the centre of the reaction zone. 

The orifice of the GS-beam valve is positioned at $(0,-62,0)$ cm and has an orifice diameter of $\sim0.5$ mm. It is operated with pure ammonia (either \ce{^{14}NH3} or \ce{^{14}ND3}) kept at a stagnation pressure of about 1 bar. The valve is connected to an ice-filled reservoir and temperature-stabilised to $274.0\pm0.5$ K. Two fast ionisation gauges (FIGs) positioned after the TOFMS are used to determine the velocity distribution of the GS beam. The selected velocity of the \ce{NH3} (\ce{ND3}) beam, $v^{\mathrm{GS}}$, was measured to be 1190 m/s (1170 m/s). The GS beam valve is triggered such that the molecules travelling at $v^{\mathrm{GS}}$ arrive at the centre of the TOFMS at the same time, designated as $t_{\mathrm{c}}$, as the Rydberg atoms. 

The background pressures in the regions of the apparatus labelled (i)-(iv) in Fig.~\ref{Fig4} rise from about $2 \times 10^{-7}$~mbar in all regions to about $2 \times 10^{-5}$~mbar in regions (i) and (ii), $7 \times 10^{-7}$~mbar in region (iii) and $4 \times 10^{-7}$~mbar in region (iv) when the pulsed valves are operated.

We use the TOFMS in two modes: (i) a product-ion-extraction mode with mass resolution and (ii) a He$(n)$ pulsed-field ionisation mode. In mode (i), the product ions are extracted by applying potentials of 1 kV and 0.5 kV to electrodes E$_1$ and E$_2$, respectively (see Fig.~\ref{Fig4}), through a commercial (home-built) high-potential switch, with a rise time of $\sim30$ ns. These potentials correspond to an electric field of 110 V/cm at the centre of the TOFMS. The ions are extracted toward an MCP detector with a phosphor screen (Photonis, diameter 4 cm), with its centre positioned at $(0,0,14.5)$ cm. The front face of the MCP is biased at a potential of $-1.96$ kV to allow for maximal sensitivity while avoiding saturation effects caused by the strong He$^+$ signal. The potentials E$_1$ and E$_2$ are optimised to achieve maximal ion-collection efficiency and mass resolution. To generate a homogeneous extraction electric field and a field-free flight tube, three meshes (M$_{1-3}$ in Fig.~\ref{Fig4}) were installed in the ion-extraction stack. The time interval $\Delta t_{r}$ during which the ion-molecule reaction is observed ($\Delta t_{\mathrm{r}} = 7$ $\mu$s in the experiments presented here) was defined by applying a prepulse of V$_1=125$ V to electrode E$_1$ (corresponding to $\sim25$ V/cm at the TOFMS centre) to sweep out all ions located in the extraction region. The value of V$_1$ was chosen to be high enough to sweep the ions formed prior to time $t_{\mathrm{c}}-\Delta t_{\mathrm{r}}$ out of the ion-extraction zone, but not high enough to field-ionise the reactant He$(n)$ atoms. 

When the TOFMS is operated in He$(n)$-detection mode [mode (ii)], a high-electric-field pulse is applied to field-ionise the Rydberg helium atoms and extract the \ce{He+} ions towards the MCP detector. The ionisation electric field [$\sim1100$ V/cm at point $(0,0,0)$] is generated by applying 5 kV to E$_1$ while keeping electrodes E$_2$ and E$_3$ grounded. This field is sufficient to ionise Rydberg atoms of principal quantum number $n\geq29$.

\subsection{Varying the velocity of the He$(n)$ beam}
We measure the dependence of the product-ion yields on the collision energy $E_{\mathrm{coll}}$ by varying the central velocity of the Rydberg atoms, $v_{\mathrm{Ryd}}$, and keeping the velocity $v^{\mathrm{GS}}$ of the GS beam constant. The short gas-pulse duration leads to a large spatial dispersion of the GS beam over the 62-cm distance between the valve orifice and the reaction zone. The particles in the GS beam thus have a well-defined velocity (with a velocity spread $\Delta v_{\mathrm{GS}} = \pm 15$ m/s) in the region where it overlaps with the He($n$) packet emerging from the chip. $v_{\mathrm{Ryd}}$ is set by applying the appropriate potentials to the surface deflector. The He$(n)$ atoms have an estimated velocity spread of full width at half maximum (FWHM) around $v_{\mathrm{Ryd}}$ of $\Delta v^{\mathrm{FWHM}}_{\mathrm{Ryd}}\approx\pm 20$ m/s.  

To ensure that the same velocity class of the ammonia GS molecules interacts with the He$(n)$ atoms regardless of the selected $v_{\mathrm{Ryd}}$ value, we perform a set of  measurements in which we determine the arrival time of the Rydberg-atom cloud at the centre of the reaction zone, $t_{\mathrm{c}}^{v_{\mathrm{Ryd}}}$, for every value of $v_{\mathrm{Ryd}}$. The He$(n)$ atoms can be detected if (i) the applied ionisation field is sufficiently high for pulsed field ionisation (PFI), and (ii) PFI takes place in a region inside the TOFMS such that the produced ions can reach the MCP and do not collide with the walls of the extraction stack. To obtain the data presented in inset (b) of Fig.~\ref{Fig4}, we operate the TOFMS in a He$(n)$-ionisation (single-pulse) mode and record the integrated He$^+$ signal as a function of the time of application of the PFI pulse relative to the laser excitation time, $t_{\mathrm{PFI}}$, for each set value of $v_{\mathrm{Ryd}}$. The dashed-dotted black curve represents the He$^+$ signal measured when operating the surface deflector in guiding mode, i.e., with the He$(n)$ atoms propagating at constant velocity over the deflector surface ($v^{\mathrm{guide}}_{\mathrm{Ryd}} = 1040$ m/s). This curve has an approximately Gaussian shape with a centre at $t_{\mathrm{c}}^{1040\;\mathrm{m/s}}=85.5$ $\mu$s and a FWHM of $\sim17$ $\mu$s. 

By applying the appropriate frequency chirp to the time-dependent electrode potentials, we can accelerate or decelerate the He$(n)$ atoms from an initial velocity of 1040 m/s to a final velocity in the $1200-750$ m/s range during deflection. The maximum (minimum) velocity in this range corresponds to an increase (decrease) of the initial He$(n)$ kinetic energy of $\sim33$\% ($\sim48$\%). The value of $t_\mathrm{c}^{v_{\mathrm{Ryd}}}$ is determined for each of the selected values of $v_{\mathrm{Ryd}}$ by determining the centres of the corresponding distributions depicted in Fig.~\ref{Fig4}(b). In addition, the spread of the measured He$(n)$ profile increases for lower values of $v_{\mathrm{Ryd}}$ because the He($n$) packet expands over a longer flight time. The GS valve opening time is then adapted so that the centre of the GS beam overlaps with the deflected Rydberg cloud at $t_{\mathrm{c}}$. 
\subsection{Normalising the measured product ion signal}
\label{subsec:norm}
In the experiment, we do not measure absolute values of the reaction rates because we do not accurately know the particle densities in the reaction zone. We measure relative values of the collision-energy-dependent rates, which we scale by a global factor for comparison with the calculated capture-rate coefficients.

The total amount of He$(n)$ that reaches the centre of the TOFMS decreases with increasing deceleration and acceleration. This decrease is illustrated by the decrease of the maximum measured height of the He$(n)$ PFI profiles
at time $t_{\mathrm{c}}$ in Fig.~\ref{Fig4}(b). For example, only 56\% and 47\% of the atoms reach the centre of the stack for $v_{\mathrm{Ryd}} = 1200$ m/s and 750 m/s, respectively, compared to $v_{\mathrm{Ryd}} = 1040$ m/s. This decrease is attributed to the opening of the quadrupole trap in the reference frame of the moving atoms,\cite{allmendinger14a} which leads to a larger fraction of the atoms escaping the trap at the highest acceleration/deceleration. For the lower values of $v_{\mathrm{Ryd}}$, additional losses result from the finite lifetimes of the Rydberg atoms. In order to correct for these effects, the integrated reaction-product signals are divided by the detected He$^+$ signal, integrated over the [$ t_{c}-\Delta t_{\mathrm{r}}/2, t_c+\Delta t_{\mathrm{r}}/2$] time interval, accounting for the times of flight of the ions to the MCP.

The different times of flight of the Rydberg cloud from the end of the deflector to the TOFMS centre correspond to different expansions in the plane perpendicular to the direction of propagation (the $xz$-plane). The ground-state-beam spatial extent is limited to 1.6 mm in the $z$-direction by the position of the baffle above the surface of the chip. The He$(n)$-expansion can thus affect the geometric overlap with the GS beam, creating a detection bias in favour of product ions generated at higher values of $v_{\mathrm{Ryd}}$. Monte-Carlo particle-trajectory simulations indicate that for the lowest values of $v_{\mathrm{Ryd}}$ ($v_{\mathrm{Ryd}}\lesssim700$ m/s), the cloud has expanded so much by the time it arrives at the stack centre that the Rydberg atoms located at its outer edge do not interact with the GS-beam molecules. This detection bias is corrected for when determining the collision-energy-dependent product yields.

\section{Experimental Results}
\label{sec:results}
Ammonia is an oblate symmetric top molecule. It has C$_{3v}$ point-group symmetry in its two equivalent \say{umbrella} equilibrium structures located at values of $\sim\pm22^{\circ}$ of the  umbrella inversion angle. These structures are separated by a relatively low energy barrier ($\sim2023$ cm$^{-1}$)\cite{herzberg91a} for inversion through the D$_{3h}$ planar structure. Tunnelling through the potential-energy barrier thus leads to a splitting of $\sim0.79$ cm$^{-1}$ ($\sim0.053$ cm$^{-1}$) of the ground vibrational state rotational levels in \ce{NH3} (\ce{ND3}), the vibrational eigenfunctions being the symmetric ($s$) and antisymmetric ($a$) superpositions of the ground-state wavefunctions in each of the potential wells.

\begin{figure*}[!h]
\includegraphics[width = 0.5\textwidth]{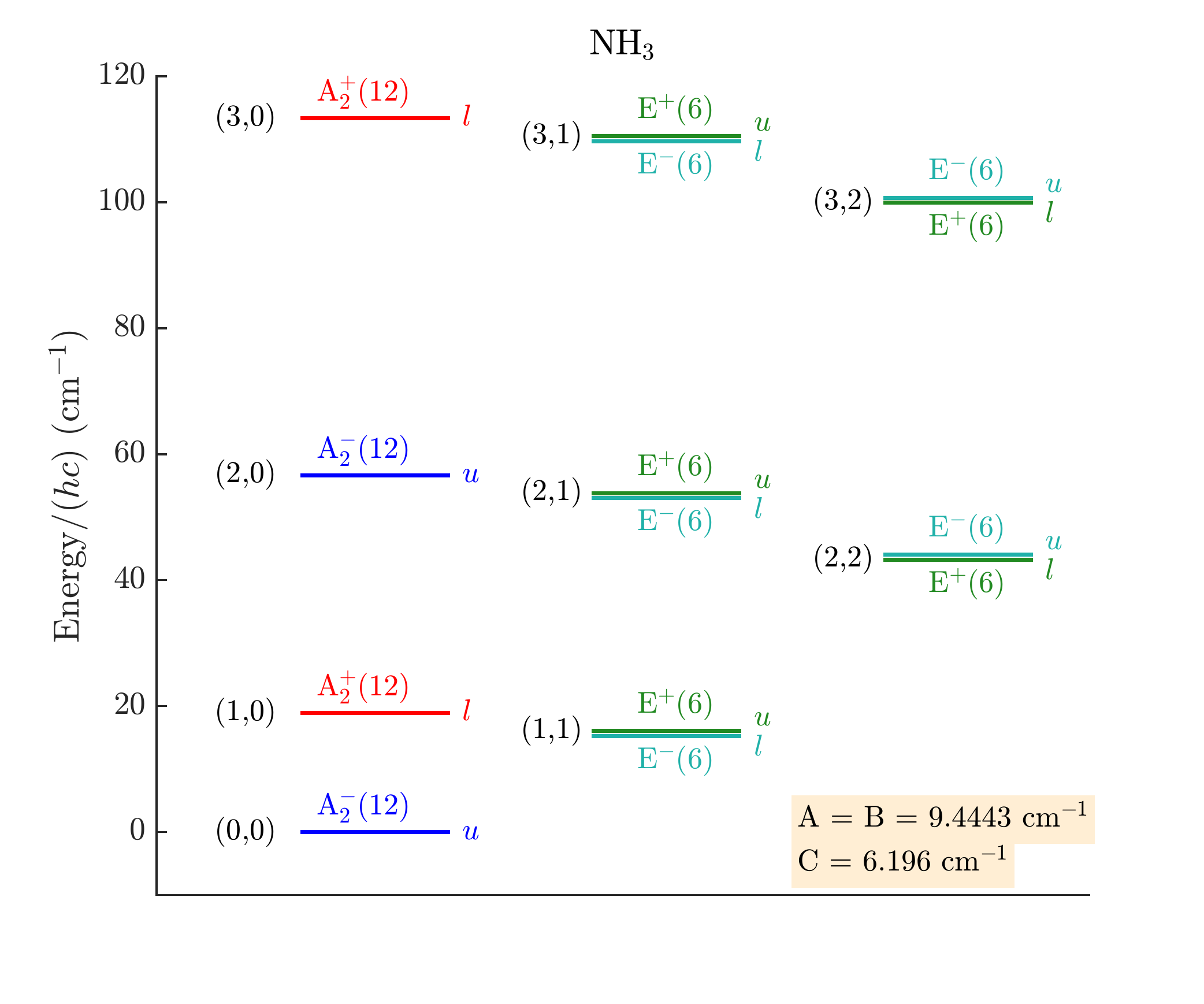}
\includegraphics[width = 0.5\textwidth]{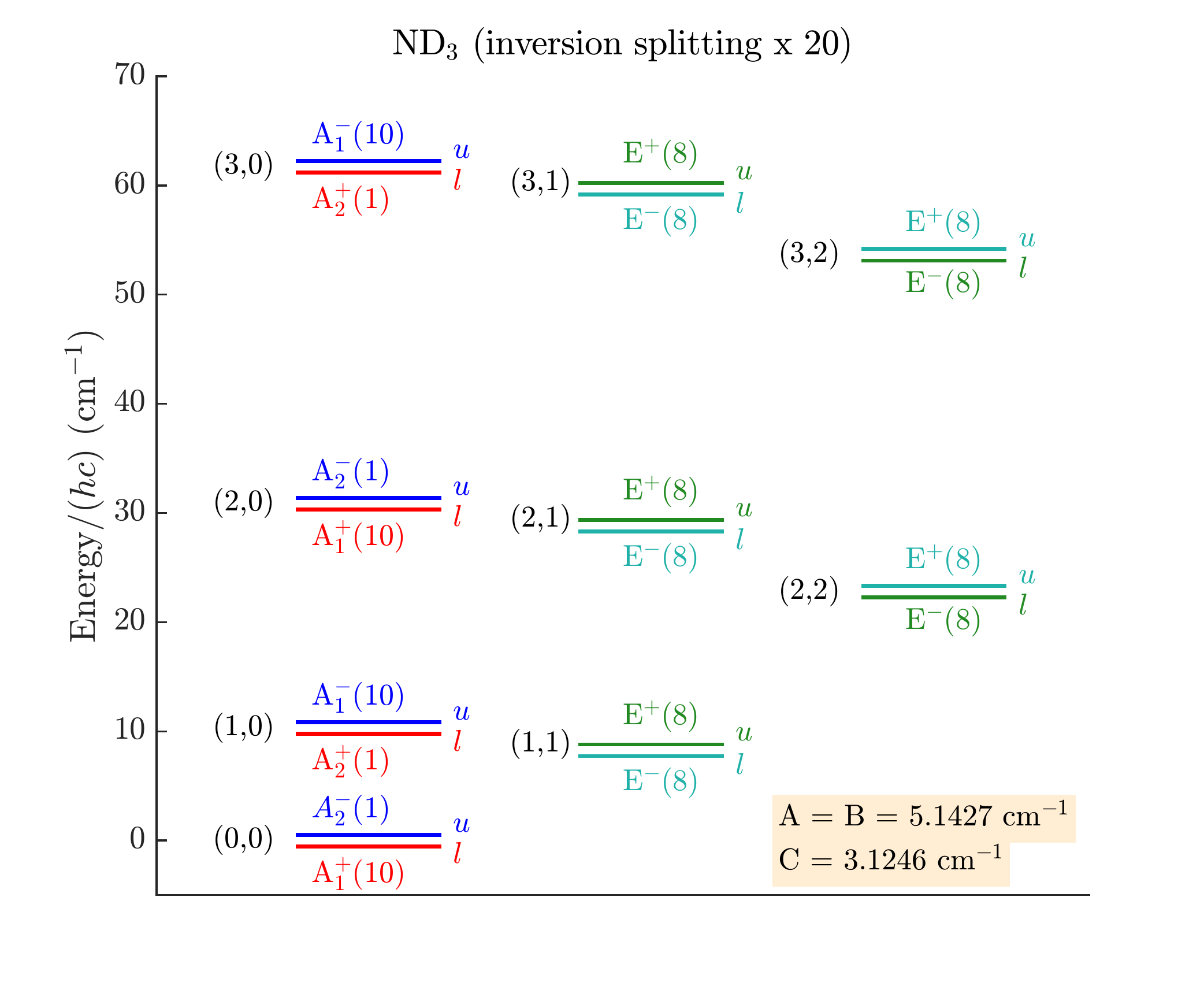}
\caption{\label{Fig1} Rotational energy levels and symmetry labels in the lowest vibrational level of the ground electronic state of \ce{NH3} and \ce{ND3} in the S$_3^*$ permutation inversion group, for $|K|\leq2$. The numbers in parentheses indicate the spin-statistical weights $g_{\mathrm{ns}}$, the $l$ and $u$ letters indicate the lower and upper components of the inversion doublets.\cite{wichmann20} The rotational constants are taken from Ref.~\cite{urban84} (\ce{NH3}) and Ref.~\cite{daniel16} (\ce{ND3}).}
\end{figure*}

The ammonia molecule has a dipole moment of 1.468 D at the C$_{3v}$ equilibrium geometry corresponding to a partial negative charge on the nitrogen side of the molecule, and a partial positive charge on the hydrogen side.
The ammonia molecule can be described in the D$_{3h}$ point group, or the isomorphic permutation-inversion group S$_3^*$, with rovibrational energy states labelled by the irreducible representations: A$_1'$/A$_1^+$, A$_2'$/A$_2^+$, A$_1''$/A$_1^-$, A$_2''$/A$_2^-$, E$'$/E$^+$ and E$''$/E$^-$ in D$_{3h}(M)$/S$_3^*$.\cite{wichmann20,snels00,urban84,bunker06} According to the Pauli exclusion principle, the total internal wavefunction in \ce{NH3} (\ce{ND3}) must have symmetry of either A$_2^+$ or A$_2^-$ (A$_1^+$ or A$_1^-$), i.e., it must be antisymmetric (symmetric) with respect to pairwise exchange of the identical fermions (bosons). The nuclear-spin states of \ce{^{14}NH3} and \ce{^{14}ND3} span the representations:
\begin{align}
\Gamma^{\mathrm{NH_3}}_{\mathrm{ns}} &= 12\mathrm{A}_2^+\oplus 6 \mathrm{E}^+ \\
\Gamma^{\mathrm{ND_3}}_{\mathrm{ns}} &= 30\mathrm{A}_1^+\oplus 3 \mathrm{A}_2^+ \oplus 24 \mathrm{E}^+, 
\label{GammaNS}
\end{align}  
and the allowed rovibronic states span the representations:
\begin{align}
\Gamma^{\mathrm{NH_3,sw}}_{\mathrm{rve}} &= 12\mathrm{A}_2^+\oplus12\mathrm{A}_2^-\oplus 6 \mathrm{E}^+\oplus 6 \mathrm{E}^- \\
\Gamma^{\mathrm{ND_3,sw}}_{\mathrm{rve}} &= 30\mathrm{A}_1^+ \oplus 30\mathrm{A}_1^- \oplus 3\mathrm{A}_2^+\oplus 3\mathrm{A}_2^-\oplus 24 \mathrm{E}^+\oplus 24 \mathrm{E}^-. 
\label{GammaNS2}
\end{align}  
Rovibrational states of symmetries A$_1^+$ and A$_1^-$ have a statistical weight of zero and are missing in \ce{NH3}. The rovibrational statistical weights of the $(J,K)$ states in \ce{NH3} and \ce{ND3} are presented in Table~\ref{tab1} (see Refs.~ \cite{wichmann20,snels00,bunker06} for details), and the rotational-level structures are presented in Fig.~\ref{Fig1}. The two components of the inversion doublet have either A$_1^+$, A$_2^+$ or E$^+$ (positive parity) or A$_1^-$, A$_2^-$ or E$^-$ (negative parity) rovibrational symmetry. In both \ce{NH3} and \ce{ND3}, states with $K\;\mathrm{mod}\;3 = 0$ have A symmetry and states with $K\;\mathrm{mod}\;3\neq 0$ have E symmetry. The lowest states of A and E symmetry are the $(J,K) = (0,0)$ and $(J,K) = (1,1)$ states, respectively. 

Because of the missing A$_1$ levels, the inversion splitting is not observed in \ce{NH3} states with $K = 0$. Instead, this manifold consists of an alternating series of negative - positive parity $J$ states (even/odd $J$ states are of A$_2^-$/A$_2^+$ symmetry).  
\begin{table}[]
\caption{\ Rovibronic spin statistical weights $g_{\mathrm{ns},i}$ for NH$_3$ and ND$_3$ in the S$_3^*$ permutation inversion group. Adapted from Refs.\cite{wichmann20,snels00}}
\label{tab1}
\begin{tabular*}{0.48\textwidth}{@{\extracolsep{\fill}}|lllll|}
\hline 
       & Symmetry (rve)       & $J$    & $K$                     & $g_{\mathrm{ns},i}$ \\[0.8ex] \hline 
NH$_3$ & $s$; A$_1^+$            & even & 0                     & 0      \\[0.5ex]
       & $a$; A$_1^-$           & odd  & 0                     & 0      \\[0.5ex]
       & $s$; A$_2^+$            & odd  & 0                     & 12     \\[0.5ex]
       & $a$; A$_2^-$            & even & 0                     & 12     \\[0.5ex]
       & $s$, $a$; E$^{\pm}$         & all  & 1, 2, 4, 5, $\ldots$  & 6      \\[0.5ex]
       & $s$, $a$; A$_1^{\pm}$ & all  & 3, 6, 9, 12, $\ldots$ & 0      \\[0.5ex]
       & $s$, $a$; A$_2^{\pm}$ & all  & 3, 6, 9, 12, $\ldots$ & 12     \\[0.5ex] \hline
ND$_3$ & $s$; A$_1^+$            & even & 0                     & 30     \\[0.8ex]
       & $a$; A$_1^-$            & odd  & 0                     & 30     \\[0.5ex]
       & $s$; A$_2^+$            & odd  & 0                     & 3      \\[0.5ex]
       & $a$; A$_2^-$            & even & 0                     & 3      \\[0.5ex]
       & $s$, $a$; E$^{\pm}$         & all  & 1, 2, 4, 5, $\dots$   & 24      \\[0.5ex]
       & $s$, $a$; A$_1^{\pm}$ & all  & 3, 6, 9, 12, $\ldots$ & 30     \\[0.5ex]
       & $s$, $a$; A$_2^{\pm}$ & all  & 3, 6, 9, 12, $\ldots$ & 3    \\ [0.5ex]\hline 
\end{tabular*}
\end{table}
At the typical rotational temperature of the molecules produced in our supersonic-beam source ($T_{\mathrm{rot}}\simeq 6$ K) and for the rotational constants of \ce{NH3} and \ce{ND3} ($A_{\mathrm{NH_3}} = B_{\mathrm{NH_3}} = 9.444$ cm$^{-1}$, $C_{\mathrm{NH_3}} = 6.196$ cm$^{-1}$; $A_{\mathrm{ND_3}} = B_{\mathrm{ND_3}} = 5.143$ cm$^{-1}$, $C_{\mathrm{ND_3}} = 3.125$ cm$^{-1}$),\cite{urban84,daniel16} only states with $K = 0$ and $|K| = 1$ are significantly populated. Taking into account the nuclear-spin symmetry, the ratio of states with $K = 0$ to states with $|K| = 1$  is $1:1$ and $11:16$ in \ce{NH3}  and \ce{ND3}, respectively. The occupation probabilities of the different ground-state rotational levels at a rotational temperature of $T_{\mathrm{rot}}=6$ K are listed in Table~\ref{tab2}. 
\begin{table}[]
\caption{Occupation probabilities of the rotational levels of the  NH$_3$ and ND$_3$ at a rotational temperature of $T_{\mathrm{rot}} = 6$ K. }
\label{tab2}
\begin{tabular*}{0.48\textwidth}{@{\extracolsep{\fill}}|ll|l|l|} 
\hline
        &           & NH$_3$                         & ND$_3$    \\ \hline
$K = 0$   & $J = 0$ ($l$) & 0                             & 0.295  \\
        & $J = 0$ ($u$) & 0.484                         & $2.94\times 10^{-2}$ \\
        & $J = 1$ ($l$) & $1.57\times 10^{-2}$  & $7.58\times 10^{-3}$  \\
        & $J = 1$ ($u$) & 0                                &   $7.42\times 10^{-2}$ \\ \hline
$|K| = 1$ & $J = 1$ ($l$) & 0.274                      & 0.295  \\
        & $J = 1$ ($u$) & 0.226                         & 0.291  \\
        & $J = 2$ ($l$) & $5.29\times 10^{-5}$ & $3.49\times 10^{-3}$  \\
        & $J = 2$ ($u$) & $4.39\times 10^{-5}$ & $3.54\times 10^{-3}$ \\ \hline
\end{tabular*}
\end{table}

\begin{figure}
	\includegraphics[trim = 1cm 0cm 1cm 0cm, clip,width = 0.5\textwidth]{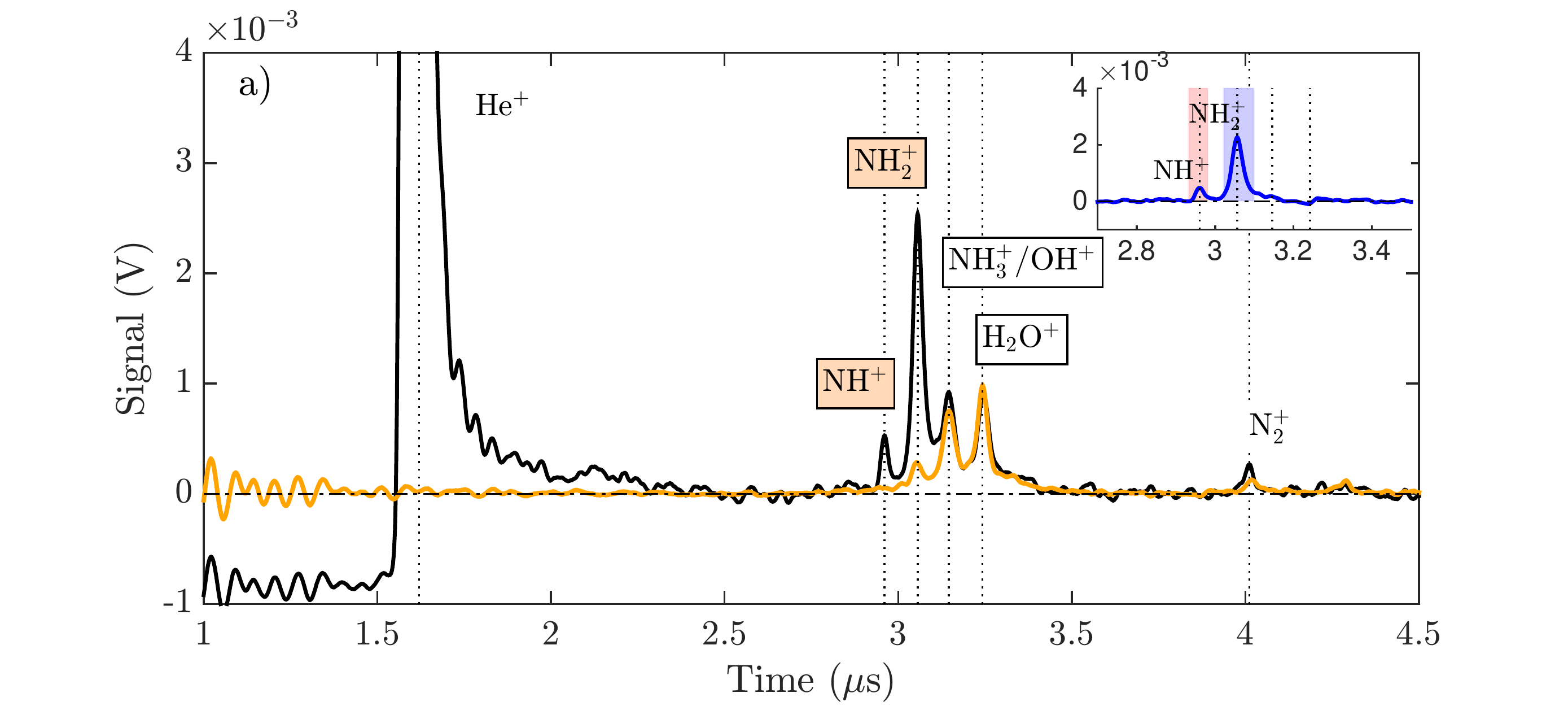}
	\includegraphics[trim = 1cm 0cm 1cm 0cm, clip,width = 0.5\textwidth]{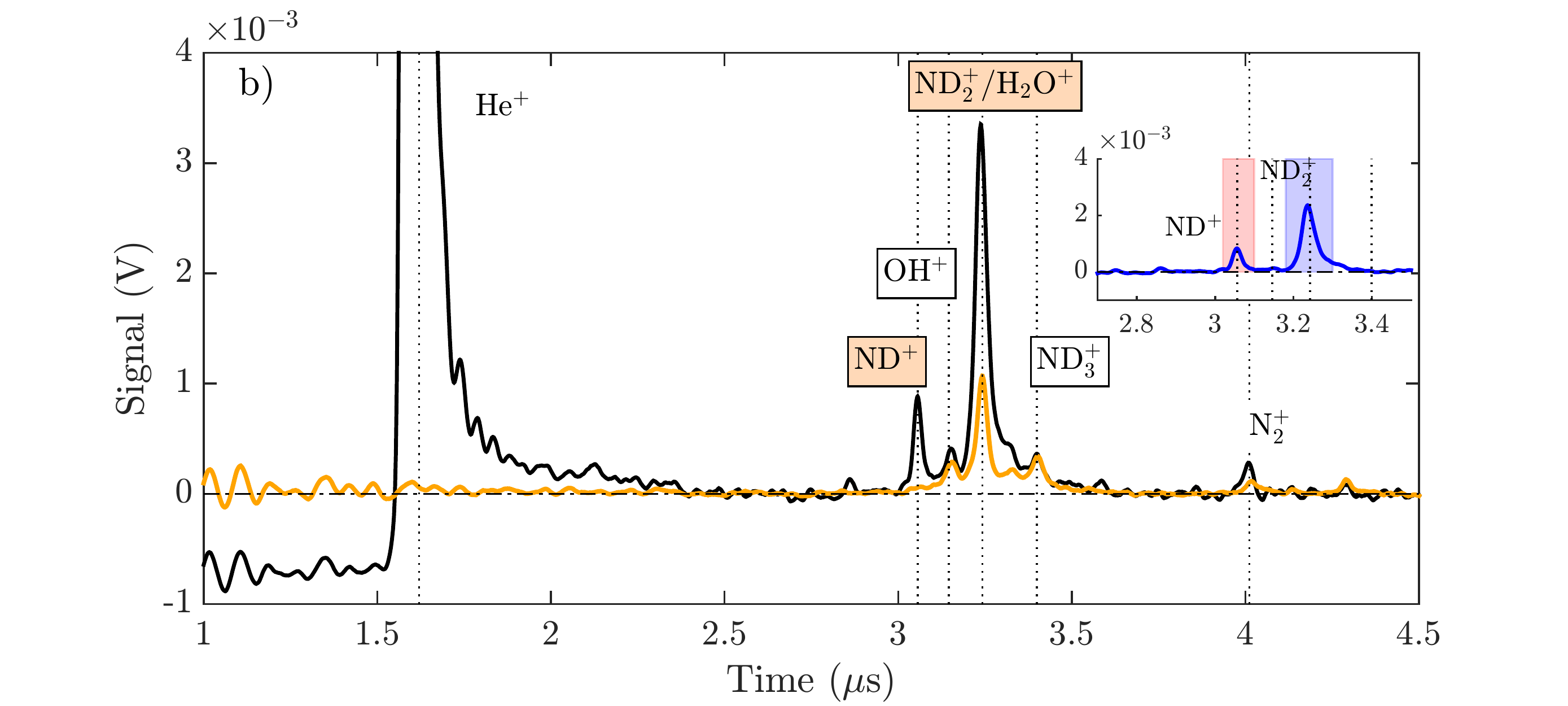}
	\caption{\label{Fig5} Measured product-ion time-of-flight spectra after a $7$-$\mu$s-long reaction-observation time for the (a) He$(n)$ + \ce{NH3} and (b) He$(n)$ + \ce{ND3} reactions. The He atoms are initially excited to the $(n,k,m) = (30,21,0)$ Rydberg-Stark state. The Rydberg-atom velocity is $v_{\mathrm{Ryd}}= 1050$ m/s. The orange traces are recorded with the Rydberg-excitation laser turned off. The coloured text boxes indicate the measured product ions which are specific to the reactions between the He($n$) atoms and the ammonia molecules (see text for details). The insets show the reaction products after subtraction of the signals generated by Penning-ionisation processes.}
\end{figure}

Displayed in Fig.~\ref{Fig5} are reaction-product time-of-flight (TOF) traces measured following reactions between He$(n)$ atoms and \ce{NH3}(a) and \ce{ND3}(b) molecules, after a reaction time of $\Delta t_{\mathrm{r}} = 7$ $\mu$s. The traces were recorded with the He$(n)$ atoms accelerated to a final velocity of $v_{\mathrm{Ryd}} = 1050$ m/s, corresponding to collisional energies $E_{\mathrm{coll}}\approx k_{\mathrm{B}}\cdot3.8$ K (a) and $k_{\mathrm{B}}\cdot2.8$ K (b). When the laser used to photoexcite He$^*$ to He$(n)$ is on (black traces in Fig.~\ref{Fig5}), a prominent peak is visible at 1.6 $\mu$s, corresponding to the \ce{He^+} ions generated by field ionisation of the He$(n)$ atoms. The extraction electric field of 110 V/cm is not sufficient to ionise the initially prepared Rydberg helium atoms with $n=30$ (this field can only ionise states with $n\gtrsim50$). However, $n$-changing blackbody-radiation-induced transitions during the $\sim85$ $\mu$s flight time of the He$(n)$ atoms from the photoexcitation region to the centre of the TOFMS populate a few higher-lying Rydberg states, which can then be field ionised with the extraction field. A constant background signal of $\sim0.7$ mV is also visible in the TOF traces in Fig.~\ref{Fig5} beyond $1.6$ $\mu$s. This background signal originates from a constant stream of He$^+$ produced either through blackbody-radiation-induced ionisation or slow tunnelling ionisation of the He$(n)$ Rydberg states in the extraction field. 

In addition to the He$^+$ peak, several other peaks are visible in the mass spectrum and can be assigned to (a) \ce{NH+}, \ce{NH2+}, \ce{NH3+}/\ce{OH+}, \ce{H2O+} and \ce{N2+}, and (b) \ce{ND+}, \ce{OH+}, \ce{ND2+}/\ce{H_2O+}, \ce{ND3+} and \ce{N2+}, as indicated in Fig.~\ref{Fig5}. The main product ions observed following the He$(n)$ + \ce{NH3} (\ce{ND3}) reaction are \ce{NH2+} and \ce{NH+} (\ce{ND2+} and \ce{ND+}). The \ce{H2O+}, \ce{OH+} and \ce{N2+} ions originate from Penning-ionisation processes involving metastable He$^*$ atoms produced in the discharge and background water and nitrogen present in the GS beam. The corresponding mass peaks in the TOF spectra are indeed observed even when the excitation laser is turned off (orange traces in Fig.~\ref{Fig5}). We have verified that the \ce{NH+}, \ce{NH2+}, \ce{ND+} and \ce{ND2+} product ions detected are generated by autoionisation of the corresponding Rydberg-molecule products. 

The comparison of the TOF mass spectra recorded with the excitation laser turned on and off illustrates that the peaks corresponding to the \ce{OH+}, \ce{NH3+}, \ce{ND3+}, \ce{H2O+} and \ce{N2+} ions originate mostly from Penning ionisation and that the \ce{NH^+}, \ce{ND^+}, \ce{NH_2^+} and \ce{ND_2^+} ions are only detected in significant quantities from reactions with He$(n)$.

\begin{figure*}
	\includegraphics[trim = 1cm 0cm 1cm 0cm, clip,width = 0.5\textwidth]{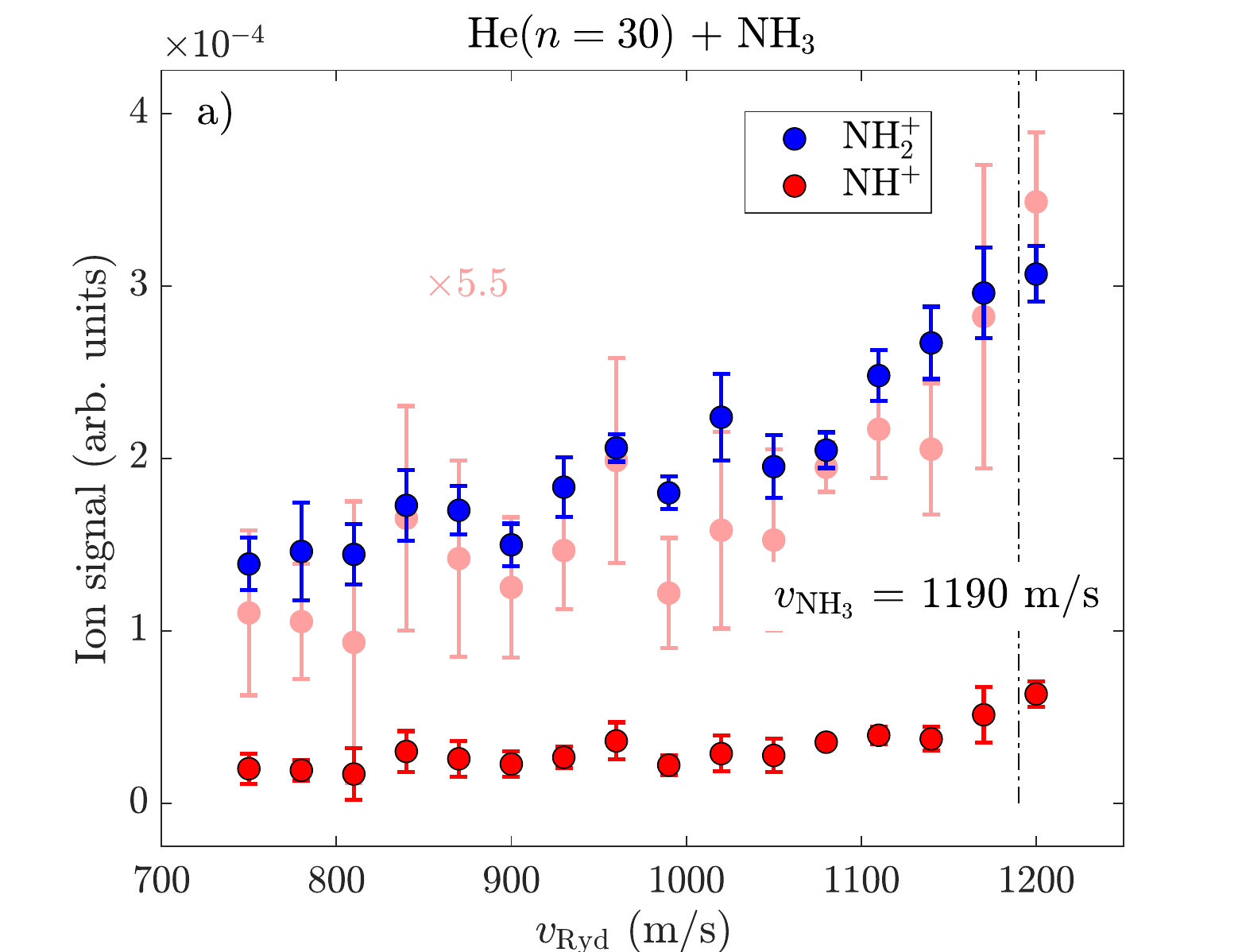}
	\includegraphics[trim = 1cm 0cm 1cm 0cm, clip,width = 0.5\textwidth]{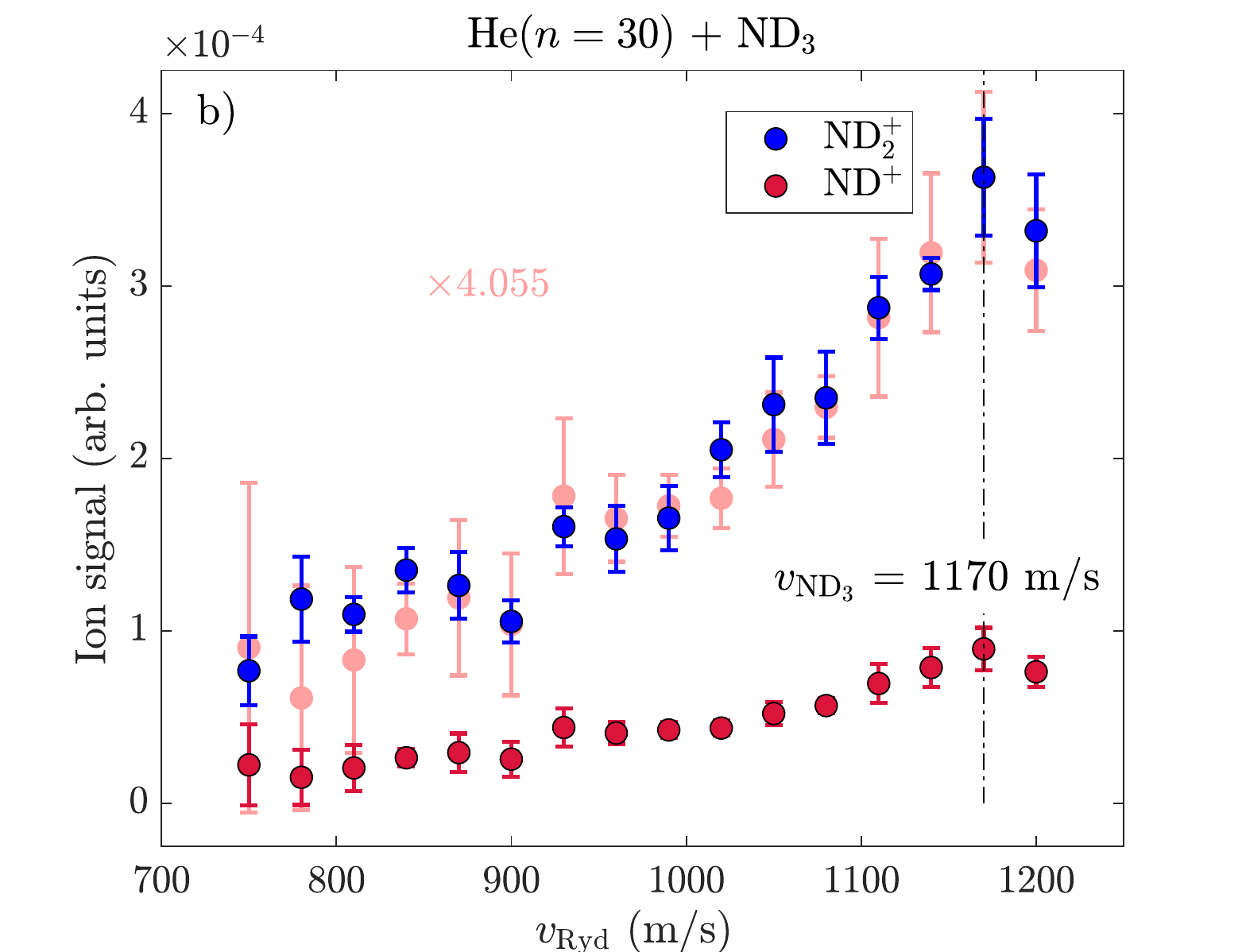}
	\includegraphics[trim = 4.0cm 2cm 4.0cm 3cm, clip,width = 0.5\textwidth]{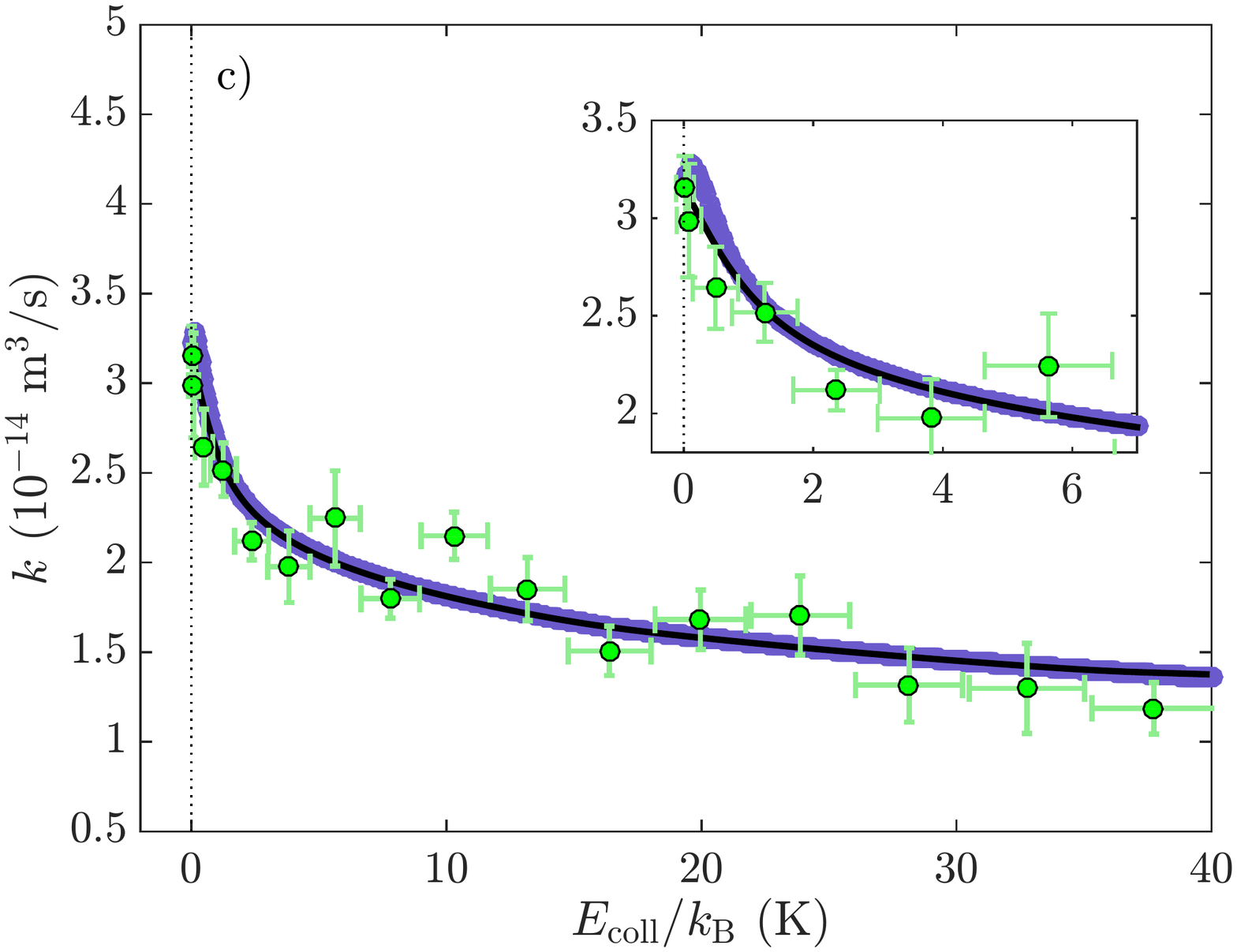}
	\includegraphics[trim = 4.0cm 2cm 4.0cm 3cm, clip,width = 0.5\textwidth]{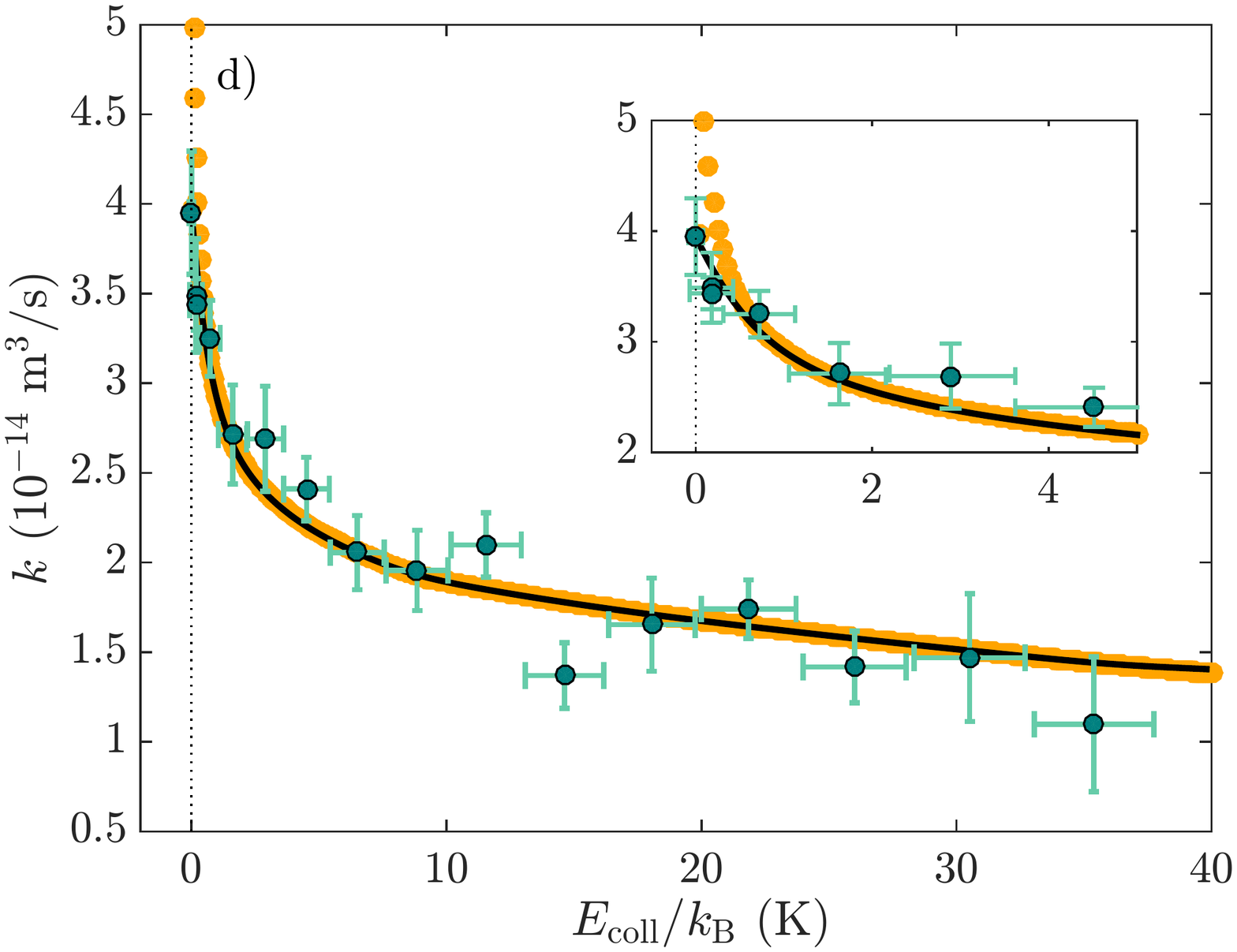}
	\caption{\label{Fig6} (a) and (b) Measured integrated product ion signals, $I_{\ce{NH+}}$, $I_{\ce{NH2+}}$, $I_{\ce{ND+}}$ and $I_{\ce{ND2+}}$ (in arbitrary units) as a function of the Rydberg velocity, $v_{\mathrm{Ryd}}$, following reactions between He$(n)$ and either \ce{NH3} or \ce{ND3}. The vertical dashed lines indicate the GS-beam velocity. (c) and (d) Comparison of the scaled total measured product-ion yields for the two reactions, $I_{\ce{NH+}+\ce{NH2+}}$ and $I_{\ce{ND+}+\ce{ND2+}}$ (green circles with error bars in arbitrary units), to the state-averaged capture rate constants $k$ calculated assuming infinite resolution (orange and purple dots) and considering the finite experimental energy resolution (black lines). In panels (a) and (b), the pink data points correspond to the red data points with intensities multiplied by factors of 5.5 and 4.055, respectively. See text for details.}
\end{figure*}

To measure the collision-energy dependence of the product yields, we record the integrated \ce{NH+}, \ce{ND+}, \ce{NH_2+} and \ce{ND_2+} product signals for several values of the He$(n)$ velocity, $v_{\mathrm{Ryd}}$, in the $750-1200$ m/s range, keeping the GS-beam velocity constant. The results of these measurements are presented in Fig.~\ref{Fig6}(a) and (b). The range of He$(n)$ velocities corresponds to collisional energies of $E_{\mathrm{coll}}/k_{\mathrm{B}}\lesssim40$ K. In order to subtract contributions from Penning-ionisation products from the He$(n)$ + \ce{NH3}/\ce{ND3} reaction products, mass spectra were recorded without the Rydberg-excitation laser and then subtracted prior to integration (see insets of Fig.~\ref{Fig5}). The integration of the light (\ce{NH+} and \ce{ND+}) and heavy (\ce{NH2^+} and \ce{ND2^+}) reaction product ions, respectively, was performed in temporal windows indicated by the red and blue shaded areas in the insets in Fig.~\ref{Fig5}. The measured integrated product-ion signals for each value of $v_{\mathrm{Ryd}}$ were further normalised following the procedure described in Sec.~\ref{subsec:norm}. The error bars are obtained from five sets of consecutive measurements, each corresponding to an average over 1000 experimental cycles.   

The integrated normalised product-ion signals $I_{\ce{NH2+}}$ and $I_{\ce{ND2+}}$ are depicted as a function of the He$(n)$ velocity using blue dots in Fig.~\ref{Fig6} (a) and (b). They show a pronounced increase as the He$(n)$ velocity, $v_{\mathrm{Ryd}}$, approaches the GS-beam velocity, which is designated by the dashed vertical lines. This increase is more pronounced for the He$(n)$ + \ce{ND3} reaction than for the  He$(n)$ + \ce{NH3} reaction: for example, $I_{\ce{NH2+}}(1200\;\mathrm{m/s})/I_{\ce{NH2+}}(750\;\mathrm{m/s})\approx 2.21$, while $I_{\ce{ND2+}}(1170\;\mathrm{m/s})/I_{\ce{ND2+}}(750\;\mathrm{m/s})\approx 4.72$. The measured values of $I_{\ce{NH+}}$ and $I_{\ce{ND+}}$ [red circles in Fig.~\ref{Fig6} (a) and (b)] are smaller than the measured values of $I_{\ce{NH2+}}$ and $I_{\ce{ND2+}}$ by a factor of $f_{\ce{NH+}}=6\pm1$ and $f_{\ce{ND+}}=4.0\pm0.6$, respectively. The measured $\ce{NH2+}:\ce{NH+}$ branching ratio is slightly lower than the one reported in Ref.~\cite{kim75}. While in previous measurements the \ce{NH3+} product ions constituted approximately 12\% of the total reaction yield,\cite{kim75} in our experiments all \ce{NH3+}/\ce{ND3+} ions are primarily generated by Penning-ionisation processes and the \ce{NH3+} and \ce{ND3+} signals obtained after subtraction of the Penning-ionisation contributions are too weak and noisy for reliable estimates of the branching ratios to be carried out. 
\section{Calculation of the capture-rate coefficients and comparison with experimental data}

\subsection{The Stark effect in NH$_3$ and ND$_3$ and the He$^++$NH$_3$/ND$_3$ interaction potentials}
\label{sec:Vint}
The interaction between the molecular dipole moment and an electric field mixes states $|JKM\rangle\leftrightarrow |J'K'M\rangle$ with $J' = J, J\pm1$ (Eq.~(\ref{Multipole_op_ME}) with $\lambda = 1$). The electric-dipole selection rules are $\Delta J = 0, \pm1\;(0\not\leftrightarrow0)$, $\Delta K = 0$ and $\Delta M = 0$. 

To describe the Stark effect in ammonia under the consideration of the inversion splitting, we expand the $|JKM\rangle$ basis set to include the inversion doublet sublevels, which have positive or negative parity. Since the rotational Hamiltonian is invariant under the parity operation, it is conventional to use the symmetrised rotational wavefunctions of well-defined parity  $|J,K,M,(+/-)\rangle$. The Stark interaction mixes states of opposite parity and the non-vanishing Stark matrix elements in the symmetrised basis are:\cite{vandemeerakker12a}  
\begin{equation}
\langle J'K'M'\pm|\widehat{H}_{\mathrm{Stark}}|JKM\mp\rangle = \langle J'K'M'|\widehat{H}_{\mathrm{Stark}}|JKM\rangle.
 \end{equation}

\begin{figure*}[!h]
\includegraphics[trim = 1.75cm 0cm 1cm 0cm, clip, width = 0.5\textwidth]{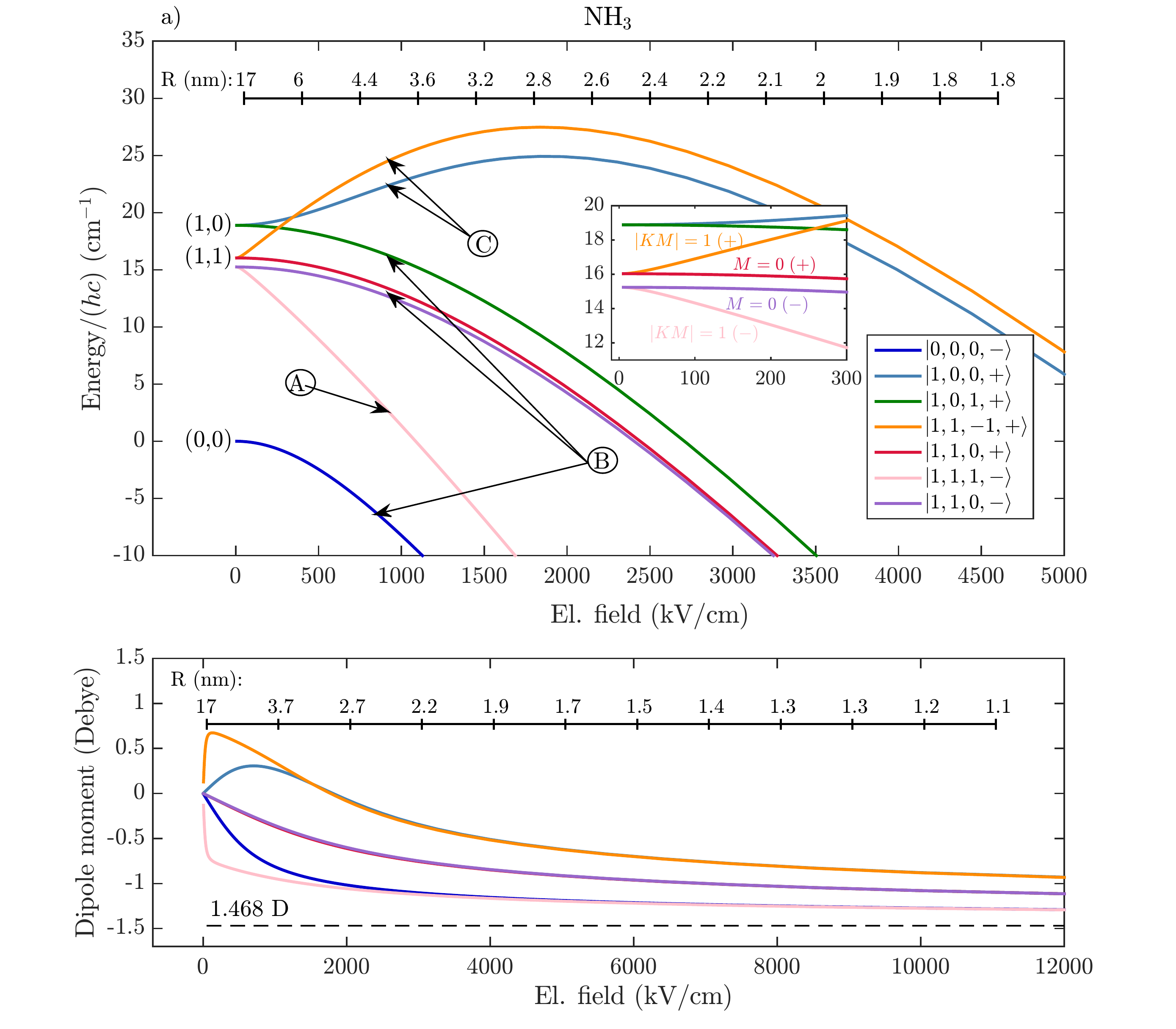}
\includegraphics[trim = 1.75cm 0cm 1cm 0cm, clip, width = 0.5\textwidth]{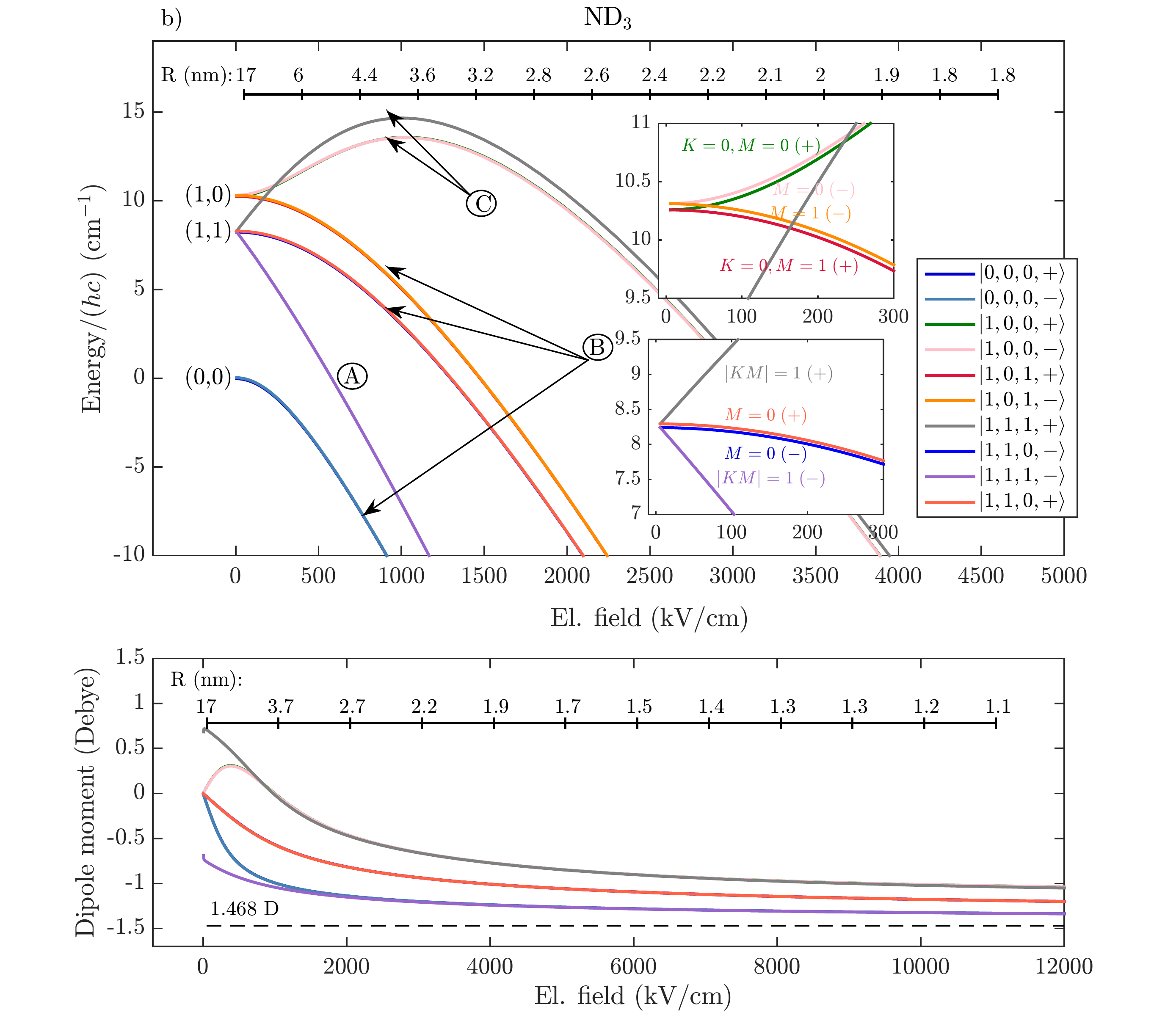}
\caption{\label{Fig2} Top panels: The calculated Stark effect in the $J = 0,1$, $|K| = 0,1$ states of \ce{NH3} (a) and \ce{ND3} (b) in electric field of up to 5000 kV/cm. The Stark shifts at low fields are depicted on an enlarged scale in the insets. Bottom panels: The corresponding state-specific electric-field-dependent dipole moments. The top scale indicates the distance between He$^+$ and the ammonia molecule at the corresponding field values. The definition of the state categories A, B and C is provided in the text.}
\end{figure*}

The Stark effect in \ce{NH3} and \ce{ND3} was calculated by determining the eigenvalues of the Hamiltonian in Eq.~(\ref{Stark_Ham}) for values of the ion-molecule separation $R$ from 0.1 nm to 50 nm in steps of 0.1 nm (corresponding to electric fields ranging from $\sim1.4\times10^{6}$ kV/cm to $\sim5.7$ kV/cm, respectively). We use the rotational constants presented in Fig.~\ref{Fig1}.\cite{urban84,daniel16} Because only $J = 0$ and 1 levels are significantly populated in our experiments, we neglect centrifugal-distortion effects. Convergence of the eigenvalues of the $J = 0,1$ levels was reached with a basis set with a maximum value of $J_{\mathrm{max}} = 6$. The states are labelled according to their adiabatic limit for $R\rightarrow\infty$ corresponding to field-free conditions. The results of the calculation for the states with $J\leq1$, $|K|\leq1$ for electric fields in the $5-5000$ kV/cm range are presented in Fig.~\ref{Fig2}. Because of the smaller value of the tunnelling splitting and the smaller (by a factor of $\sim2$) values of the rotational constants, the Stark shifts are more pronounced in \ce{ND3} at lower fields. In addition, the Stark effect of the $J =1, |K|=1$ levels becomes linear earlier in \ce{ND3}.

We classify the Stark states of interest in ammonia in three types according to the energy shift they experience with increasing electric field: {\bf type A} --  strongly high-field-seeking states with a linear Stark shift, i.e., the lower-energy components of the inversion doublets with $J = 1$ and $|KM| = 1$;  {\bf type B} --  high-field-seeking states with a quadratic Stark shift at low fields, e.g., the $|J = 0, K = 0,M=0,-\rangle$ and $|J = 1, K=0, M = 1,+ \rangle$ states; and {\bf type C} -- low-field-seeking states at relatively low fields which exhibit an energy maximum at intermediate values of the electric field before turning into high-field-seeking states at high fields, e.g., selected components of the inversion doublets with $J = 1$ and $|KM| = 1$. 

States of type A comprise the lower components of the $J =1, |KM|=1$ inversion doublets in \ce{NH3} and \ce{ND3} (see Fig.~\ref{Fig2}). These states experience the strongest Stark shifts, $\Delta E_i$, in both \ce{NH3} and \ce{ND3}: $\Delta E^{\mathrm{(A)}}\sim-88$ cm$^{-1}$ in \ce{NH3} and $\sim-95$ cm$^{-1}$ in \ce{ND3} at 5000 kV/cm. States of type B with $J = 0$ (i.e., the $|0,0,0,-\rangle$ state in \ce{NH3} and the $|0,0,0,+\rangle$ and $|0,0,0,-\rangle$ states in \ce{ND3}) do not have a dipole moment at low fields, but acquire one through mixing with the $|1,0,0,\pm\rangle$ states in an electric field. In states of type C, the potential energy barriers occur at lower values of the electric field and have lower heights in \ce{ND3} than in \ce{NH3} (see Table~\ref{tab3}). At the lowest electric fields ($\lesssim80$ kV/cm), the inversion splitting of $\sim0.79$ cm$^{-1}$ in \ce{NH3} causes the $J =1, |KM| = 1$ states to undergo a quadratic Stark shift [inset of Fig.~\ref{Fig2}(a)]. In contrast, because of the much smaller inversion splitting of $\sim0.053$ cm$^{-1}$ in \ce{ND3}, the corresponding states experience a linear Stark shift, even at the lowest fields [first inset of Fig.~\ref{Fig2}(b)]. The bottom panels in Fig.~\ref{Fig2} show the electric-field-dependent dipole moments of the states depicted in the top panels. 
\begin{table}[]
\caption{\label{tab3} Potential energy barriers in the Stark shifts of the $J = 1, |K| = 0,1$ low-field-seeking Stark states of NH$_3$ and ND$_3$.}
\begin{tabular*}{0.50\textwidth}{m{0.5cm}m{3.0cm}|m{1.0cm}m{2.75cm}} %@{\extracolsep{\fill}}|m{1.0cm}l|m{0.8cm}|ll|l|l
\hline 
         &                          &  Height    &  Position   \\ 
         &.                         &(cm$^{-1}$) & (kV/cm) \\ \hline
         
NH$_3$ & $J = 1, |KM| = 1\;(+)$   & 11.4                                                                  & 1836 ($R = 2.80$ nm)                                                                \\ [0.5ex]
       & $J = 1, K = M = 0\; (+)$ & 6.03                                                                  & 1837 ($R = 2.79$ nm)                                                               \\[0.5ex] \hline
ND$_3$ & $J = 1, |KM| = 1\; (+)$   & 6.37                                                                  & 997.2 ($R = 3.80$ nm)                                                              \\[0.5ex]
       & $J = 1, K = M = 0\; (+)$ & 3.32                                                                  & 1052  ($R = 3.70$ nm)                                                              \\[0.5ex]
       & $J = 1, K = M = 0\; (-)$ & 3.25                                                                  & 997.2 ($R = 3.80$ nm)                                                              \\ [0.5ex] \hline\end{tabular*}
\end{table}

\begin{figure*}[!h]
\includegraphics[trim = 0.5cm 0cm 1cm 0cm, clip, width = 0.5\textwidth]{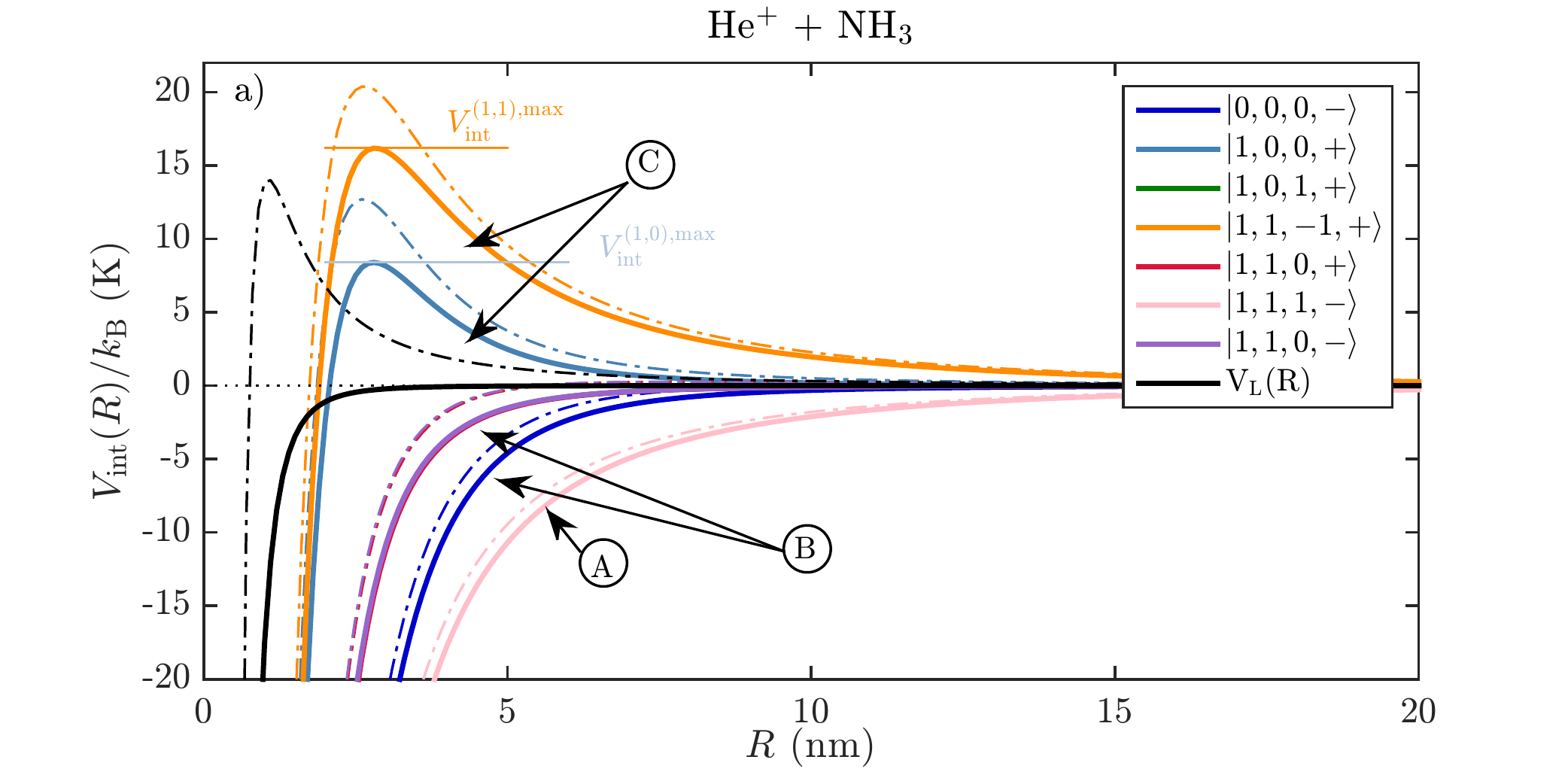}
\includegraphics[trim = 0.5cm 0cm 1cm 0cm, clip, width = 0.5\textwidth]{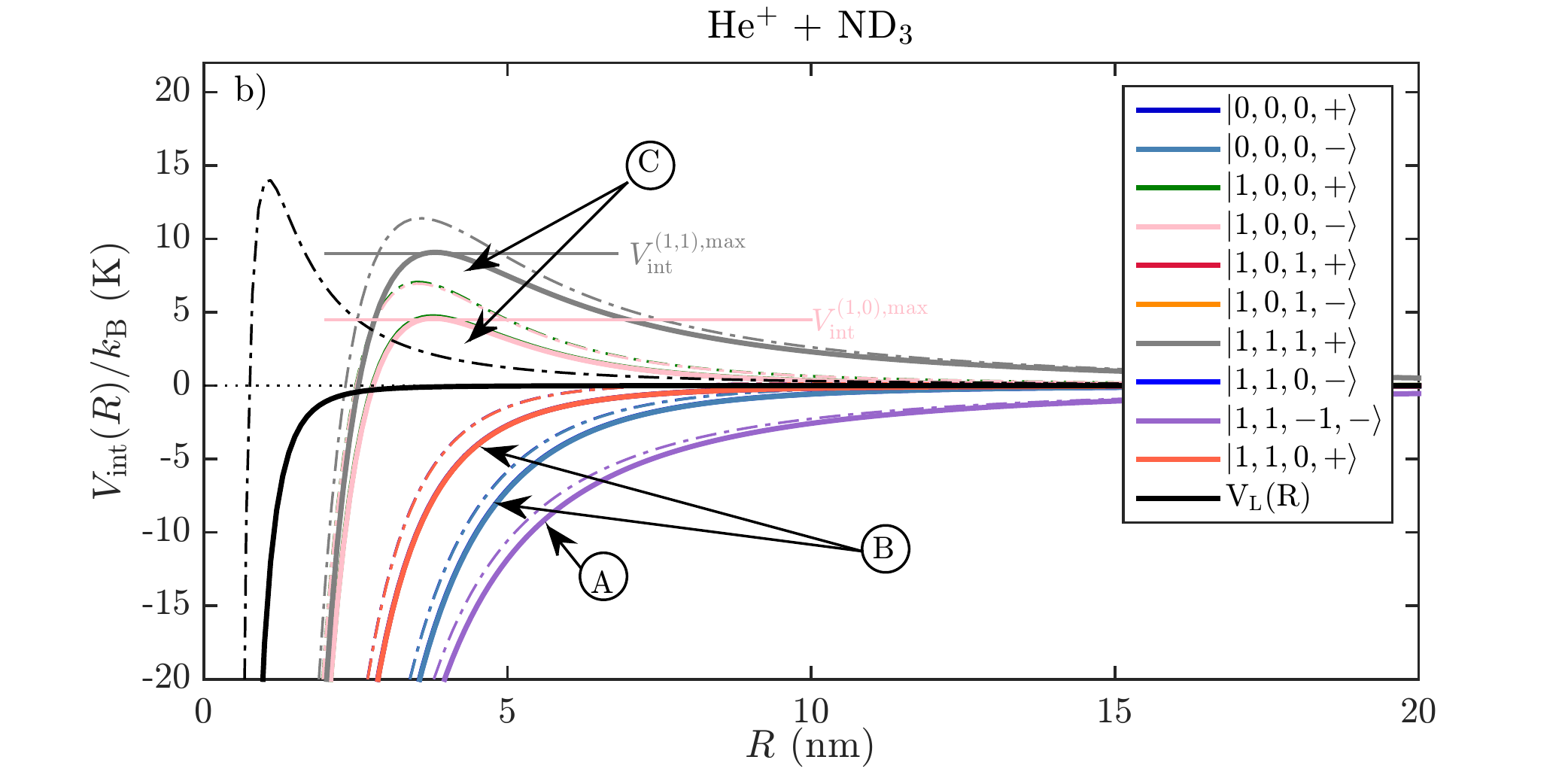}
\includegraphics[trim = 0.5cm 6.5cm 1cm 6.5cm, clip, width = 0.5\textwidth]{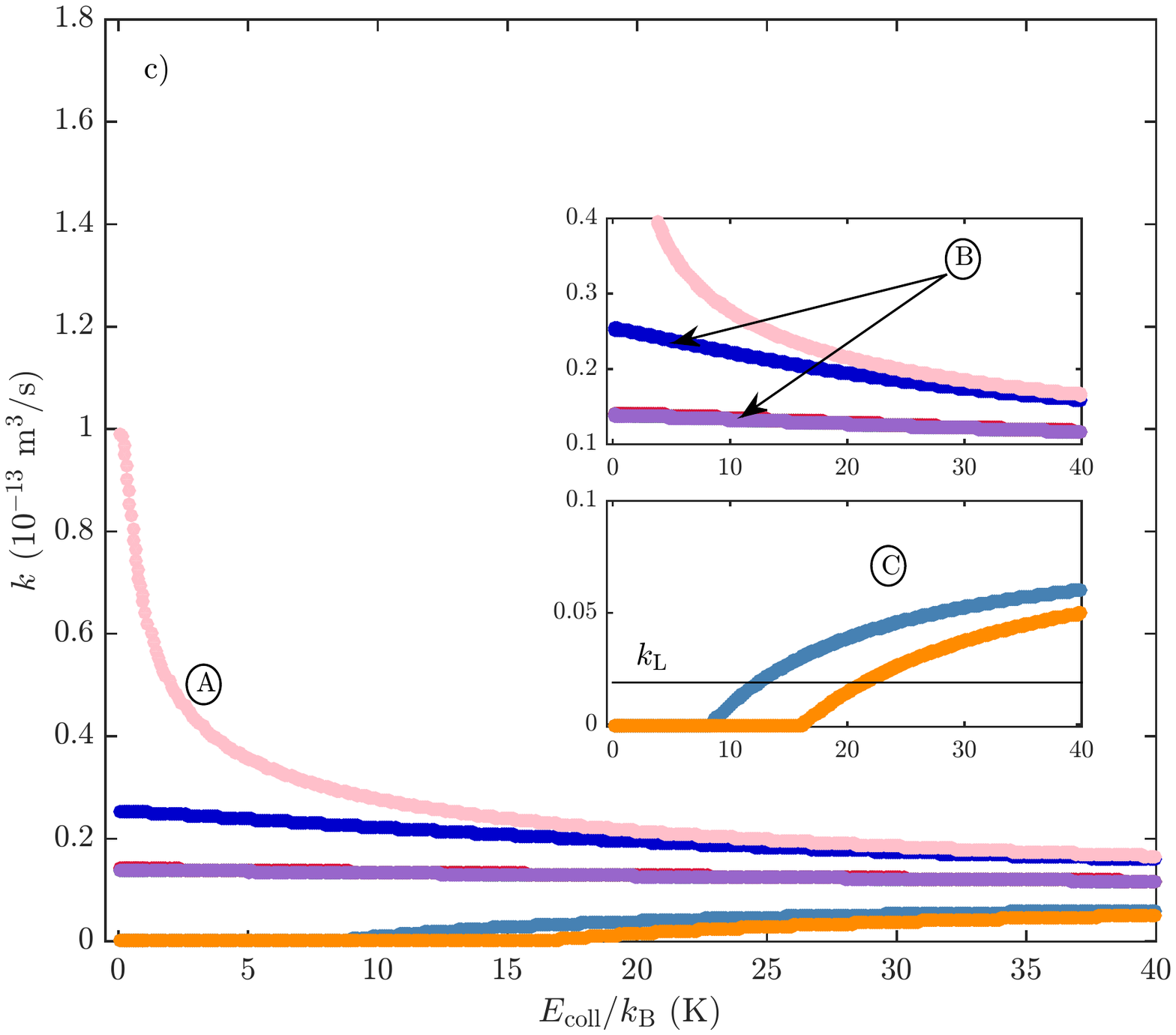}
\includegraphics[trim = 0.5cm 6.5cm 1cm 6.5cm, clip, width = 0.5\textwidth]{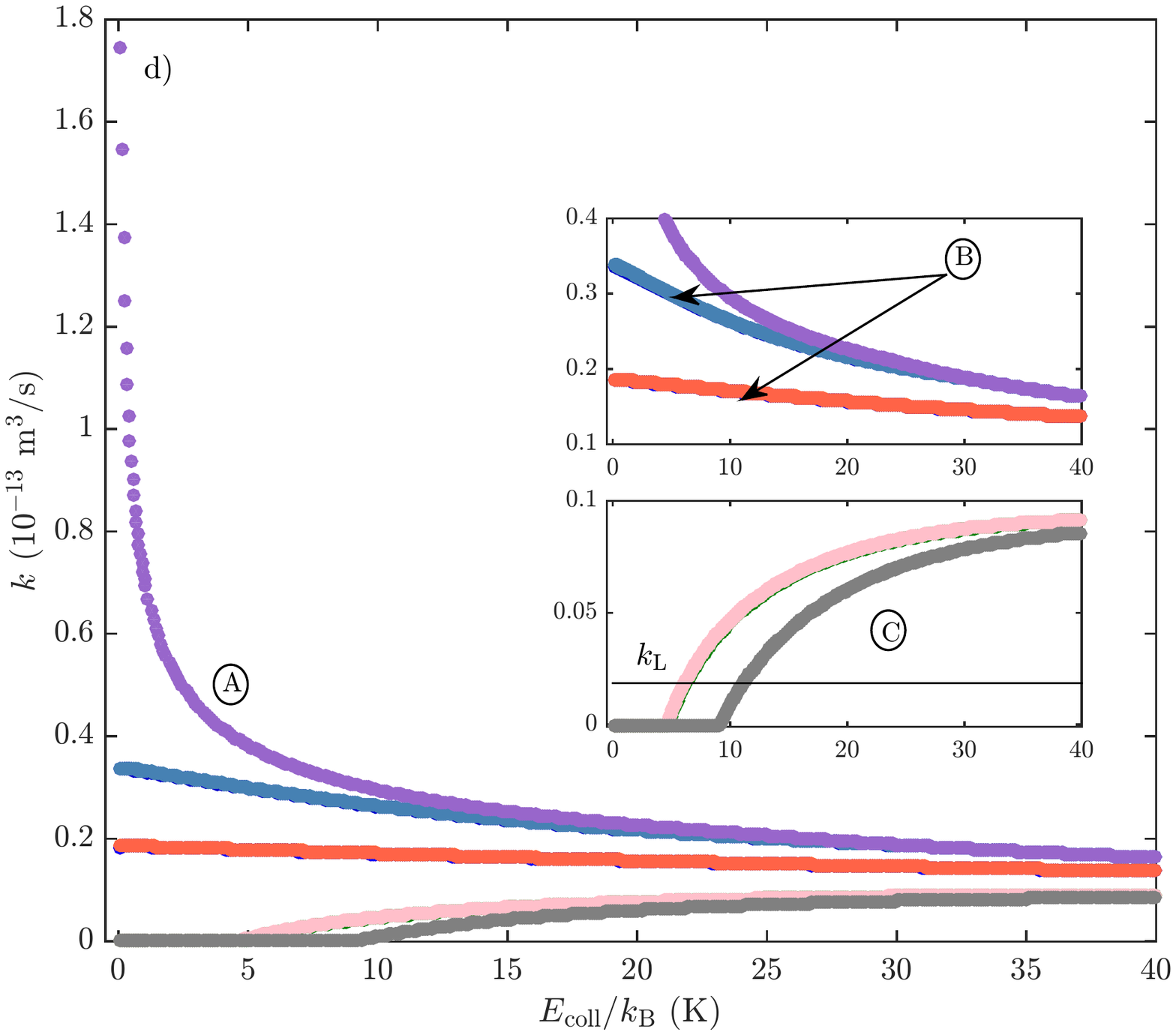}
\includegraphics[trim = 0.5cm 7.5cm 1cm 8.5cm, clip, width = 0.5\textwidth]{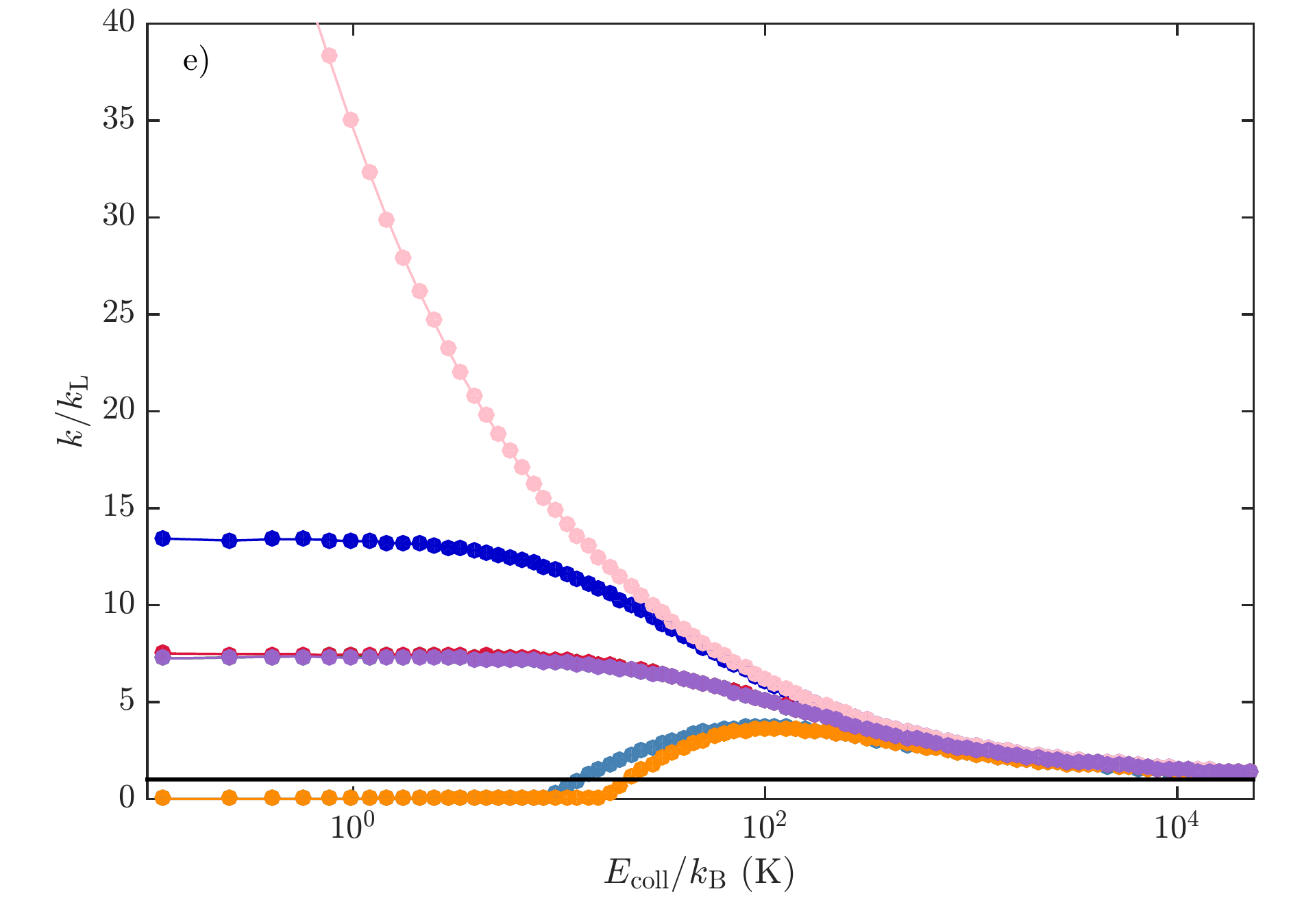}
\includegraphics[trim = 0.5cm 7.5cm 1cm 8.5cm, clip, width = 0.5\textwidth]{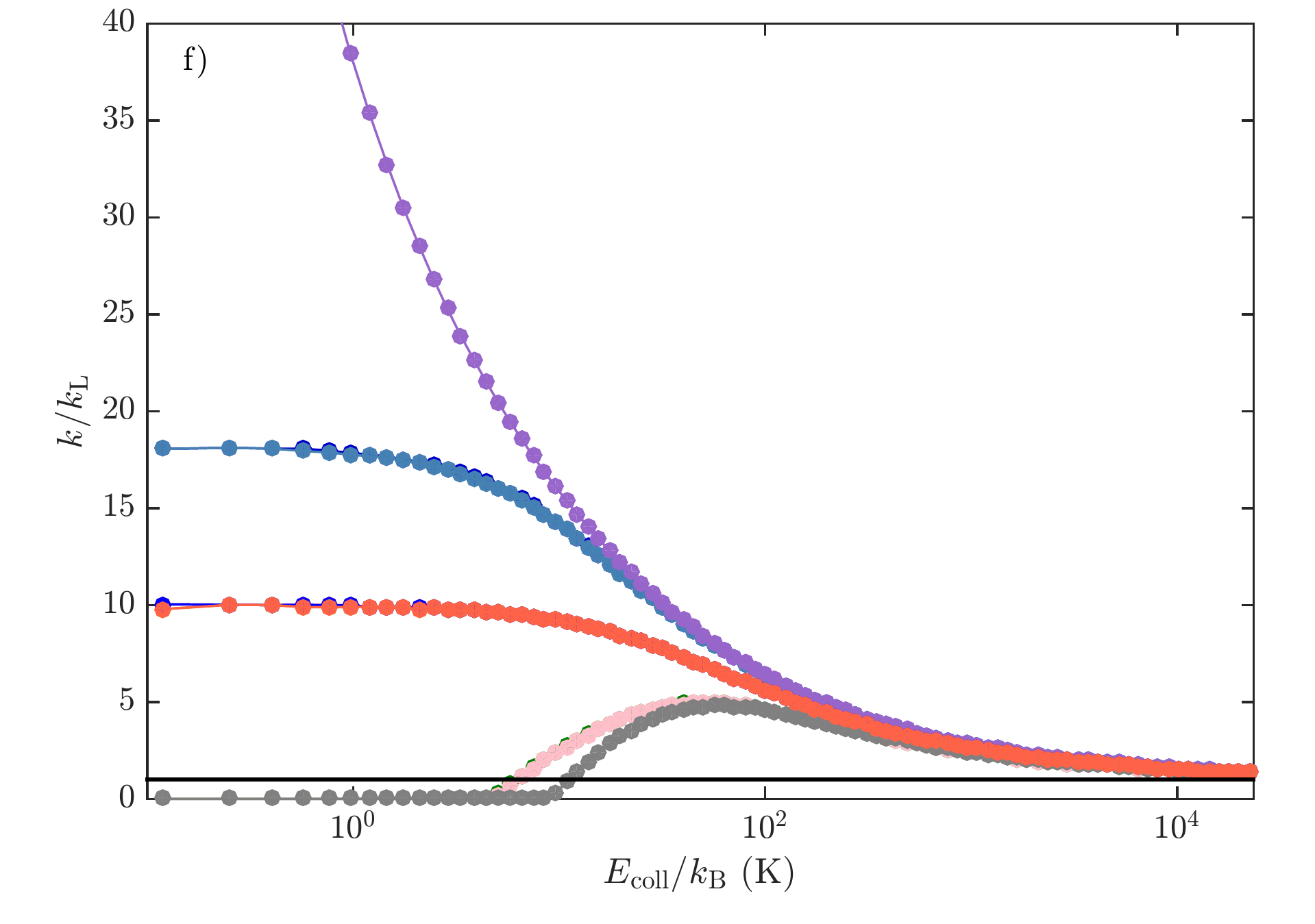}
\caption{\label{Fig3} Top panels: The interaction potential, $V_{\mathrm{int},i}(R)$, of the \ce{He+ + NH3}(a) and \ce{He+ + ND3}(b) reactions, for the molecular states with $J = 0,1$ and $|K| = 0,1$. The solid and dash-dotted coloured lines correspond to values of the angular momentum of the collision $L = \hslash\sqrt{\ell(\ell+1)}$ with $\ell = 0$ and $\ell = 20$, respectively. Black solid and dash-dotted lines represent the Langevin interaction potential, $V_{\mathrm{L}}(R)$. Middle and bottom panels: Calculated rotational-state-specific capture rate coefficients for $E_{\mathrm{coll}}/k_{\mathrm{B}}$ in the $0.05-40$ K (c,d) and $0.1-2.4\times10^{4}$ K (e,f) range. The horizontal black lines in (c-f) represent the value of the Langevin rate coefficient, $k_{\mathrm{L}}$, for the two reactions.} 
\end{figure*}

Because of the $R^{-2}$ dependence of the ion-dipole interaction, the $\Delta E_i(R)$ terms in Eq.~(\ref{eq:Vint}) dominate over the $R^{-4}$-dependent charge--induced-dipole interaction term for all states, as illustrated in Fig.~\ref{Fig3}. This figure compares the total interaction potentials, $V_{\mathrm{int},i}(R)$, for the states presented in Fig.~\ref{Fig2} for an s-wave collision ($L = 0$, solid lines) and a collision with $\ell = 20$ ($L^2 = \hslash^2\ell(\ell+1)$, dash-dotted lines), with $V_{\mathrm{L}}$ (black solid and dash-dotted lines, respectively). 

The Stark shifts significantly modify the potential barriers, and thus affect the value of $L_{\mathrm{max},i}$ for each collisional energy (see Eq.~(\ref{expr_k})). In states of type A and B, the centrifugal barrier is substantially lowered compared to the centrifugal barrier in $V_{\mathrm{L}}$, the effect being significantly more pronounced in states of type A. For example, the $\ell = 20$ term creates a centrifugal barrier of $\sim k_{\mathrm{B}}\cdot14$ K in the Langevin interaction potential [dash-dotted black line in Fig.~\ref{Fig3}(a)], whereas the total interaction potentials for the same value of $\ell$ in the A states are strongly attractive. The Stark effect in states of type A and B thus effectively suppresses the centrifugal barriers, which leads to significantly larger value of $L_{\mathrm{max},i}^{\mathrm{(A,B)}}$ and to much higher capture rate coefficients $k^{(\mathrm{A},\mathrm{B})}_i$. As the collisional energy increases, the $E_{\mathrm{coll}}^{-1/2}$-dependence of $k_i(E_{\mathrm{coll}})$ [see Eq.~(\ref{expr_k})] leads to a decrease of the rate coefficients. States of type C, in contrast, are characterised by a non-zero potential-energy barrier even for $\ell = 0$. This $\ell = 0$ potential-energy barrier is designated as $V_{\mathrm{int}}^{\mathrm{(C),\;max}}$ in Fig.~\ref{Fig3}(a) and (b). For molecules in states of type C, the ion-molecule pair experiences a repulsive potential and the capture rate coefficients vanish for $E_{\mathrm{coll}}<V_{\mathrm{int}}^{\mathrm{(C),\;max}}$.

\subsection{The rotational-state-dependent capture rate coefficients}
\label{sec:krates}
The calculated state-dependent capture rate coefficients, $k_i(E_{\mathrm{coll}})$, for the \ce{He+}+$\mathrm{NH_3}(J,K,M,\pm)$ and \ce{He+}+$\mathrm{ND_3}(J,K,M,\pm)$ reactions with $J = 0,1$ and $|K| = 0,1$ are displayed in Fig.~\ref{Fig3}(c) and (d). The calculations were performed for collisional energies $E_{\mathrm{coll}}/k_{\mathrm{B}}$ in the $0.05-40$ K range, with a step size of 50 mK. For states of type A and B, the rate coefficients are significantly higher than the Langevin rate constant $k_{\mathrm{L}}$ (black horizontal lines in the lower insets) and increase with decreasing collisional energy. The differences resulting from the quadratic Stark shift of B-type states and the linear Stark shift of the A-type states become apparent at collisional energies $E_{\mathrm{coll}}/k_{\mathrm{B}}\lesssim20$ K. Both A- and B-type states experience a gradual increase of $k$ in the $E_{\mathrm{coll}}/k_{\mathrm{B}}=20-40$ K range. However, below $E_{\mathrm{coll}}/k_{\mathrm{B}}\approx20$ K the $k^{\mathrm{(A)}}$ rate coefficients exhibit a much faster increase than the $k^{\mathrm{(B)}}$ rate coefficients. At the lowest collisional energies of $E_{\mathrm{coll}}/k_{\mathrm{B}}= 50$ mK, the rate coefficients of A-type states reach values of $k^{\mathrm{(A),He^++ND_3}} = 93.7\;k_{\mathrm{L}}$ and $k^{\mathrm{(A),He^++NH_3}} = 50.9\;k_{\mathrm{L}}$. The enhancement of the capture rate coefficients at $E_{\mathrm{coll}}/k_{\mathrm{B}}=50$ mK compared to $20$ K is $\sim1.1-1.3$ ($\sim1.2-1.6$) for type B states and $\sim4.5$ ($\sim6.8$) for type A states in \ce{NH3}  (\ce{ND3}).

The $k^{\mathrm{(C)}}$ coefficients are zero for $E_{\mathrm{coll}}<V^{\mathrm{(C),max}}_{\mathrm{int}}$. For $E_{\mathrm{coll}}\geq V^{\mathrm{(C),max}}_{\mathrm{int}}$, the rate coefficients rapidly increase with $E_{\mathrm{coll}}$, and typically already exceed the value of $k_{\mathrm{L}}$ at $E_{\mathrm{coll}}/k_{\mathrm{B}}\approx10-20$ K [see the insets in Fig.~\ref{Fig3}(c) and (d)].    

The behaviour of the capture rate coefficients in the high-collision-energy regime up to $E_{\mathrm{coll}}/k_{\mathrm{B}} = 2.4\times 10^4$ K is presented in Fig.~\ref{Fig3}(e) and (f). States of type A and B have rate coefficients which decrease monotonously with $E_{\mathrm{coll}}$, and approach the values of $k_{\mathrm{L}}$ for $E_{\mathrm{coll}}/k_{\mathrm{B}}\gtrsim10^4$ K. In the case of states of type C, in contrast, the rates reach a maximum at $E_{\mathrm{coll}}/k_{\mathrm{B}}\approx 100$ K and 60 K in the \ce{He+ +NH3} and \ce{He+ + ND3} reactions, respectively. Then they decrease again, eventually converging to the value of $k_{\mathrm{L}}$ at similar collision energies as the type-A and type-B states. Apparent in Fig.~\ref{Fig3}(e) and (f) is that the $k^{\mathrm{(B)}}$ rate coefficients increase with decreasing $E_{\mathrm{coll}}$, but eventually stabilise below $E_{\mathrm{coll}}/k_{\mathrm{B}}\lesssim5$ K at a value of $\sim7.5\;k_{\mathrm{L}}$ (for the $J = 1, KM = 0$ states) and $\sim13.5\;k_{\mathrm{L}}$ (for the $J=0$ states) in \ce{NH3} and $\sim10\;k_{\mathrm{L}}$ and $\sim18\;k_{\mathrm{L}}$ in \ce{ND3}, respectively. The greater increase of the $k^{\mathrm{(B)}}$ rates relative to $k_\mathrm{L}$ in \ce{ND3} is a consequence of the larger quadratic Stark-shifts of the B-type states at low fields. 

In the experiments, we measure the relative collision-energy-dependent product yields, which reflect the initial molecular population distribution over the $J = 0,1$ $|K |=0,1$ rotational states of NH$_3$ and ND$_3$. However, because the A-type states experience a linear Stark shift and a large increase over $k_\mathrm{L}$ even at relatively high collisional energies (e.g., below $E_{\mathrm{coll}}/k_{\mathrm{B}}<40$ K), the overall rate coefficient is dominated by the contribution from these states, even though they are occupied only by a small fraction of the molecular population at $T_{\mathrm{rot}}=6$ K ($\sim 0.182$ in \ce{NH3} and $\sim 0.196$ in \ce{ND3}, see Fig.~\ref{Fig7} below).  

\subsection{Modelling of the measured rate coefficients}
To model the experimentally measured relative product-ion signals as a function of the collision energy for the two reactions, $I_{\ce{NH+}+\ce{NH2+}}$ and $I_{\ce{ND+}+\ce{ND2+}}$, we proceed in the following way: In a first step, we calculate the weighted sum of the rotational-state-dependent capture rate constants, $k_i(E_{\mathrm{coll}})$, for a given rotational temperature, $T_{\mathrm{rot}}$, with weights determined by the corresponding state population. In a second step, we perform a convolution with a Gaussian distribution accounting for the collisional-energy-dependent energy-resolution, as described in Ref.~\cite{zhelyazkova20}, using the equation:
\begin{equation}
\frac{\Delta E_{\mathrm{coll}}}{k_{\mathrm{B}}} = \Delta T_{\mathrm{res}} + 2\sqrt{\Delta T_{\mathrm{res}}}\sqrt{\frac{E_{\mathrm{coll}}}{k_{\mathrm{B}}}},
\label{eq:ExpRes}
\end{equation}
where $\Delta E_{\mathrm{coll}}/k_{\mathrm{B}}$ is the FWHM of a Gaussian function. In Eq.~(\ref{eq:ExpRes}), $\Delta T_{\mathrm{res}}$ is an effective experimental temperature corresponding to the distribution of relative velocities of the He$(n)$ atoms and the ammonia molecules at zero nominal collisional energy (i.e., when $v_{\mathrm{Ryd}}-v_{\mathrm{c}}^{\mathrm{GS}}=0$). The value of $\Delta T_{\mathrm{res}}$ is mostly determined by the temperature of the Rydberg-atom cloud released from the quadrupole trap.\cite{zhelyazkova20} In a third step, we multiply the experimentally measured values of $I_{\ce{NH+}+\ce{NH2+}}$ and $I_{\ce{ND+}+\ce{ND2+}}$ by a global scaling factor such that the total integrated signals measured at the lowest collision energy probed experimentally matches the average rate constant determined in the second step at $E_{\mathrm{coll}}/k_{\mathrm{B}}=\Delta T_{\mathrm{res}}$. Finally, we vary $T_{\mathrm{rot}}$ and $\Delta T_{\mathrm{res}}$ until the best agreement between the experimental data and the model is obtained over the whole range of collisional energies probed. 

The comparison between the experimental data and the ion-molecule capture-rate model is presented in Fig.~\ref{Fig6}(c) and (d). The best agreement was achieved for $T_{\mathrm{rot}} = 6$ K and $\Delta T_{\mathrm{res}} = 200$ mK for both reactions. The coloured (orange and purple) dots and the black curves are the rates obtained in the second and third steps of the procedure described above, respectively. The insets display the behaviour at low collisional energies ($E_{\mathrm{coll}}/k_{\mathrm{B}}\lesssim6$ K) on an enlarged scale. The agreement between the data and the model taking into account the finite resolution [black lines in Fig.~\ref{Fig6}(c) and (d)] is excellent for both reactions. The model thus quantitatively describes the experimentally observed increase of the product-ion signals with decreasing collisional energy. The overall behaviour exhibits a gradual, slow increase over the $E_{\mathrm{coll}}/k_{\mathrm{B}}\approx5-40$ K range and a much steeper increase for $E_{\mathrm{coll}}/k_{\mathrm{B}}\lesssim5$ K. The latter effect is significantly more pronounced for the \ce{He+ + ND3} reaction than for the \ce{He+ + NH3} reaction. 

The calculated rate coefficients averaged over the rotational population of the molecules [purple and orange dots in Fig.~\ref{Fig6}(c) and (d)] and taking the finite experimental collision-energy resolution into account [black lines in Fig.~\ref{Fig6}(c) and (d)] are enhanced by factors of $\sim2.29$ ($\sim2.81$) at a collision energy of $k_{\mathrm{B}}\cdot200$ mK compared to the value at $E_{\mathrm{coll}}=k_{\mathrm{B}}\cdot40$ K and of $\sim1.71$ ($\sim2.09$) compared to the value at $E_{\mathrm{coll}}=k_{\mathrm{B}}\cdot 10$ K for the \ce{He+ + NH3} (\ce{He+ + ND3}) reaction. 
\begin{figure*}
\includegraphics[trim = 6cm 3.4cm 6.5cm 3.7cm, clip,width = 0.5\textwidth]{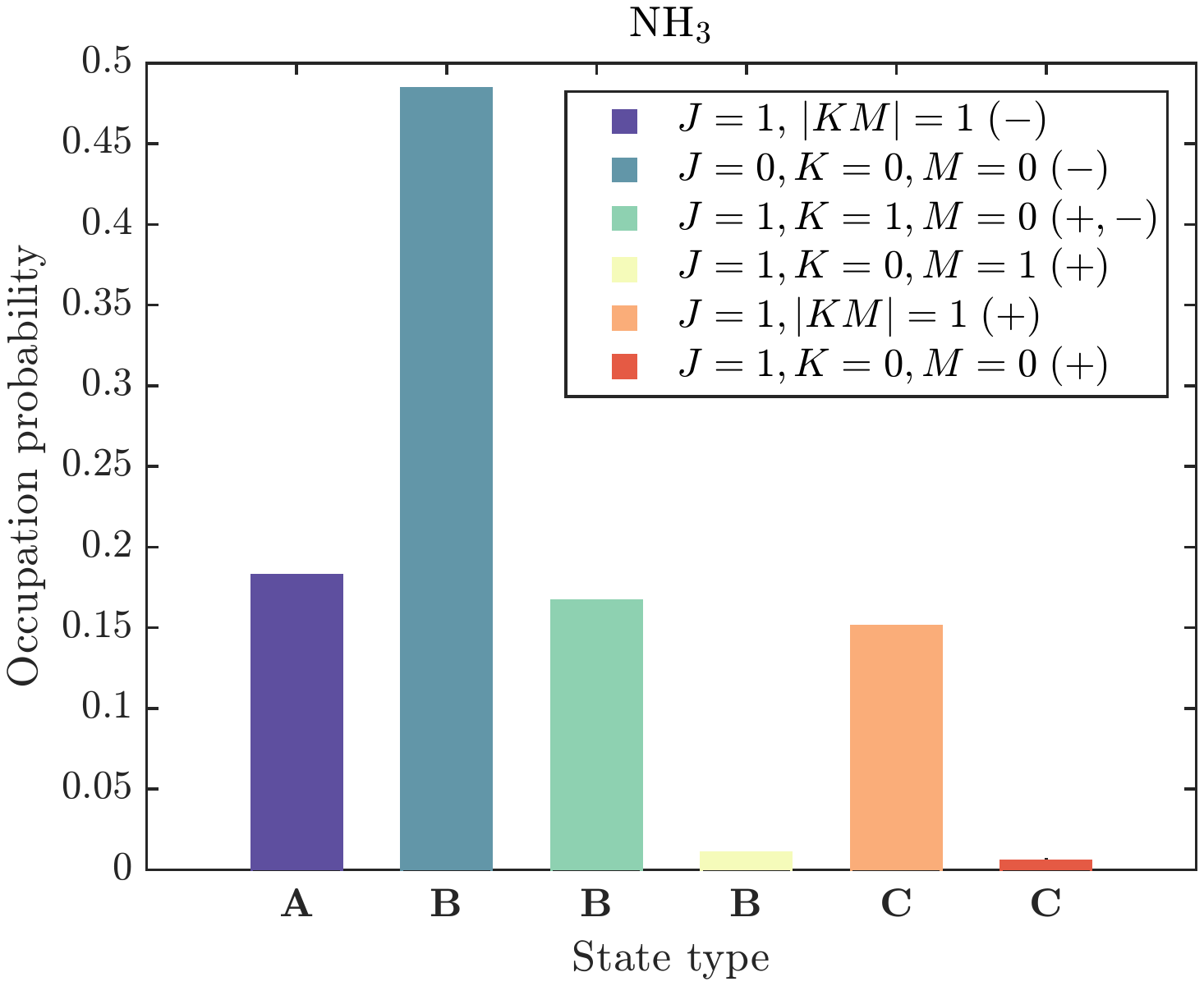}
\includegraphics[trim = 6cm 3.4cm 6.5cm 3.7cm, clip,width = 0.5\textwidth]{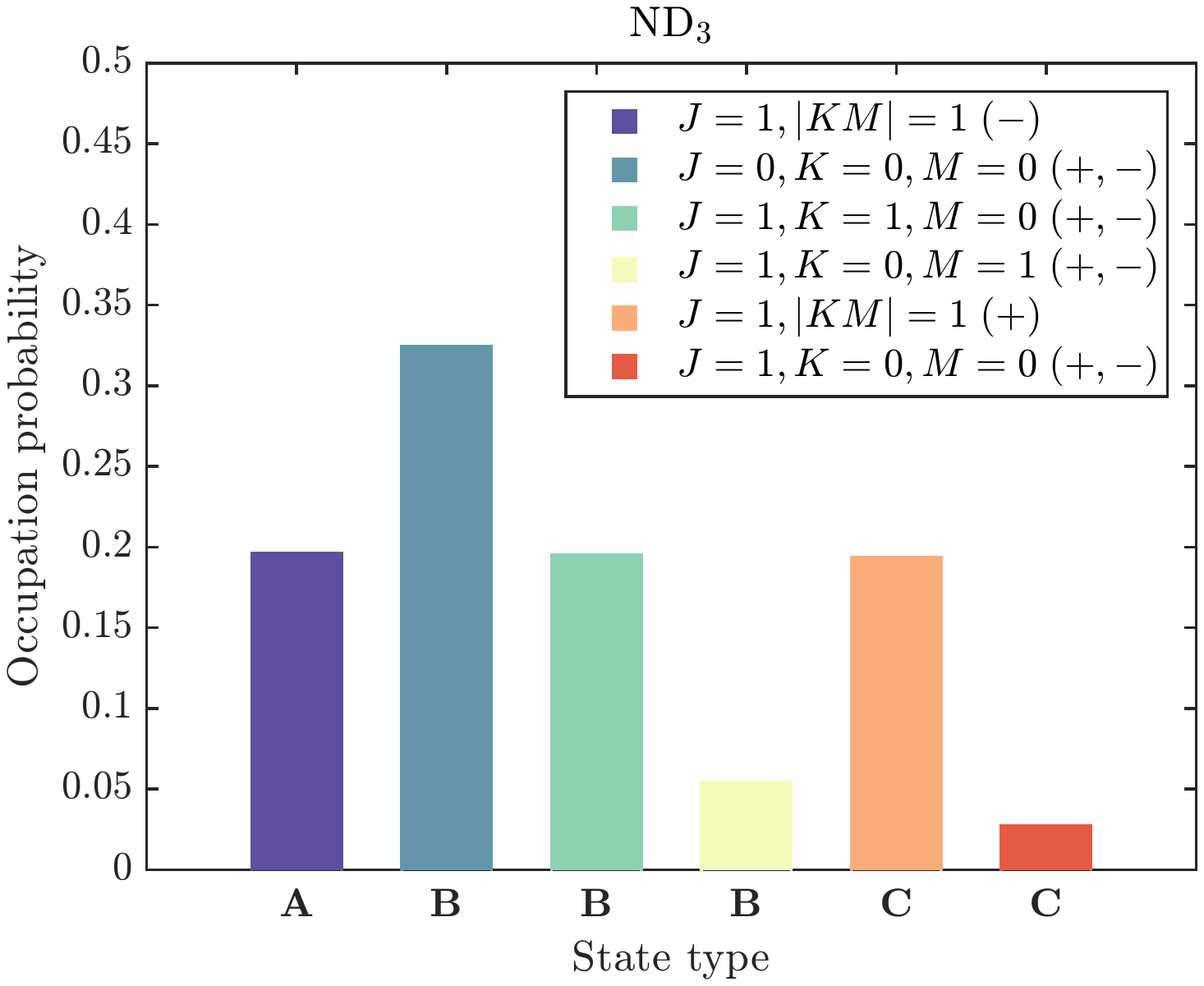}
\caption{\label{Fig7} Occupation probabilities of the A-, B- and C-type of states in \ce{NH3} and \ce{ND3}, at a rotational temperature of $T_{\mathrm{rot}}=6$ K. See text for details.}
\end{figure*}
The observed larger increase with decreasing collision energy of the rates of the \ce{He+ + ND3} reaction is perfectly reproduced by the calculations and originates from (i) the larger rotational-state-dependent rate constants compared to the \ce{He+ + NH3} reaction, as discussed in Sec.~\ref{sec:krates}, and (ii) the higher contribution to the rate coefficients from type-A states in \ce{ND3} resulting from the nuclear-spin statistical factors, as is now explained. 

The occupation probabilities $p$ of the A-, B- and C-type states in \ce{NH3} and \ce{ND3} at a rotational temperature of $T_{\mathrm{rot}}=6$ K are presented in Fig.~\ref{Fig7} and are  $p^{\mathrm{(A),\ce{NH3}}} = 0.182$, $p^{\mathrm{(B),\ce{NH3}}} = 0.662$ and $p^{\mathrm{(C),\ce{NH3}}} = 0.156$ for \ce{NH3} and $p^{\mathrm{(A),\ce{ND3}}} = 0.196$, $p^{\mathrm{(B),\ce{ND3}}} = 0.580$ and $p^{\mathrm{(C),\ce{ND3}}} = 0.224$ for \ce{ND3}. The experimentally measured increase of product-ion signals with decreasing collisional energy can thus be almost entirely attributed to states of type A and B. The observed steeper increase below $E_{\mathrm{coll}}/k_{\mathrm{B}}\lesssim5$ K of the rate of the \ce{He+ + ND3} reaction compared to the rate of the \ce{He+ + NH3} reaction results from: (i) the more pronounced Stark effect in \ce{ND3} and (ii) the larger fraction of molecules in high-field-seeking states of type A: $p^{\mathrm{(A),\ce{ND3}}}/(p^{\mathrm{(A),\ce{ND3}}}+p^{\mathrm{(B),\ce{ND3}}}) = 0.255$ in \ce{ND3} compared to $p^{\mathrm{(A),\ce{NH3}}}/(p^{\mathrm{(A),\ce{NH3}}}+p^{\mathrm{(B),\ce{NH3}}}) = 0.216$ in \ce{NH3}.   

\subsection{Thermal rate constants}
 The good agreement between the experimental and calculated collision-energy-dependent rates (see Fig.~\ref{Fig6}) validates the model used to calculate the rotational-state-specific rate coefficients $k_i(E_{\rm coll})$ with $i = |JKM\pm1\rangle$. These coefficients can in turn be used to determine the thermal rate constants $k(T)$ with the relations~\cite{levine05}
 \begin{equation}\label{thermal_rates1}
 k(T) = \sum_i g_{{\rm ns},i} \exp{\left(-{\frac{E_{{\rm rot},i}-E_0}{k_{\rm B}T}}\right)}k_i(T),
 \end{equation}
 and
\begin{equation}\label{thermal_rates2}
k_i(T) = \sqrt{\frac{8k_{\rm B}T}{\pi\mu}} \int_0^\infty \sigma_i(E_{\rm coll}) \left(\frac{E_{\rm coll}}{k_{\rm B}T}\right)\exp{\left(-{\frac{E_{\rm coll}}{k_{\rm B}T}}\right)} {\rm d}\left(\frac{E_{\rm coll}}{k_{\rm B}T}\right),
\end{equation}
where the state-specific reaction cross sections $\sigma_i(E_{\rm coll})$ are given by $k_i(E_{\rm coll})/|v_{\rm rel}|$ and $v_{\rm rel}$ is the relative velocity $v_{\rm Ryd}-v^{\rm GS}$. In Eq.~(\ref{thermal_rates1}), $g_{{\rm ns},i}$ represents the nuclear-spin-statistical factors (see Table~\ref{tab1}) and $E_{{\rm rot},i}$ and $E_0$ are the rotational energy of state $i$ and the zero-point energy, respectively.

\begin{figure*}
\includegraphics[trim = 1cm 0cm 1cm 0cm, clip,width = 0.5\textwidth]{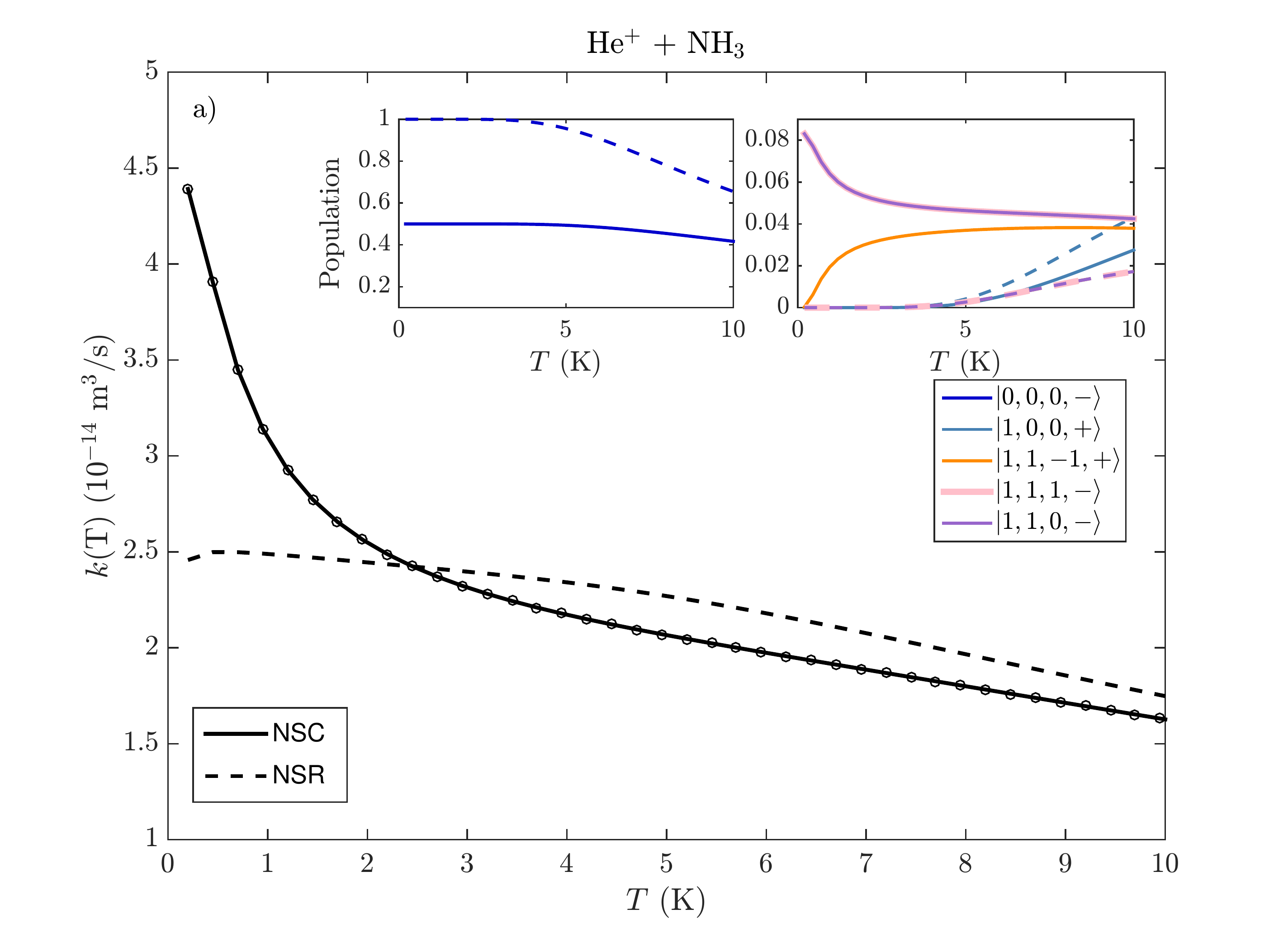}
\includegraphics[trim = 1cm 0cm 1cm 0cm,width = 0.5\textwidth]{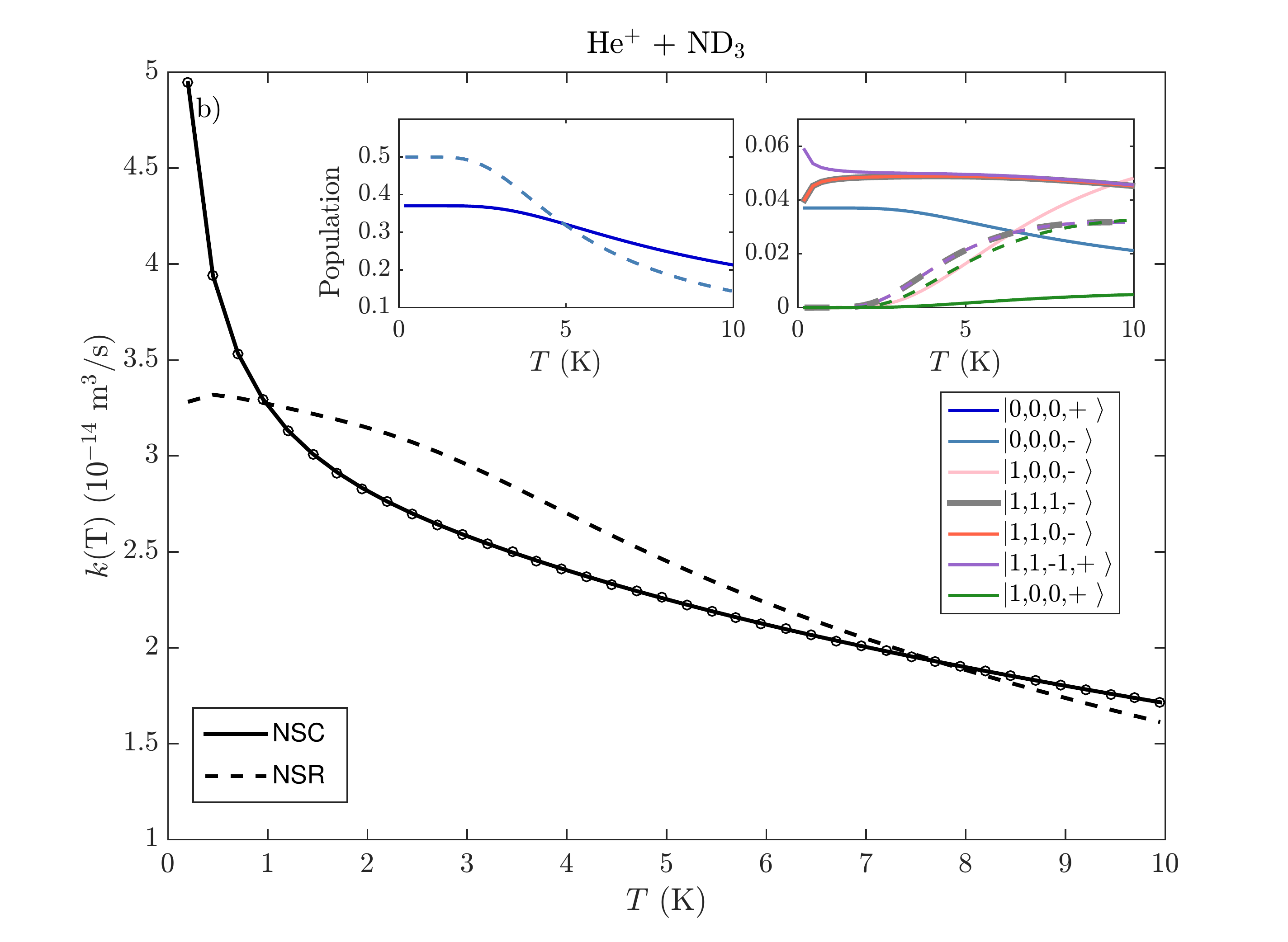}
\caption{\label{Fig8} Calculated thermal rate coefficients for the \ce{He^+ + NH3}(a) and \ce{He^+ + ND3}(b) reactions in the temperature range between 0.2~K and 10~K. The calculations with nuclear-spin-symmetry conservation (NSC) and nuclear-spin-symmetry relaxation (NSR) are presented as solid  and dashed black curves, respectively. Insets: rotational states population of selected states in \ce{NH3} and \ce{ND3} as a function of the temperature, assuming nuclear-spin-symmetry conservation (solid coloured lines) and nuclear-spin-symmetry relaxation effects (dashed coloured lines).}
\end{figure*}

In an environment where conversion between the two nuclear-spin isomers is possible, $E_0$ is taken as the energy of the $J=0,K=0$ rotational ground state. If the nuclear-spin-symmetry is conserved, $E_0$ represents the rotational energy of the ground state of the respective nuclear-spin isomer, i.e., $E_0$ is the energy of the $(J=0,K=0)^u$ ground state for ortho NH$_3$ and ND$_3$, the energy of the lower tunnelling component of the $(J=1,K=1)$ rotational level for para
NH$_3$ and ND$_3$, and the energy of the $(J = 0, K = 0)^l$ for meta \ce{ND3} (see Fig.~\ref{Fig1}). The rate constants calculated assuming that the relative populations of the ortho, para and meta nuclear-spin isomers correspond to the nuclear-spin-statistical factors are presented in
panels (a) and (b) of Fig.~\ref{Fig8} for NH$_3$ and ND$_3$, respectively. In this figure, the full lines correspond to the case where the nuclear spin is assumed to be conserved and the dashed lines to the case where ortho-para conversion is assumed to be possible.

Whereas the nuclear-spin symmetry is known to be conserved to an excellent approximation in low-density gases and supersonic expansions,\cite{wichmann20,tanner12} conversion between the different nuclear-spin isomers may take place if the ammonia molecules are adsorbed at the surface of certain materials and dust grains. The insets in Fig.~\ref{Fig8} give the relative population of the different rotational levels for these two cases, labelled nuclear-spin-conservation (NSC) and nuclear-spin-relaxation (NSR), respectively. In both cases, the thermal capture rate is enhanced as the temperature decreases. However, the enhancement is more pronounced if the nuclear-spin symmetry is conserved because, in this case, there is always a significant population of the high-field-seeking tunnelling component of the $J=1, K=1$  state, even at 0~K.

Equations~(\ref{thermal_rates1}) and~(\ref{thermal_rates2}) can be used to calculate the thermal capture rate coefficients at the temperatures of 27, 68, and 300~K for which the rate of the He$^+$ + NH$_3$ reaction was measured in uniform supersonic flows.\cite{marquette85a} The results are presented in Table~\ref{table4}, where the first and second rows of the body of the table list the rates calculated assuming that the nuclear-spin symmetry is conserved (NSC) or relaxed (NSR), respectively. Beyond 10~K, the state populations do not strongly depend any more on whether NSR or NSC is assumed, so that the thermal rate constant with both assumption are almost identical. The third row gives the experimental values reported by Marquette {\it et al.}\cite{marquette85a} The calculated rates are larger than the experimental ones, which is expected given that the calculated capture rates correspond to 100\% reaction probability upon close encounter of the reactants whereas the experiments only monitor the reactive collisions. The comparison of experimental and calculated values of the thermal rate coefficient suggests that about 40 to 50\% of the Langevin collisions are reactive. The last two columns of Table~\ref{table4} present the ratios $r_1$ and $r_2$ of the thermal rate coefficient obtained at 27~K to those obtained at 68 and 300~K. The calculated $r_1$ value is in perfect agreement with the experimental result, whereas the experimental and calculated values of $r_2$ differ by about 30\%. 
\begin{table}[]
	\caption{\label{table4} Comparison of the thermal capture rate coefficients of the He$^+$ + NH$_3$ reaction calculated with Eqs.~(\ref{thermal_rates1}) and~(\ref{thermal_rates2}) assuming nuclear-spin conservation (NSC) or nuclear-spin relaxation (NSR) during the supersonic expansion and the experimental reaction rate coefficients results of Marquette {\it et al.} \cite{marquette85a} The last two columns give the ratios $r_1$ and $r_2$ of the value obtained at 27~K to the values obtained at 68 and 300~K, respectively.}
	%\begin{ruledtabular} 
	\begin{tabular}{l| c c c c c}
		& $k$(27~K) 	& $k$(68~K) &  $k$(300~K) & $r_1$  & $r_2$\\
		& (m$^3$s$^{-1}$) 	& (m$^3$s$^{-1}$) &  (m$^3$s$^{-1}$) & & \\ \hline 
		NSC  & $1.13\cdot 10^{-14}$  & $7.39\cdot 10^{-15}$    	&  $3.18\cdot 10^{-15}$  & 1.5 & 	3.6	      \\ 	
		NSR  & $1.14\cdot 10^{-14}$  & $7.48\cdot 10^{-15}$    	&  $3.11\cdot 10^{-15}$  & 1.52 & 	3.7	      \\ 
		Exp. \cite{marquette85a} & $4.5\cdot 10^{-15}$ & $3.0\cdot 10^{-15}$  &     $1.65\cdot 10^{-15}$	&  1.5  & 2.7   \\ 
	\end{tabular}
	%\end{ruledtabular}
\end{table}

\section{Conclusions}
\label{sec:concl}
In this article, which is the first in a series of three articles dedicated to the role of long-range electrostatic interactions in ion-molecule reactions at low temperatures, we have presented the results of measurements of the collision-energy-dependent rates of the reactions between He$^+$ and ammonia (NH$_3$ and ND$_3$) in the range of collision energies between $k_{\mathrm{B}}\cdot 0.2$~K and $k_{\mathrm{B}}\cdot 40$~K. To reach such low collision energies, we used a merged-beam approach and observed the reactions within the orbit of a highly excited Rydberg electron. In this range, we observed that the rates increase strongly with decreasing collision energy, the increase being particularly pronounced below $k_{\mathrm{B}}\cdot5 $~K. We have also observed that the rates of reactions involving ND$_3$ increase more strongly than the rates of the reactions involving NH$_3$ at low collision energies. To interpret the experimental results, we have calculated the rates using an adiabatic capture model based on the calculation of the Stark shifts of the rotational levels of \ce{NH3} and \ce{ND3} in the field of the He$^+$ ion.

The calculations provided a quantitatively accurate description of the experimental observations and could be used to rationalise the observed collision-energy-dependent reaction yields in terms of capture rate coefficient for rotationally state-selected ($|JKM\pm \rangle$) ammonia molecules. These states could be classified into three categories: (A) States with $|KM|\ge 1$ correlating at low fields with the lower components of the inversion-tunnelling doublets. These states undergo linear negative Stark shifts in the field of the colliding ion, which suppresses the centrifugal barriers in the effective intermolecular potential and greatly enhances the capture rates at low collision energies; (B) States with $KM=0$, which only become sensitive to the fields at high fields through the quadratic and higher-order Stark shifts. The corresponding capture rates are only weakly enhanced at low collision energies. (C) States with positive linear Stark shifts at low fields which generate large potential energy barriers and completely suppress the reactions at low collision energies (below $\sim k_{\mathrm{B}}\cdot15$ K). At high collision energies, beyond $k_{\mathrm{B}}\cdot300$~K, the capture rates for all three categories gradually converge to the Langevin capture rates. These observations are a direct consequence of the dipolar nature of ammonia.

The model also provided an explanation for the different behaviour of the reactions involving NH$_3$ and ND$_3$ at low collision energies. The stronger enhancements of the rates observed below $k_{\mathrm{B}}\cdot5$ K in the reactions of ND$_3$ was interpreted as arising from the smaller tunnelling splittings of the inversion mode and from the larger population of the rotational states belonging to category A at the 6~K rotational temperature of the supersonic beams. The latter effect results primarily from the different nuclear-spin statistical weights in NH$_3$ and ND$_3$. From the calculated rotational-state-specific collision-energy-dependent captures rates, {\it thermal} capture rate coefficients were determined in the temperature range between 0.2 and 10~K of relevance for modelling reactions in the interstellar medium. Assuming that the relative populations of the ortho, para and meta nuclear-spin isomers reflect the nuclear-spin-statistical factors, as can be expected for dilute gases and supersonic expansions, leads to a large nonlinear increase of the thermal capture rates below 5~K. The thermal capture rate constant reach the value of $4.4\times 10^{-14}$~m$^3$/s at 0.2~K, which is more than 20 times larger than the Langevin rate constant.
Assuming that ortho-para-meta conversion processes can take place, as might be the case if the molecules have thermalised at the surface of certain materials and dust grains, leads to slightly larger rate enhancements above 2~K, but to smaller rate enhancements at the lowest temperatures (see Fig.~\ref{Fig8}).

In future work, it will be important to obtain data at even lower temperatures and improved energy resolution. Moreover, the preparation of rotationally state-selected samples would enable more extensive tests of the model predictions. The generation of pure samples of molecules in high-field-seeking states of the category A seems particularly attractive in this regard, and recent experiments have demonstrated ways of generating such samples~\cite{ploenes21}. Calculations of the (nonadiabatic) dynamics on the multidimensional potential-energy surfaces involved would be needed to interpret the observed branching ratios and the fact that only 40\% of the close collisions appear to be reactive.

\section{Acknowledgements}
We thank M. \v{Z}e\v{s}ko for his contributions to the initial phase of this project and K. H\"{o}veler and J. Deiglmayr for fruitful discussions. This work is supported financially by the Swiss National Science Foundation (Grant No. 200020B-200478) and by the European Research Council through the ERC advanced grant (Grant No. 743121) under the European Union's Horizon 2020 research and innovation programme. 
%%%END OF MAIN TEXT%%%


\balance

%\newpage

%\bibliography{article_final_pruned}

\begin{mcitethebibliography}{81}
	\providecommand*{\natexlab}[1]{#1}
	\providecommand*{\mciteSetBstSublistMode}[1]{}
	\providecommand*{\mciteSetBstMaxWidthForm}[2]{}
	\providecommand*{\mciteBstWouldAddEndPuncttrue}
	{\def\EndOfBibitem{\unskip.}}
	\providecommand*{\mciteBstWouldAddEndPunctfalse}
	{\let\EndOfBibitem\relax}
	\providecommand*{\mciteSetBstMidEndSepPunct}[3]{}
	\providecommand*{\mciteSetBstSublistLabelBeginEnd}[3]{}
	\providecommand*{\EndOfBibitem}{}
	\mciteSetBstSublistMode{f}
	\mciteSetBstMaxWidthForm{subitem}
	{(\emph{\alph{mcitesubitemcount}})}
	\mciteSetBstSublistLabelBeginEnd{\mcitemaxwidthsubitemform\space}
	{\relax}{\relax}
	
	\bibitem[Bowers(1979)]{bowers79a}
	\emph{Gas Phase Ion Chemistry: Vol. 1 and 2}, ed. M.~T. Bowers, Academic Press,
	New York, 1979\relax
	\mciteBstWouldAddEndPuncttrue
	\mciteSetBstMidEndSepPunct{\mcitedefaultmidpunct}
	{\mcitedefaultendpunct}{\mcitedefaultseppunct}\relax
	\EndOfBibitem
	\bibitem[Ausloos(1978)]{ausloos78a}
	\emph{Kinetics of Ion-Molecule Reactions}, ed. P.~Ausloos, Plenum Press, New
	York, 1978\relax
	\mciteBstWouldAddEndPuncttrue
	\mciteSetBstMidEndSepPunct{\mcitedefaultmidpunct}
	{\mcitedefaultendpunct}{\mcitedefaultseppunct}\relax
	\EndOfBibitem
	\bibitem[Clary(1985)]{clary85a}
	D.~C. Clary, \emph{Mol. Phys.}, 1985, \textbf{54}, 605--618\relax
	\mciteBstWouldAddEndPuncttrue
	\mciteSetBstMidEndSepPunct{\mcitedefaultmidpunct}
	{\mcitedefaultendpunct}{\mcitedefaultseppunct}\relax
	\EndOfBibitem
	\bibitem[Troe(1987)]{troe87a}
	J.~Troe, \emph{Chem. Phys.}, 1987, \textbf{87}, 2773--2780\relax
	\mciteBstWouldAddEndPuncttrue
	\mciteSetBstMidEndSepPunct{\mcitedefaultmidpunct}
	{\mcitedefaultendpunct}{\mcitedefaultseppunct}\relax
	\EndOfBibitem
	\bibitem[Clary(1990)]{clary90a}
	D.~C. Clary, \emph{Ann. Rev. Phys. Chem.}, 1990, \textbf{41}, 61--90\relax
	\mciteBstWouldAddEndPuncttrue
	\mciteSetBstMidEndSepPunct{\mcitedefaultmidpunct}
	{\mcitedefaultendpunct}{\mcitedefaultseppunct}\relax
	\EndOfBibitem
	\bibitem[Troe(1996)]{troe96a}
	J.~Troe, \emph{J. Chem. Phys.}, 1996, \textbf{105}, 6249--6262\relax
	\mciteBstWouldAddEndPuncttrue
	\mciteSetBstMidEndSepPunct{\mcitedefaultmidpunct}
	{\mcitedefaultendpunct}{\mcitedefaultseppunct}\relax
	\EndOfBibitem
	\bibitem[Ng and Baer(1992)]{ng92a}
	\emph{State-selected and state-to-state ion-molecule reaction dynamics: Part 1.
		Experiment and Part 2. Theory}, ed. C.-Y. Ng and M.~Baer, John Wiley \& Sons,
	Inc., New York, 1992\relax
	\mciteBstWouldAddEndPuncttrue
	\mciteSetBstMidEndSepPunct{\mcitedefaultmidpunct}
	{\mcitedefaultendpunct}{\mcitedefaultseppunct}\relax
	\EndOfBibitem
	\bibitem[Gerlich(2008)]{gerlich08a}
	D.~Gerlich, \emph{The study of cold collisions using ion guides and traps, in:
		Low Temperatures and Cold Molecules}, Ed. I. M. W. Smith, Imperial College
	Press, London, 2008, ch.~3, pp. 121--174\relax
	\mciteBstWouldAddEndPuncttrue
	\mciteSetBstMidEndSepPunct{\mcitedefaultmidpunct}
	{\mcitedefaultendpunct}{\mcitedefaultseppunct}\relax
	\EndOfBibitem
	\bibitem[Willitsch \emph{et~al.}(2008)Willitsch, Bell, Gingell, and
	Softley]{willitsch08a}
	S.~Willitsch, M.~T. Bell, A.~D. Gingell and T.~P. Softley, \emph{Phys. Chem.
		Chem. Phys.}, 2008, \textbf{10}, 7200--7210\relax
	\mciteBstWouldAddEndPuncttrue
	\mciteSetBstMidEndSepPunct{\mcitedefaultmidpunct}
	{\mcitedefaultendpunct}{\mcitedefaultseppunct}\relax
	\EndOfBibitem
	\bibitem[Heazlewood and Softley(2021)]{heazlewood21a}
	B.~R. Heazlewood and T.~P. Softley, \emph{Nat. Rev. Chem.}, 2021, \textbf{5},
	125--140\relax
	\mciteBstWouldAddEndPuncttrue
	\mciteSetBstMidEndSepPunct{\mcitedefaultmidpunct}
	{\mcitedefaultendpunct}{\mcitedefaultseppunct}\relax
	\EndOfBibitem
	\bibitem[Marshall \emph{et~al.}(1997)Marshall, Izgi, and Muenter]{marshall97a}
	M.~D. Marshall, K.~C. Izgi and J.~S. Muenter, \emph{J. Chem. Phys.}, 1997,
	\textbf{107}, 1037--1044\relax
	\mciteBstWouldAddEndPuncttrue
	\mciteSetBstMidEndSepPunct{\mcitedefaultmidpunct}
	{\mcitedefaultendpunct}{\mcitedefaultseppunct}\relax
	\EndOfBibitem
	\bibitem[Graham \emph{et~al.}(1998)Graham, Imrie, and Raab]{graham98}
	C.~Graham, D.~A. Imrie and R.~E. Raab, \emph{Mol. Phys.}, 1998, \textbf{93},
	49--56\relax
	\mciteBstWouldAddEndPuncttrue
	\mciteSetBstMidEndSepPunct{\mcitedefaultmidpunct}
	{\mcitedefaultendpunct}{\mcitedefaultseppunct}\relax
	\EndOfBibitem
	\bibitem[Birnbaum and Cohen(1975)]{birnbaum75}
	G.~Birnbaum and E.~R. Cohen, \emph{J. Chem. Phys.}, 1975, \textbf{62},
	3807--3812\relax
	\mciteBstWouldAddEndPuncttrue
	\mciteSetBstMidEndSepPunct{\mcitedefaultmidpunct}
	{\mcitedefaultendpunct}{\mcitedefaultseppunct}\relax
	\EndOfBibitem
	\bibitem[Albritton(1978)]{albritton78}
	D.~Albritton, \emph{At. Data Nucl. Data Tables}, 1978, \textbf{22}, 1--89\relax
	\mciteBstWouldAddEndPuncttrue
	\mciteSetBstMidEndSepPunct{\mcitedefaultmidpunct}
	{\mcitedefaultendpunct}{\mcitedefaultseppunct}\relax
	\EndOfBibitem
	\bibitem[Anicich and Huntress(1986)]{anicich86}
	V.~G. Anicich and W.~T. Huntress, \emph{Astrophys. J. Suppl. Ser.}, 1986,
	\textbf{62}, 553--672\relax
	\mciteBstWouldAddEndPuncttrue
	\mciteSetBstMidEndSepPunct{\mcitedefaultmidpunct}
	{\mcitedefaultendpunct}{\mcitedefaultseppunct}\relax
	\EndOfBibitem
	\bibitem[Stoecklin \emph{et~al.}(1992)Stoecklin, Clary, and
	Palma]{stoecklin92a}
	T.~Stoecklin, D.~C. Clary and A.~Palma, \emph{J. Chem. Soc. Faraday Trans.},
	1992, \textbf{88}, 901--908\relax
	\mciteBstWouldAddEndPuncttrue
	\mciteSetBstMidEndSepPunct{\mcitedefaultmidpunct}
	{\mcitedefaultendpunct}{\mcitedefaultseppunct}\relax
	\EndOfBibitem
	\bibitem[Dashevskaya \emph{et~al.}(2005)Dashevskaya, Litvin, Nikitin, and
	Troe]{dashevskaya05a}
	E.~I. Dashevskaya, I.~Litvin, E.~E. Nikitin and J.~Troe, \emph{J. Chem. Phys.},
	2005, \textbf{122}, 184311\relax
	\mciteBstWouldAddEndPuncttrue
	\mciteSetBstMidEndSepPunct{\mcitedefaultmidpunct}
	{\mcitedefaultendpunct}{\mcitedefaultseppunct}\relax
	\EndOfBibitem
	\bibitem[Auzinsh \emph{et~al.}(2013)Auzinsh, Dashevskaya, Litvin, Nikitin, and
	Troe]{auzinsh13a}
	M.~Auzinsh, E.~I. Dashevskaya, I.~Litvin, E.~E. Nikitin and J.~Troe, \emph{J.
		Chem. Phys.}, 2013, \textbf{139}, 084311\relax
	\mciteBstWouldAddEndPuncttrue
	\mciteSetBstMidEndSepPunct{\mcitedefaultmidpunct}
	{\mcitedefaultendpunct}{\mcitedefaultseppunct}\relax
	\EndOfBibitem
	\bibitem[Auzinsh \emph{et~al.}(2013)Auzinsh, Dashevskaya, Litvin, Nikitin, and
	Troe]{auzinsh13b}
	M.~Auzinsh, E.~I. Dashevskaya, I.~Litvin, E.~E. Nikitin and J.~Troe, \emph{J.
		Chem. Phys.}, 2013, \textbf{139}, 144315\relax
	\mciteBstWouldAddEndPuncttrue
	\mciteSetBstMidEndSepPunct{\mcitedefaultmidpunct}
	{\mcitedefaultendpunct}{\mcitedefaultseppunct}\relax
	\EndOfBibitem
	\bibitem[Dashevskaya \emph{et~al.}(2016)Dashevskaya, Litvin, Nikitin, and
	Troe]{dashevskaya16a}
	E.~I. Dashevskaya, I.~Litvin, E.~E. Nikitin and J.~Troe, \emph{J. Chem. Phys.},
	2016, \textbf{145}, 244315\relax
	\mciteBstWouldAddEndPuncttrue
	\mciteSetBstMidEndSepPunct{\mcitedefaultmidpunct}
	{\mcitedefaultendpunct}{\mcitedefaultseppunct}\relax
	\EndOfBibitem
	\bibitem[Hougen(1976)]{hougen76a}
	J.~T. Hougen, \emph{{"Methane Symmetry operations" in "Spectroscopy"}}, ed. D.
	A. Ramsay, Butterworths, London, 1976, vol.~3, pp. 75--125\relax
	\mciteBstWouldAddEndPuncttrue
	\mciteSetBstMidEndSepPunct{\mcitedefaultmidpunct}
	{\mcitedefaultendpunct}{\mcitedefaultseppunct}\relax
	\EndOfBibitem
	\bibitem[Bunker and Jensen(2006)]{bunker06}
	P.~R. Bunker and P.~Jensen, \emph{Molecular Symmetry and Spectroscopy}, NRC
	research press, Ottawa, Canada, 2006\relax
	\mciteBstWouldAddEndPuncttrue
	\mciteSetBstMidEndSepPunct{\mcitedefaultmidpunct}
	{\mcitedefaultendpunct}{\mcitedefaultseppunct}\relax
	\EndOfBibitem
	\bibitem[Snels \emph{et~al.}(2000)Snels, Fusina, Hollenstein, and
	Quack]{snels00}
	M.~Snels, L.~Fusina, H.~Hollenstein and M.~Quack, \emph{Mol. Phys.}, 2000,
	\textbf{98}, 837--854\relax
	\mciteBstWouldAddEndPuncttrue
	\mciteSetBstMidEndSepPunct{\mcitedefaultmidpunct}
	{\mcitedefaultendpunct}{\mcitedefaultseppunct}\relax
	\EndOfBibitem
	\bibitem[Wichmann \emph{et~al.}(2020)Wichmann, Miloglyadov, Seyfang, and
	Quack]{wichmann20}
	G.~Wichmann, E.~Miloglyadov, G.~Seyfang and M.~Quack, \emph{Mol. Phys.}, 2020,
	\textbf{118}, e1752946\relax
	\mciteBstWouldAddEndPuncttrue
	\mciteSetBstMidEndSepPunct{\mcitedefaultmidpunct}
	{\mcitedefaultendpunct}{\mcitedefaultseppunct}\relax
	\EndOfBibitem
	\bibitem[Herbst and Klemperer(1973)]{herbst73a}
	E.~Herbst and W.~Klemperer, \emph{Astrophys. J.}, 1973, \textbf{185}, 505\relax
	\mciteBstWouldAddEndPuncttrue
	\mciteSetBstMidEndSepPunct{\mcitedefaultmidpunct}
	{\mcitedefaultendpunct}{\mcitedefaultseppunct}\relax
	\EndOfBibitem
	\bibitem[Roueff(1987)]{roueff87}
	E.~Roueff, \emph{Phys. Scr.}, 1987, \textbf{36}, 319--322\relax
	\mciteBstWouldAddEndPuncttrue
	\mciteSetBstMidEndSepPunct{\mcitedefaultmidpunct}
	{\mcitedefaultendpunct}{\mcitedefaultseppunct}\relax
	\EndOfBibitem
	\bibitem[Smith(1992)]{smith92a}
	D.~Smith, \emph{Chem. Rev.}, 1992, \textbf{92}, 1473--1485\relax
	\mciteBstWouldAddEndPuncttrue
	\mciteSetBstMidEndSepPunct{\mcitedefaultmidpunct}
	{\mcitedefaultendpunct}{\mcitedefaultseppunct}\relax
	\EndOfBibitem
	\bibitem[Herbst(2001)]{herbst01a}
	E.~Herbst, \emph{{Spectroscopy from space, in: NATO science series. Series II,
			Mathematics, physics, and chemistry}}, Ed. J. Demaison \emph{et al.}, Kluwer
	Academic Publishers, Dordrecht, 2001, vol.~20\relax
	\mciteBstWouldAddEndPuncttrue
	\mciteSetBstMidEndSepPunct{\mcitedefaultmidpunct}
	{\mcitedefaultendpunct}{\mcitedefaultseppunct}\relax
	\EndOfBibitem
	\bibitem[Snow and Bierbaum(2008)]{snow08a}
	T.~P. Snow and V.~M. Bierbaum, \emph{Annu. Rev. Anal. Chem.}, 2008, \textbf{1},
	229--259\relax
	\mciteBstWouldAddEndPuncttrue
	\mciteSetBstMidEndSepPunct{\mcitedefaultmidpunct}
	{\mcitedefaultendpunct}{\mcitedefaultseppunct}\relax
	\EndOfBibitem
	\bibitem[Smith(2011)]{smith11a}
	I.~W.~M. Smith, \emph{Annu. Rev. Astron. Astrophys.}, 2011, \textbf{49},
	29--66\relax
	\mciteBstWouldAddEndPuncttrue
	\mciteSetBstMidEndSepPunct{\mcitedefaultmidpunct}
	{\mcitedefaultendpunct}{\mcitedefaultseppunct}\relax
	\EndOfBibitem
	\bibitem[Wakelam \emph{et~al.}(2010)Wakelam, Smith, Herbst, Troe, Geppert,
	Linnartz, {\"O}berg, Roueff, Ag{\'u}ndez, Pernot, Cuppen, Loison, and
	Talbi]{wakelam10a}
	V.~Wakelam, I.~W.~M. Smith, E.~Herbst, J.~Troe, W.~Geppert, H.~Linnartz,
	K.~{\"O}berg, E.~Roueff, M.~Ag{\'u}ndez, P.~Pernot, H.~M. Cuppen, J.~C.
	Loison and D.~Talbi, \emph{Space Sci. Rev.}, 2010, \textbf{156}, 13--72\relax
	\mciteBstWouldAddEndPuncttrue
	\mciteSetBstMidEndSepPunct{\mcitedefaultmidpunct}
	{\mcitedefaultendpunct}{\mcitedefaultseppunct}\relax
	\EndOfBibitem
	\bibitem[van Dishoeck(2017)]{dishoeck17}
	E.~F. van Dishoeck, \emph{Proc. Int. Astron. Union}, 2017, \textbf{13},
	3--22\relax
	\mciteBstWouldAddEndPuncttrue
	\mciteSetBstMidEndSepPunct{\mcitedefaultmidpunct}
	{\mcitedefaultendpunct}{\mcitedefaultseppunct}\relax
	\EndOfBibitem
	\bibitem[Wester(2009)]{wester09a}
	R.~Wester, \emph{J. Phys. B: At. Mol. Opt. Phys.}, 2009, \textbf{42},
	154001\relax
	\mciteBstWouldAddEndPuncttrue
	\mciteSetBstMidEndSepPunct{\mcitedefaultmidpunct}
	{\mcitedefaultendpunct}{\mcitedefaultseppunct}\relax
	\EndOfBibitem
	\bibitem[Markus \emph{et~al.}(2020)Markus, Asvany, Salomon, Schmid,
	Br{\"u}nken, Lipparini, Gauss, and Schlemmer]{markus20a}
	C.~Markus, O.~Asvany, T.~Salomon, P.~Schmid, S.~Br{\"u}nken, F.~Lipparini,
	J.~Gauss and S.~Schlemmer, \emph{Phys. Rev. Lett.}, 2020, \textbf{124},
	233401\relax
	\mciteBstWouldAddEndPuncttrue
	\mciteSetBstMidEndSepPunct{\mcitedefaultmidpunct}
	{\mcitedefaultendpunct}{\mcitedefaultseppunct}\relax
	\EndOfBibitem
	\bibitem[Marquette \emph{et~al.}(1985)Marquette, Rowe, Dupeyrat, Poissant, and
	Rebrion]{marquette85a}
	J.~B. Marquette, B.~R. Rowe, G.~Dupeyrat, G.~Poissant and C.~Rebrion,
	\emph{Chem. Phys. Lett.}, 1985, \textbf{122}, 431--435\relax
	\mciteBstWouldAddEndPuncttrue
	\mciteSetBstMidEndSepPunct{\mcitedefaultmidpunct}
	{\mcitedefaultendpunct}{\mcitedefaultseppunct}\relax
	\EndOfBibitem
	\bibitem[Rowe \emph{et~al.}(1995)Rowe, Canosa, and Page]{rowe95a}
	B.~Rowe, A.~Canosa and V.~L. Page, \emph{Int. J. Mass Spectrom. Ion Processes},
	1995, \textbf{149}, 573--596\relax
	\mciteBstWouldAddEndPuncttrue
	\mciteSetBstMidEndSepPunct{\mcitedefaultmidpunct}
	{\mcitedefaultendpunct}{\mcitedefaultseppunct}\relax
	\EndOfBibitem
	\bibitem[Willitsch(2017)]{willitsch17a}
	S.~Willitsch, \emph{{Chemistry with Controlled Ions, Adv. Chem. Phys., eds. S.
			A. Rice and A. R. Dinner}}, 2017, vol. 162, pp. 307--340\relax
	\mciteBstWouldAddEndPuncttrue
	\mciteSetBstMidEndSepPunct{\mcitedefaultmidpunct}
	{\mcitedefaultendpunct}{\mcitedefaultseppunct}\relax
	\EndOfBibitem
	\bibitem[Petralia \emph{et~al.}(2020)Petralia, Tsikritea, Loreau, Softley, and
	Heazlewood]{petralia20a}
	L.~S. Petralia, A.~Tsikritea, J.~Loreau, T.~P. Softley and B.~R. Heazlewood,
	\emph{Nat. Commun.}, 2020, \textbf{11}, 173\relax
	\mciteBstWouldAddEndPuncttrue
	\mciteSetBstMidEndSepPunct{\mcitedefaultmidpunct}
	{\mcitedefaultendpunct}{\mcitedefaultseppunct}\relax
	\EndOfBibitem
	\bibitem[Toscano \emph{et~al.}(2020)Toscano, Lewandowski, and
	Heazlewood]{toscano20}
	J.~Toscano, H.~J. Lewandowski and B.~R. Heazlewood, \emph{Phys. Chem. Chem.
		Phys.}, 2020, \textbf{22}, 9180--9194\relax
	\mciteBstWouldAddEndPuncttrue
	\mciteSetBstMidEndSepPunct{\mcitedefaultmidpunct}
	{\mcitedefaultendpunct}{\mcitedefaultseppunct}\relax
	\EndOfBibitem
	\bibitem[Allmendinger \emph{et~al.}(2016)Allmendinger, Deiglmayr, Schullian,
	H{\"o}veler, Agner, Schmutz, and Merkt]{allmendinger16a}
	P.~Allmendinger, J.~Deiglmayr, O.~Schullian, K.~H{\"o}veler, J.~A. Agner,
	H.~Schmutz and F.~Merkt, \emph{ChemPhysChem}, 2016, \textbf{17},
	3596--3608\relax
	\mciteBstWouldAddEndPuncttrue
	\mciteSetBstMidEndSepPunct{\mcitedefaultmidpunct}
	{\mcitedefaultendpunct}{\mcitedefaultseppunct}\relax
	\EndOfBibitem
	\bibitem[Allmendinger \emph{et~al.}(2016)Allmendinger, Deiglmayr, H{\"o}veler,
	Schullian, and Merkt]{allmendinger16b}
	P.~Allmendinger, J.~Deiglmayr, K.~H{\"o}veler, O.~Schullian and F.~Merkt,
	\emph{J. Chem. Phys.}, 2016, \textbf{145}, 244316\relax
	\mciteBstWouldAddEndPuncttrue
	\mciteSetBstMidEndSepPunct{\mcitedefaultmidpunct}
	{\mcitedefaultendpunct}{\mcitedefaultseppunct}\relax
	\EndOfBibitem
	\bibitem[Zhelyazkova \emph{et~al.}(2020)Zhelyazkova, Martins, Agner, Schmutz,
	and Merkt]{zhelyazkova20}
	V.~Zhelyazkova, F.~B.~V. Martins, J.~A. Agner, H.~Schmutz and F.~Merkt,
	\emph{Phys. Rev. Lett.}, 2020, \textbf{125}, 263401\relax
	\mciteBstWouldAddEndPuncttrue
	\mciteSetBstMidEndSepPunct{\mcitedefaultmidpunct}
	{\mcitedefaultendpunct}{\mcitedefaultseppunct}\relax
	\EndOfBibitem
	\bibitem[H{\"o}veler \emph{et~al.}(2021)H{\"o}veler, Deiglmayr, Agner, Schmutz,
	and Merkt]{hoeveler21a}
	K.~H{\"o}veler, J.~Deiglmayr, J.~A. Agner, H.~Schmutz and F.~Merkt, \emph{Phys.
		Chem. Chem. Phys.}, 2021, \textbf{23}, 2676--2683\relax
	\mciteBstWouldAddEndPuncttrue
	\mciteSetBstMidEndSepPunct{\mcitedefaultmidpunct}
	{\mcitedefaultendpunct}{\mcitedefaultseppunct}\relax
	\EndOfBibitem
	\bibitem[Pratt \emph{et~al.}(1994)Pratt, Dehmer, Dehmer, and Chupka]{pratt94a}
	S.~T. Pratt, J.~L. Dehmer, P.~M. Dehmer and W.~A. Chupka, \emph{J. Chem.
		Phys.}, 1994, \textbf{101}, 882--890\relax
	\mciteBstWouldAddEndPuncttrue
	\mciteSetBstMidEndSepPunct{\mcitedefaultmidpunct}
	{\mcitedefaultendpunct}{\mcitedefaultseppunct}\relax
	\EndOfBibitem
	\bibitem[Wrede \emph{et~al.}(2005)Wrede, Schnieder, Seekamp-Schnieder,
	Niederjohann, and Welge]{wrede05a}
	E.~Wrede, L.~Schnieder, K.~Seekamp-Schnieder, B.~Niederjohann and K.~H. Welge,
	\emph{Phys. Chem. Chem. Phys.}, 2005, \textbf{7}, 1577--1582\relax
	\mciteBstWouldAddEndPuncttrue
	\mciteSetBstMidEndSepPunct{\mcitedefaultmidpunct}
	{\mcitedefaultendpunct}{\mcitedefaultseppunct}\relax
	\EndOfBibitem
	\bibitem[Matsuzawa(2010)]{matsuzawa10a}
	M.~Matsuzawa, \emph{Phys. Rev. A}, 2010, \textbf{82}, 054701\relax
	\mciteBstWouldAddEndPuncttrue
	\mciteSetBstMidEndSepPunct{\mcitedefaultmidpunct}
	{\mcitedefaultendpunct}{\mcitedefaultseppunct}\relax
	\EndOfBibitem
	\bibitem[Hogan \emph{et~al.}(2012)Hogan, Allmendinger, Sassmannshausen,
	Schmutz, and Merkt]{hogan12b}
	S.~D. Hogan, P.~Allmendinger, H.~Sassmannshausen, H.~Schmutz and F.~Merkt,
	\emph{Phys. Rev. Lett.}, 2012, \textbf{108}, 063008\relax
	\mciteBstWouldAddEndPuncttrue
	\mciteSetBstMidEndSepPunct{\mcitedefaultmidpunct}
	{\mcitedefaultendpunct}{\mcitedefaultseppunct}\relax
	\EndOfBibitem
	\bibitem[Allmendinger \emph{et~al.}(2014)Allmendinger, Deiglmayr, Agner,
	Schmutz, and Merkt]{allmendinger14a}
	P.~Allmendinger, J.~Deiglmayr, J.~A. Agner, H.~Schmutz and F.~Merkt,
	\emph{Phys. Rev. A}, 2014, \textbf{90}, 043403\relax
	\mciteBstWouldAddEndPuncttrue
	\mciteSetBstMidEndSepPunct{\mcitedefaultmidpunct}
	{\mcitedefaultendpunct}{\mcitedefaultseppunct}\relax
	\EndOfBibitem
	\bibitem[Hogan(2016)]{hogan16a}
	S.~D. Hogan, \emph{EPJ Techniques and Instrumentation}, 2016, \textbf{3},
	2\relax
	\mciteBstWouldAddEndPuncttrue
	\mciteSetBstMidEndSepPunct{\mcitedefaultmidpunct}
	{\mcitedefaultendpunct}{\mcitedefaultseppunct}\relax
	\EndOfBibitem
	\bibitem[Zhelyazkova \emph{et~al.}(2019)Zhelyazkova, {\v{Z}}e{\v{s}}ko,
	Schmutz, Agner, and Merkt]{zhelyazkova19a}
	V.~Zhelyazkova, M.~{\v{Z}}e{\v{s}}ko, H.~Schmutz, J.~A. Agner and F.~Merkt,
	\emph{Molecular Physics}, 2019, \textbf{117}, 2980--2989\relax
	\mciteBstWouldAddEndPuncttrue
	\mciteSetBstMidEndSepPunct{\mcitedefaultmidpunct}
	{\mcitedefaultendpunct}{\mcitedefaultseppunct}\relax
	\EndOfBibitem
	\bibitem[Shagam and Narevicius(2013)]{shagam13a}
	Y.~Shagam and E.~Narevicius, \emph{J. Phys. Chem. C}, 2013, \textbf{117},
	22454--22461\relax
	\mciteBstWouldAddEndPuncttrue
	\mciteSetBstMidEndSepPunct{\mcitedefaultmidpunct}
	{\mcitedefaultendpunct}{\mcitedefaultseppunct}\relax
	\EndOfBibitem
	\bibitem[Shagam \emph{et~al.}(2015)Shagam, Klein, Skomorowski, Yun, Averbukh,
	Koch, and Narevicius]{shagam15a}
	Y.~Shagam, A.~Klein, W.~Skomorowski, R.~Yun, V.~Averbukh, {\relax Ch.}.~P. Koch
	and E.~Narevicius, \emph{Nat. Chem.}, 2015, \textbf{7}, 921--926\relax
	\mciteBstWouldAddEndPuncttrue
	\mciteSetBstMidEndSepPunct{\mcitedefaultmidpunct}
	{\mcitedefaultendpunct}{\mcitedefaultseppunct}\relax
	\EndOfBibitem
	\bibitem[Bibelnik \emph{et~al.}(2019)Bibelnik, Gersten, Henson, Lavert-Ofir,
	Shagam, Skomorowski, Koch, and Narevicius]{bibelnik19a}
	N.~Bibelnik, S.~Gersten, A.~B. Henson, E.~Lavert-Ofir, Y.~Shagam,
	W.~Skomorowski, {\relax Ch.}.~P. Koch and E.~Narevicius, \emph{Mol. Phys.},
	2019, \textbf{117}, 2128--2137\relax
	\mciteBstWouldAddEndPuncttrue
	\mciteSetBstMidEndSepPunct{\mcitedefaultmidpunct}
	{\mcitedefaultendpunct}{\mcitedefaultseppunct}\relax
	\EndOfBibitem
	\bibitem[Jankunas \emph{et~al.}(2014)Jankunas, Bertsche, Jachymski, Hapka, and
	Osterwalder]{jankunas14a}
	J.~Jankunas, B.~Bertsche, K.~Jachymski, M.~Hapka and A.~Osterwalder, \emph{J.
		Chem. Phys.}, 2014, \textbf{140}, 244302\relax
	\mciteBstWouldAddEndPuncttrue
	\mciteSetBstMidEndSepPunct{\mcitedefaultmidpunct}
	{\mcitedefaultendpunct}{\mcitedefaultseppunct}\relax
	\EndOfBibitem
	\bibitem[Jankunas and Osterwalder(2015)]{jankunas15c}
	J.~Jankunas and A.~Osterwalder, \emph{Ann. Rev. Phys. Chem.}, 2015,
	\textbf{66}, 241--262\relax
	\mciteBstWouldAddEndPuncttrue
	\mciteSetBstMidEndSepPunct{\mcitedefaultmidpunct}
	{\mcitedefaultendpunct}{\mcitedefaultseppunct}\relax
	\EndOfBibitem
	\bibitem[Gordon and Osterwalder(2020)]{gordon20a}
	S.~D.~S. Gordon and A.~Osterwalder, \emph{Intl. Rev. Phys. Chem.}, 2020,
	\textbf{39}, 109--134\relax
	\mciteBstWouldAddEndPuncttrue
	\mciteSetBstMidEndSepPunct{\mcitedefaultmidpunct}
	{\mcitedefaultendpunct}{\mcitedefaultseppunct}\relax
	\EndOfBibitem
	\bibitem[Langevin(1905)]{langevin05a}
	P.~Langevin, \emph{Annales de Chimie et de Physique}, 1905, \textbf{T5},
	245--288\relax
	\mciteBstWouldAddEndPuncttrue
	\mciteSetBstMidEndSepPunct{\mcitedefaultmidpunct}
	{\mcitedefaultendpunct}{\mcitedefaultseppunct}\relax
	\EndOfBibitem
	\bibitem[Cheung \emph{et~al.}(1968)Cheung, Rank, Townes, Thornton, and
	Welch]{cheung68}
	A.~C. Cheung, D.~M. Rank, C.~H. Townes, D.~D. Thornton and W.~J. Welch,
	\emph{Phys. Rev. Lett.}, 1968, \textbf{21}, 1701--1705\relax
	\mciteBstWouldAddEndPuncttrue
	\mciteSetBstMidEndSepPunct{\mcitedefaultmidpunct}
	{\mcitedefaultendpunct}{\mcitedefaultseppunct}\relax
	\EndOfBibitem
	\bibitem[Roueff \emph{et~al.}(2005)Roueff, Lis, van~der Tak, G{\'e}rin, and
	Goldsmith]{roueff05}
	E.~Roueff, D.~C. Lis, F.~F.~S. van~der Tak, M.~G{\'e}rin and P.~F. Goldsmith,
	\emph{Astron. Astrophys.}, 2005, \textbf{438}, 585--598\relax
	\mciteBstWouldAddEndPuncttrue
	\mciteSetBstMidEndSepPunct{\mcitedefaultmidpunct}
	{\mcitedefaultendpunct}{\mcitedefaultseppunct}\relax
	\EndOfBibitem
	\bibitem[Hermsen \emph{et~al.}(1985)Hermsen, Wilson, Walmsley, and
	Batrla]{hermsen85}
	W.~Hermsen, T.~L. Wilson, C.~M. Walmsley and W.~Batrla, \emph{Astron.
		Astrophys.}, 1985, \textbf{146}, 134--138\relax
	\mciteBstWouldAddEndPuncttrue
	\mciteSetBstMidEndSepPunct{\mcitedefaultmidpunct}
	{\mcitedefaultendpunct}{\mcitedefaultseppunct}\relax
	\EndOfBibitem
	\bibitem[Martin and Ho(1979)]{martin79}
	R.~N. Martin and P.~T.~P. Ho, \emph{Astron. Astrophys.}, 1979, \textbf{74},
	L7--L9\relax
	\mciteBstWouldAddEndPuncttrue
	\mciteSetBstMidEndSepPunct{\mcitedefaultmidpunct}
	{\mcitedefaultendpunct}{\mcitedefaultseppunct}\relax
	\EndOfBibitem
	\bibitem[Salinas \emph{et~al.}(2016)Salinas, Hogerheijde, Bergin, Cleeves,
	Brinch, Blake, Lis, Melnick, Pani{\'c}, Pearson, Kristensen, Y{\i}ld{\i}z,
	and van Dishoeck]{salinas16}
	V.~N. Salinas, M.~R. Hogerheijde, E.~A. Bergin, L.~I. Cleeves, C.~Brinch, G.~A.
	Blake, D.~C. Lis, G.~J. Melnick, O.~Pani{\'c}, J.~C. Pearson, L.~Kristensen,
	U.~A. Y{\i}ld{\i}z and E.~F. van Dishoeck, \emph{Astron. Astrophys.}, 2016,
	\textbf{591}, A122\relax
	\mciteBstWouldAddEndPuncttrue
	\mciteSetBstMidEndSepPunct{\mcitedefaultmidpunct}
	{\mcitedefaultendpunct}{\mcitedefaultseppunct}\relax
	\EndOfBibitem
	\bibitem[Irwin \emph{et~al.}(2018)Irwin, Bowles, Braude, Garland, and
	Calcutt]{irwin18}
	P.~G.~J. Irwin, N.~Bowles, A.~S. Braude, R.~Garland and S.~Calcutt,
	\emph{Icarus}, 2018, \textbf{302}, 426--436\relax
	\mciteBstWouldAddEndPuncttrue
	\mciteSetBstMidEndSepPunct{\mcitedefaultmidpunct}
	{\mcitedefaultendpunct}{\mcitedefaultseppunct}\relax
	\EndOfBibitem
	\bibitem[Danby \emph{et~al.}(1988)Danby, Flower, Valiron, Schilke, and
	Walmsley]{danby88}
	G.~Danby, D.~R. Flower, P.~Valiron, P.~Schilke and M.~C. Walmsley, \emph{Mon.
		Not. R. astr. Soc.}, 1988, \textbf{235}, 229--238\relax
	\mciteBstWouldAddEndPuncttrue
	\mciteSetBstMidEndSepPunct{\mcitedefaultmidpunct}
	{\mcitedefaultendpunct}{\mcitedefaultseppunct}\relax
	\EndOfBibitem
	\bibitem[Izotov and Thuan(2010)]{izotov10}
	Y.~I. Izotov and T.~X. Thuan, \emph{Astrophys. J.}, 2010, \textbf{710},
	L67--L71\relax
	\mciteBstWouldAddEndPuncttrue
	\mciteSetBstMidEndSepPunct{\mcitedefaultmidpunct}
	{\mcitedefaultendpunct}{\mcitedefaultseppunct}\relax
	\EndOfBibitem
	\bibitem[Bolden \emph{et~al.}(1970)Bolden, Hemsworth, Shaw, and
	Twiddy]{bolden70}
	R.~C. Bolden, R.~S. Hemsworth, M.~J. Shaw and N.~D. Twiddy, \emph{J. Phys. B:
		At. Mol. Opt. Phys.}, 1970, \textbf{3}, 45--60\relax
	\mciteBstWouldAddEndPuncttrue
	\mciteSetBstMidEndSepPunct{\mcitedefaultmidpunct}
	{\mcitedefaultendpunct}{\mcitedefaultseppunct}\relax
	\EndOfBibitem
	\bibitem[Lindinger \emph{et~al.}(1975)Lindinger, Albritton, and
	Fehsenfeld]{lindinger75}
	W.~Lindinger, D.~L. Albritton and F.~C. Fehsenfeld, \emph{J. Chem. Phys.},
	1975, \textbf{62}, 4957--4958\relax
	\mciteBstWouldAddEndPuncttrue
	\mciteSetBstMidEndSepPunct{\mcitedefaultmidpunct}
	{\mcitedefaultendpunct}{\mcitedefaultseppunct}\relax
	\EndOfBibitem
	\bibitem[Kim and Huntress(1975)]{kim75}
	J.~K. Kim and W.~T. Huntress, \emph{Int. J. Mass Spectrom. Ion Physics}, 1975,
	\textbf{16}, 451--454\relax
	\mciteBstWouldAddEndPuncttrue
	\mciteSetBstMidEndSepPunct{\mcitedefaultmidpunct}
	{\mcitedefaultendpunct}{\mcitedefaultseppunct}\relax
	\EndOfBibitem
	\bibitem[Ruscic \emph{et~al.}(2005)Ruscic, Pinzon, von Laszewski, Kodeboyina,
	Burcat, Leahy, Montoy, and Wagner]{ruscic05}
	B.~Ruscic, R.~E. Pinzon, G.~von Laszewski, D.~Kodeboyina, A.~Burcat, D.~Leahy,
	D.~Montoy and A.~F. Wagner, \emph{J. Phys. Conf. Ser.}, 2005, \textbf{16},
	561--570\relax
	\mciteBstWouldAddEndPuncttrue
	\mciteSetBstMidEndSepPunct{\mcitedefaultmidpunct}
	{\mcitedefaultendpunct}{\mcitedefaultseppunct}\relax
	\EndOfBibitem
	\bibitem[Wong(1998)]{wong98}
	S.~S.~M. Wong, \emph{Introductory Nuclear Physics}, John Wiley \& Sons,
	1998\relax
	\mciteBstWouldAddEndPuncttrue
	\mciteSetBstMidEndSepPunct{\mcitedefaultmidpunct}
	{\mcitedefaultendpunct}{\mcitedefaultseppunct}\relax
	\EndOfBibitem
	\bibitem[Zare(1988)]{zare88a}
	R.~N. Zare, \emph{{Angular Momentum}}, John Wiley \& Sons, New York, 1988\relax
	\mciteBstWouldAddEndPuncttrue
	\mciteSetBstMidEndSepPunct{\mcitedefaultmidpunct}
	{\mcitedefaultendpunct}{\mcitedefaultseppunct}\relax
	\EndOfBibitem
	\bibitem[\L{}ach and Pachucki(2001)]{lach01}
	G.~\L{}ach and K.~Pachucki, \emph{Phys. Rev. A}, 2001, \textbf{64},
	042510\relax
	\mciteBstWouldAddEndPuncttrue
	\mciteSetBstMidEndSepPunct{\mcitedefaultmidpunct}
	{\mcitedefaultendpunct}{\mcitedefaultseppunct}\relax
	\EndOfBibitem
	\bibitem[Halfmann \emph{et~al.}(2000)Halfmann, Koensgen, and
	Bergmann]{halfmann00a}
	T.~Halfmann, J.~Koensgen and K.~Bergmann, \emph{Meas. Sci. Technol.}, 2000,
	\textbf{11}, 1510--1514\relax
	\mciteBstWouldAddEndPuncttrue
	\mciteSetBstMidEndSepPunct{\mcitedefaultmidpunct}
	{\mcitedefaultendpunct}{\mcitedefaultseppunct}\relax
	\EndOfBibitem
	\bibitem[Gallagher(1994)]{gallagher94a}
	T.~F. Gallagher, \emph{{Rydberg Atoms}}, Cambridge University Press, Cambridge,
	1994\relax
	\mciteBstWouldAddEndPuncttrue
	\mciteSetBstMidEndSepPunct{\mcitedefaultmidpunct}
	{\mcitedefaultendpunct}{\mcitedefaultseppunct}\relax
	\EndOfBibitem
	\bibitem[Herzberg(1991)]{herzberg91a}
	G.~Herzberg, \emph{{Molecular Spectra and Molecular Structure, Volume II,
			Infrared and Raman Spectra of Polyatomic Molecules}}, Krieger Publishing
	Company, Malabar, 1991\relax
	\mciteBstWouldAddEndPuncttrue
	\mciteSetBstMidEndSepPunct{\mcitedefaultmidpunct}
	{\mcitedefaultendpunct}{\mcitedefaultseppunct}\relax
	\EndOfBibitem
	\bibitem[\v{S}. Urban \emph{et~al.}(1984)\v{S}. Urban, D'Cunha, {Narahari Rao},
	and {Papou\v{s}ek}]{urban84}
	\v{S}. Urban, R.~D'Cunha, K.~{Narahari Rao} and D.~{Papou\v{s}ek}, \emph{Can.
		J. Phys.}, 1984, \textbf{62}, 1775--1791\relax
	\mciteBstWouldAddEndPuncttrue
	\mciteSetBstMidEndSepPunct{\mcitedefaultmidpunct}
	{\mcitedefaultendpunct}{\mcitedefaultseppunct}\relax
	\EndOfBibitem
	\bibitem[Daniel \emph{et~al.}(2016)Daniel, Rist, Faure, Roueff, G{\'e}rin, Lis,
	Hily-Blant, Bacmann, and Wiesenfeld]{daniel16}
	F.~Daniel, C.~Rist, A.~Faure, E.~Roueff, M.~G{\'e}rin, D.~C. Lis,
	P.~Hily-Blant, A.~Bacmann and L.~Wiesenfeld, \emph{Mon. Not. R. Astron.
		Soc.}, 2016, \textbf{457}, 1535--1549\relax
	\mciteBstWouldAddEndPuncttrue
	\mciteSetBstMidEndSepPunct{\mcitedefaultmidpunct}
	{\mcitedefaultendpunct}{\mcitedefaultseppunct}\relax
	\EndOfBibitem
	\bibitem[{van de Meerakker} \emph{et~al.}(2012){van de Meerakker}, Bethlem,
	Vanhaecke, and Meijer]{vandemeerakker12a}
	S.~Y.~T. {van de Meerakker}, H.~L. Bethlem, N.~Vanhaecke and G.~Meijer,
	\emph{Chem. Rev.}, 2012, \textbf{112}, 4828--4878\relax
	\mciteBstWouldAddEndPuncttrue
	\mciteSetBstMidEndSepPunct{\mcitedefaultmidpunct}
	{\mcitedefaultendpunct}{\mcitedefaultseppunct}\relax
	\EndOfBibitem
	\bibitem[Levine(2005)]{levine05}
	R.~D. Levine, \emph{Molecular Reaction Dynamics}, Cambridge University Press,
	2005\relax
	\mciteBstWouldAddEndPuncttrue
	\mciteSetBstMidEndSepPunct{\mcitedefaultmidpunct}
	{\mcitedefaultendpunct}{\mcitedefaultseppunct}\relax
	\EndOfBibitem
	\bibitem[{Manca Tanner} and Quack(2012)]{tanner12}
	C.~{Manca Tanner} and M.~Quack, \emph{Mol. Phys.}, 2012, \textbf{110},
	2111--2135\relax
	\mciteBstWouldAddEndPuncttrue
	\mciteSetBstMidEndSepPunct{\mcitedefaultmidpunct}
	{\mcitedefaultendpunct}{\mcitedefaultseppunct}\relax
	\EndOfBibitem
	\bibitem[Ploenes \emph{et~al.}(2021)Ploenes, Stra{\v n}{\'a}k, Gao, K{\"u}pper,
	and Willitsch]{ploenes21}
	L.~Ploenes, P.~Stra{\v n}{\'a}k, H.~Gao, J.~K{\"u}pper and S.~Willitsch,
	\emph{Molecular Physics}, 2021, \textbf{119}, e1965234\relax
	\mciteBstWouldAddEndPuncttrue
	\mciteSetBstMidEndSepPunct{\mcitedefaultmidpunct}
	{\mcitedefaultendpunct}{\mcitedefaultseppunct}\relax
	\EndOfBibitem
\end{mcitethebibliography}
%\bibliographystyle{rsc} %the RSC's .bst file

\providecommand*{\mcitethebibliography}{\thebibliography}
\csname @ifundefined\endcsname{endmcitethebibliography}
{\let\endmcitethebibliography\endthebibliography}{}

\end{document}